\documentclass[english,aps,prb,reprint,superscriptaddress,longbibliography]{revtex4-2}
\usepackage[T1]{fontenc}
\usepackage[latin9]{inputenc}
\usepackage{amsmath}
\usepackage{amssymb}
\usepackage{graphicx}

\makeatletter
\usepackage{color}
\usepackage{hyperref}

\makeatother

\usepackage{babel}
\begin{document}
\title{Dynamical characterization of $Z_{2}$ Floquet topological phases
via quantum quenches}
\author{Lin Zhang}
\thanks{lin.zhang@icfo.eu}
\affiliation{ICFO-Institut de Ciencies Fotoniques, The Barcelona Institute of Science
and Technology, Av. Carl Friedrich Gauss 3, 08860 Castelldefels (Barcelona),
Spain}
\begin{abstract}
The complete characterization of a generic $d$-dimensional Floquet topological phase is usually
hard for the requirement of information about the micromotion throughout
the entire driving period. In a recent work [L. Zhang {\it et al}., Phys. Rev. Lett. {\bf 125}, 183001 (2020)], 
an experimentally feasible dynamical detection scheme was proposed to characterize the integer Floquet topological phases using quantum quenches. 
However, this theory is still far away from completion, especially for free-fermion Floquet topological phases, where the states can also be characterized by $Z_{2}$ invariants.
Here we develop the first full and unified dynamical
characterization theory for the $Z_{2}$ Floquet topological phases
of different dimensionality and tenfold-way symmetry classes
by quenching the system from a trivial and static initial state to
the Floquet topological regime through suddenly changing the parameters
and turning on the periodic driving. By measuring the minimal information
of Floquet bands via the stroboscopic time-averaged spin polarizations,
we show that the topological spin texture patterns emerging on certain
discrete momenta of Brillouin zone called the $0$ or $\pi$ gap highest-order
band-inversion surfaces provide a measurable dynamical $Z_{2}$ Floquet
invariant, which uniquely determines the Floquet boundary modes in
the corresponding quasienergy gap and characterizes the $Z_{2}$ Floquet topology. The applications of our theory
are illustrated via one- and two-dimensional models that are accessible
in current quantum simulation experiments. Our work provides a highly
feasible way to detect the $Z_{2}$ Floquet topology 
and completes the dynamical characterization for the full tenfold classes of Floquet topological phases,
which shall advance the research in theory and experiments.
\end{abstract}
\maketitle

\section{Introduction}

The discovery of topological quantum states~\citep{Laughlin1981,Tsui1982,Laughlin1983}
has revolutionized the classification of fundamental phases of quantum
matter. One of the most famous examples is the integer quantum Hall
effect in a two-dimensional (2D) electron gas~\citep{Klitzing1980},
where the Hall conductance is quantized and is proportional to the
Thouless-Kohmoto-Nightingale-den Nijs (TKNN) number~\citep{Thouless1982},
a topological invariant that depends only on the global property of
the equilibrium ground state. This is sharply distinct from the celebrated
Landau-Ginzburg-Wilson framework~\citep{Landau2013}, where the quantum
phases are classified by symmetry breaking and are characterized by
local order parameters. The studies of topological quantum phases
have further become a mainstream of research in condensed matter physics
since the discovery of topological insulators~\citep{Kane2005,Kane2005a,Bernevig2006,Konig2007},
and a full classification of free-fermion topological systems has
been achieved through the tenfold way periodic table~\citep{Altland1997,Schnyder2008,Kitaev2009,Ryu2010},
where each topological phase is either characterized by an integer
invariant or by a $Z_{2}$ topological index. The most salient and
ubiquitous feature of these states is the so-called bulk-boundary
correspondence, which states that a system with nontrivial bulk topology
shall host protected gapless modes at the boundary~\citep{Hasan2010,Qi2011,Chiu2016}.
This correspondence is the foundation for many experimental detection
schemes in quantum materials, such as the transport measurement~\citep{Konig2007,Chang2013,He2017}
and angle resolved photoemission spectroscopy~\citep{Hsieh2008,Chen2009,Xia2009}.

Not only restricted in equilibrium systems, the notion of topological
phases has also been proposed and realized in Floquet systems~\citep{Oka2009,Kitagawa2010,Kitagawa2010a,Inoue2010,Jiang2011,Lindner2011,Kitagawa2011,Wang2013,Mahmood2016,Jotzu2014,Flaeschner2016,Wintersperger2020,Slager2022},
where the periodic driving provides a versatile approach to create
and engineer topological matters~\citep{Cayssol2013,Nag2019,Harper2020,Rudner2020,Jangjan2020,Molignini2018,Molignini2020,Molignini2020a,Molignini2021,Nag2021,Ghosh2022,Jangjan2022}.
Like the static systems, the periodically driven matters can be characterized
by quasienergy states with Floquet bands obtained from the effective
Floquet Hamiltonian~\citep{Bukov2015,Eckardt2017}. However, the
Floquet topological phases cannot be fully understood via the topological
indices from Floquet Hamiltonian due to the additional periodicity
in time domain, and anomalous Floquet topological phases have been
observed, where the number of edge modes within each bulk band gap
is not uniquely determined by the total topological invariants of
Floquet bands below this gap~\citep{Kitagawa2010a,Jiang2011,Kitagawa2012}.
To characterize the Floquet topological phases, new topological invariants
involving the micromotion throughout the entire driving period, captured
either by the unitary evolution operator~\citep{Rudner2013,Yao2017} (equivalently, 
we can also consider a set of micromotion-parameterized effective Floquet Hamiltonians~\citep{Xu2022})
or by the topologically protected singularities in the so-called phase
bands~\citep{Nathan2015}, have been proposed, which further provide
a tenfold way classification of the free-fermion Floquet topological
phases~\citep{Roy2017,Yao2017}. However, it is still challenging
to detect a generic Floquet topological state in practical experiments
due to the complexity of these invariants.

Recently, considerable efforts have been devoted to detecting topological
quantum phases via quench dynamics~\citep{Wang2017,Tarnowski2019,Uenal2020,Mizoguchi2021}
and studying dynamical topological phenomena therein~\citep{Budich2016,Yang2018,Gong2018,McGinley2018,Qiu2019,Wang2019,McGinley2019,Hu2020,Sim2022}. In particular, a dynamical detection scheme was proposed in Ref.~\citep{Zhang2020} and following studies~\citep{Zhang2021,Wang2023} to characterize the Floquet topological phases using quantum quenches, which is built on the Floquet generalization of the so-called dynamical bulk-surface correspondence for equilibrium topological phases~\citep{Zhang2018,Zhang2019,Zhang2019a,Yu2021,Zhang2022}. This correspondence shows that the bulk topology of a generic $d$D system has a universal one-to-one correspondence to the dynamical topological
patterns emerging on certain lower-dimensional momentum subspace called
(higher-order) band-inversion surfaces (BISs) after quenching the
system from a trivial phase to the topological regime, providing the dynamical characterization of topological phases, which can be even generalized to non-Hermitian~\citep{Zhou2018,Zhu2020} and higher-order~\citep{Li2021,Niu2021,Lei2022,Jia2022} topological systems. As the dynamical topological patterns on BISs can be measured directly~\citep{Sun2018,Yi2019,Song2019,Wang2019a,Ji2020,Xin2020,Wang2021,Liang2021,Yu2021a}, the dynamical characterization scheme of Floquet topological phases proposed in Ref.~\citep{Zhang2020} is highly feasible and has been demonstrated in experiments very recently~\citep{Chen2020,Zhang2023}. Nevertheless, this scheme so far is only applicable to periodically driven systems with integer invariants and is still far away from completion, especially
for the free-fermion Floquet topological systems, where the states can also be characterized by
$Z_{2}$ invariants. Whether a general framework of dynamical characterization theory can be established
for these Floquet topological phases is an open question, whose resolution is crucial to advance this research topic 
and benefits the experimental studies.

In this work, we establish the first full and unified dynamical characterization theory for $Z_{2}$ Floquet topological phases of 
different dimensionality and tenfold-way symmetry classes
based on a recent work for equilibrium $Z_{2}$ topological phases~\citep{Zhang2022}.
By introducing the highest-order BISs in the $0$ and $\pi$ quasienergy gaps, we
show that a generic $d$D $Z_{2}$ Floquet topological phase can be
characterized by the $0$D topological patterns emerging in the stroboscopic
time-averaged spin textures after quenching the system from a trivial
and static initial state to the topological regime through suddenly
changing the parameters and turning on the periodic driving. The dynamical
invariants on these two types of BISs uniquely determine the Floquet
edge modes in the $0$ and $\pi$ quasienergy gaps, respectively,
hence providing a full and feasible dynamical characterization of
the $Z_{2}$ Floquet topological phases both in the conventional and anomalous sense. We illustrate the applications
of this theory via the one- and two-dimensional models.
Together with Ref.~\citep{Zhang2020}, our work completes the
dynamical characterization for the full tenfold classes of Floquet topological phases, which shall 
advance the research in theory and experiments.

The remaining part of this article is organized as follows. In Sec.~\ref{sec:dynamical_characterization_for_1D_Floquet_topological_phases},
we first illustrate the characterization scheme via a simple 1D periodically
driven model. Then the central result of this work is presented in Sec.~\ref{sec:generic dynamical characterization schemes},
where we provide the dynamical characterization theory
for a generic $d$D $Z_{2}$ Floquet topological phase of the tenfold-way symmetry classes and further show its application in two dimensions.
We also discuss the validity and modifications of our theory for driven systems symmetric
about a generic reference time $t_{*}$ instead of the trigger time
$t=0$ of quantum quenches in Sec.~\ref{sec:floquet topological systems with generic reference time}.
Finally, the conclusion is provided in Sec.~\ref{sec:conclusion}.
More details are shown in the appendices.

\section{illustration of dynamical characterization via 1D Floquet topological phases\label{sec:dynamical_characterization_for_1D_Floquet_topological_phases}}

In this section, we start with a simple $1$D periodically driven
model to illustrate the dynamical characterization scheme. The corresponding
Hamiltonian is given by
\begin{equation}
H(k,t)=[\mu(t)-2t_{0}\cos k]\sigma_{z}+2\Delta(\sin k\sigma_{x}+\sin2k\sigma_{y})\label{eq:1D_Hamiltonian_of_class_D}
\end{equation}
with the driving $\mu(t)=\mu_{0}+\mu_{\mathrm{d}}\cos\omega t$ of
period $T=2\pi/\omega$. Here $t_{0}$ is the hopping coefficient,
and $\mu_{0,\mathrm{d}}$ (or $\Delta$) resembles the Zeeman potential
(spin-orbit coupling coefficient). The Pauli matrices are denoted
by $\sigma_{x,y,z}$ and are referred to as pseudo spins. To facilitate
the discussions, we divide the above Hamiltonian into the static part
$H_{\mathrm{s}}=\boldsymbol{h}\cdot\boldsymbol{\sigma}$ with $\mu_{\mathrm{d}}=0$
and the periodically driven part $V(t)=\mu_{\mathrm{d}}\cos\omega t\sigma_{z}$.
Due to the particle-hole symmetry $\Xi H(k,t)\Xi^{-1}=-H(-k,t)$ with
$\Xi=\sigma_{x}K$, where $K$ is the complex conjugation operator,
the Floquet topological phase \eqref{eq:1D_Hamiltonian_of_class_D}
belongs to class D and is characterized by $Z_{2}$ invariants~\citep{Nathan2015,Roy2017,Yao2017}.
Usually, detecting such a phase is quite involved and requires the
information in the whole momentum-time space, which is hard to access
in practical experiments. To address this issue, here we introduce
a highly feasible scheme to characterize the bulk $Z_{2}$ Floquet
topology based on the higher-order BISs. 

\subsection{Band-inversion surfaces for Floquet systems}

The concept of (higher-order) BISs was at first proposed to characterize
the equilibrium integer topological phases~\citep{Zhang2018,Yu2021},
where the $n$-th order BISs (denoted as $n\text{-BISs}$) are defined
as the momentum subspaces with $n$ vanishing Hamiltonian components.
Recently, this concept has also been adopted to describe the static
$Z_{2}$ topological phase by connecting it to a higher-dimensional
integer topological phase of the same symmetry class via dimension
extension~\citep{Zhang2022}, and the BISs for the $Z_{2}$ topology
are inherited from the corresponding parent integer phase. For the
static topological phase $H_{{\rm s}}$ without periodic driving,
the $0$D highest-order $2\text{-BIS}$~\citep{2-BIS} exists if
$|\mu_{0}|<2t_{0}$ and corresponds to the momenta of Brillouin zone (BZ) with $h_{z}(k)=0$,
on which the decoupled $h_{z}$-bands cross with each other while
the nonzero field $h_{x,y}$ opens a topological gap and characterizes
the static $Z_{2}$ topology.

\begin{figure}
\includegraphics{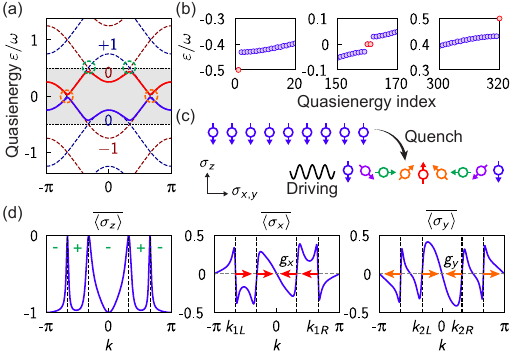}

\caption{Dynamical scheme to characterize the $1$D $Z_{2}$ Floquet topological
phase. (a) Quasienergy band structure and higher-order BISs. The Floquet
bands (solid lines) in the FBZ (shadow region) can be obtained by
opening the finite gaps at the crossings of the copied and shifted
decoupled $h_{z}$-bands labeled by $m=0$, $\pm1$, $\dots$ (red and
blue dashed lines represent $\sigma_{z}=\pm1$, respectively) for the cases with weak spin-orbit coupling and periodic driving, which
define the $2$-BIS points in the $0$ (orange dashed circles) or
$\pi$ (green dashed circles) quasienergy gaps. (b) Quasienergy spectrum
under open boundary conditions with $160$ lattice sites. Both the
$0$ and $\pi$ quasienergy gaps support nontrivial edge modes (red
dots). (c) Quantum quench protocol. At time $t=0$, the static initial
state fully polarized in the $\sigma_{z}$ axis is quenched to the
Floquet topological regime by suddenly changing the parameters and
turning on the periodic driving. (d) Stroboscopic time-averaged spin
textures. The vanishing spin polarizations in all directions determine
the $2\text{-BIS}_{0}$ points $k_{1L(R)}$ and the $2\text{-BIS}_{\pi}$
points $k_{2L(R)}$, where the nonzero dynamical fields $g_{x,y}$
are shown in red (or orange) arrows. In the left panel, the regions
with $h_{F,z}\gtrless0$ are labeled by the symbols ``$\pm$'', respectively.
Here we set $\mu_{0}=3t_{0}$, $\mu_{\mathrm{d}}=3t_{0}$, $\Delta=0.2t_{0}$,
and $\omega=4t_{0}$.\label{fig:figure1}}
\end{figure}

Here we generalize the higher-order BIS characterization into periodically
driven $Z_{2}$ topological systems. {Since the Floquet bands are captured by the Floquet Hamiltonian defined as $H_{F}\equiv(\mathrm{i}/T)\log U(T)$ with
$U(t)=\mathcal{T}\exp[-\mathrm{i}\int_{0}^{t}\mathrm{d}\tau\,H(\tau)]$
and $\mathcal{T}$ denoting the time ordering, which describes the spectra in the Floquet Brillouin zone (FBZ) $[-\pi/T, \pi/T]$ and takes the form of $H_{F}=\boldsymbol{h}_{F}\cdot\boldsymbol{\sigma}$ for time-periodic Hamiltonian \eqref{eq:1D_Hamiltonian_of_class_D}, we define the $0$D highest-order $\text{2-BISs}$ for Floquet systems as $\text{2-BIS}\equiv \{k\in\mathrm{BZ}\vert h_{F,z}(k)=0\}$.
These BISs can be intuitively understood from the quasienergy operator $Q(t)\equiv H(t)-\mathrm{i}\partial_{t}$~\citep{Eckardt2015}, which gives the quasienergy spectra and takes a block-tridiagonal matrix form
\begin{equation}\label{eq:quasienergy operator}
  Q=\begin{pmatrix}\ddots\\
  \tilde{V}^{\dagger} & H_{\mathrm{s}}+\omega & \tilde{V}\\
   & \tilde{V}^{\dagger} & H_{\mathrm{s}} & \tilde{V}\\
   &  & \tilde{V}^{\dagger} & H_{\mathrm{s}}-\omega & \tilde{V}\\
   &  &  &  & \ddots
  \end{pmatrix}
\end{equation}
with $\tilde{V}\equiv(1/T)\int_{0}^{T}\mathrm{d}t\,e^{-\mathrm{i}\omega t}V(t)$ in the extended Hilbert space $\mathcal{F}=\mathcal{H}\otimes\mathcal{L}_{T}$,
where $\mathcal{H}$ is the physical Hilbert space and $\mathcal{L}_{T}$
is the space of $T$-periodic functions with bases $e^{\mathrm{i}m\omega t}$
labeled by integer $m$. In this picture, the Floquet system is considered as a multiband system, where the static Hamiltonian $H_{\mathrm{s}}$ is
copied and shifted by energies $m\omega$ ($m=0,\pm1,\pm2,\dots$),
leading to the band crossings and giving the BISs; see Fig.~\ref{fig:figure1}(a) for an example with weak spin-orbit coupling and periodic driving, for which the Floquet bands can be obtained from the copied and shifted decoupled $h_{z}$-bands perturbatively by opening a gap at the band crossings (i.e., BISs) via the off-diagonal blocks $\tilde{V}$ and finite $h_{x,y}$. As the Floquet bands are repeated in the quasienergy domain, it is possible for the Floquet systems to have band crossings both in the center and border of FBZ. For this, we shall divide the BISs into two categories according to the positions of band crossings, i.e., the $0$ gap BISs and $\pi$ gap BISs, and denote them as $2\text{-BISs}_{0(\pi)}$, respectively.

To determine which quasienergy gap a BIS belongs to, we first consider the cases with weak spin-orbit coupling and periodic driving. In this case, the $0$ and $\pi$ gap BISs can be identified according to the locations of band crossings for the copied and shifted decoupled bands, as the weak spin-orbit coupling and periodic driving, which open a small gap at these band crossings, will not change their relative positions; see Fig.~\ref{fig:figure1}(a). This scheme is also valid for the cases with weak spin-orbit coupling but strong periodic driving. In this case, the static bands in most regions of Brillouin zone resemble the decoupled ones, for which the driving Zeeman field has very limited influence; cf. Eq.~\eqref{eq:quasienergy operator}. Only near the band crossings of original decoupled bands, the spin-orbit coupling mixes different bands and the driving Zeeman field can have remarkable influence on the positions of BISs. As this region is quite small for weak spin-orbit coupling, the above scheme based on decoupled bands still identifies the relative positions of $0$ and $\pi$ gap BISs; see Appendix~\ref{subsec:strong periodic driving} for an example.

On the other hand, we also notice that the strong spin-orbit coupling can strongly deforms the decoupled bands and may induce additional BISs; see Appendix~\ref{subsec:strong spin-orbit coupling} for an example. To identify the $0$ and $\pi$ gap BISs induced by the periodic driving, we can consider the band crossings of the copied and shifted static bands with spin-orbit coupling taken into account, while the $0$ gap BISs for the static Hamiltonian and opened by the spin-orbit coupling are still identified by the decoupled bands. We note that for the cases with strong spin-orbit coupling, our characterization theory based on BISs in general only works in the weak periodic driving regime, since the strong spin-orbit coupling mixes the decoupled bands in a large region of Brillouin zone and a strong periodic driving can strongly deforms the static bands, which may cause the coincidence and annihilation of different types of BISs, while leaving the Floquet band topology and edge modes unchanged. However, this limitation will not affect the application of our theory in practical situations, as a strong spin-orbit coupling is not realistic in real experiments.
}

\subsection{Floquet $Z_{2}$ invariant}

{With these $0$ and $\pi$ gap BISs, we now introduce our Floquet $Z_{2}$ invariant. To this end, we note that although the Floquet edge modes cannot be captured by the global topological indices of Floquet bands in the FBZ, it is still possible to utilize the Floquet Hamiltonian to characterize the Floquet topological phases, since the latter contains more information beyond the global band topology. An important type of such information is the local topological structure formed on each BIS, which can uniquely determine the full features of edge modes, as proved in Ref.~\cite{Zhang2021}. It is possible that the global band topology is trivial, while the local topological structures on BISs are still nontrivial. Here we use these local topological structures to characterize the $Z_{2}$ Floquet topological phases.

Our $Z_{2}$ Floquet invariants defined on BISs are similar to those introduced for the static topological phases in Ref.~\citep{Zhang2022}. But for the Floquet Hamiltonian of periodically driven system \eqref{eq:1D_Hamiltonian_of_class_D}, we need to define two $Z_{2}$ invariants associated with the
$q=0$ and $\pi$ gap BISs as
\begin{equation}\label{eq:1D Floquet Z2 invariant}
\nu_{q}=e^{\mathrm{i}\pi(\mathrm{sgn}[h_{F,\alpha}(k_{R})]-\mathrm{sgn}[h_{F,\alpha}(k_{L})])/2}\vert_{2\text{-BIS}_{q}},
\end{equation}
where $h_{F,\alpha}$ ($\alpha=x$ or $y$) should be nonzero on the
left (right) point $k_{L(R)}$ of the $2\text{-BIS}_{q}$ with $h_{F,z}(k)=0$.
Then the Floquet invariant $\nu_{q}$ uniquely determines the edge modes in the corresponding quasienergy gap {[}Fig.~\ref{fig:figure1}(b){]}, and the Floquet band topology is given by $\nu_{0}\cdot\nu_{\pi}$. 

To validate the above statement, we consider the thought experiment proposed in Ref.~\cite{Rudner2013}. Suppose that we start with a system, for which the Floquet bands are all trivial and there is no edge modes in the $0$ and $\pi$ quasienergy gaps. We also assume a weak spin-orbit coupling and periodic driving, such that the local topological structures formed on BISs will not affect each other. As we change the parameters, one of the gaps, e.g., the $\pi$ quasienergy gap, may close and reopen at certain BISs [i.e., the $\pi$ gap BISs; see Fig.~\ref{fig:figure1}(a)], in such a way that the topological number of Floquet bands becomes nontrivial and there exist edge modes crossing the $\pi$ quasienergy gap. This case is similar to the static $Z_{2}$ topological phases and can be characterized by the $Z_{2}$ invariant $\nu_{\pi}$ defined on the $\pi$ gap BISs, as proved in Ref.~\citep{Zhang2022}. 

As the parameters are further varied, the $0$ quasienergy gap may close and reopen at other momentum points (i.e., the $0$ gap BISs), bringing the Floquet band back to trivial. During this process, the edge modes in the $\pi$ quasienergy gap cannot disappear, since the $\pi$ quasienergy gap remains open throughout it. Therefore, after reopening the $0$ quasienergy gap, another edge modes must appear around $\varepsilon=0$. We note that the gap close and reopening at the $0$ gap BISs will not affect the local topological structures of $\pi$ gap BISs for the weak spin-orbit coupling and periodic driving. Hence the characterization of edge modes in the $\pi$ quasienergy gap via $\nu_{\pi}$ remains unchanged, while the edge modes in the $0$ quasienergy gap is captured by the changes of Floquet band topology, which is indeed the $Z_{2}$ invariant $\nu_{0}$ defined on the $0$ gap BISs, since the summation of these two kinds of BISs gives the total topology of Floquet bands. This demonstrates the validity of our $Z_{2}$ Floquet invariant \eqref{eq:1D Floquet Z2 invariant} defined on BISs. The above arguments are also valid for the cases with weak (strong) spin-orbit coupling but strong (weak) periodic driving; see Appendix~\ref{sec:strong periodic driving or spin-orbit coupling} for examples. On the other hand, as the combination of strong spin-orbit coupling and periodic driving can strongly affect the local topological structures of BISs, our theory may be not applicable in this case. But this will not affect the application of our theory in practical experiments, as clarified before. Without loss of generality, in the following we mainly focus on the cases with weak spin-orbit coupling and periodic driving.

We note that there could be multiple band crossings in each quasienergy gap for certain topological systems, and the above $Z_{2}$ invariant can be easily generalized to this situation; see Eq.~\eqref{eq:generic_Z2_Floquet_invariant}. Moreover, as the BISs can be easily identified from the quantum quenches, our BIS characterization further facilitates the dynamical detection
of $Z_{2}$ Floquet topological phases, as presented below.
}

\subsection{Dynamical detection}

We consider the quantum dynamics induced by quenching an initially
fully polarized state {in the $\sigma_{z}$ axis} with density matrix $\rho_{0}$ {(i.e., $\mu_{0}\gg t_{0},\omega$ for the initial state)}
to the Floquet topological phase \eqref{eq:1D_Hamiltonian_of_class_D}
at time $t=0$ through suddenly changing the parameters and turning
on the periodic driving {[}Fig.~\ref{fig:figure1}(c){]}. The BISs
and Floquet topology can be characterized dynamically based on the
stroboscopic time-averaged spin textures~\citep{Zhang2020}
\begin{equation}
\overline{\langle\sigma_{i}(k)\rangle}\equiv\lim_{N\to\infty}\frac{1}{N}\sum_{n=0}^{N-1}\langle\sigma_{i}(k,t=nT)\rangle,
\end{equation}
where $\langle\sigma_{i}(k,t)\rangle=\mathrm{Tr}[\rho_{0}(k)U^{\dagger}(k,t)\sigma_{i}U(k,t)]$
for $i=x,y,z$ is the spin polarization measured at time $t$. The
numerical results are shown in Fig.~\ref{fig:figure1}(d), from which
two pairs of momenta with $\overline{\langle\sigma_{x,y,z}(k)\rangle}=0$
can be identified. Due to $U(nT)=\exp(-\mathrm{i}H_{F}\cdot nT)$,
we can readily obtain $\overline{\langle\sigma_{i}\rangle}=-h_{F,z}h_{F,i}/|\boldsymbol{h}_{F}|^{2}$,
which vanishes at the momenta with $h_{F,z}$ or $h_{F,i}=0$. Thus
the above two pairs of momenta indeed measure the $2\text{-BIS}_{0,\pi}$,
respectively. The $Z_{2}$ Floquet invariant in each quasienergy gap
can be further detected through the dynamical field $g_{i}(k)\equiv-(1/\mathcal{N}_{k})\partial_{k_{\perp}}\overline{\langle\sigma_{i}(k)\rangle}\vert_{2\text{-BIS}}$,
which quantifies the variation slope of $\overline{\langle\sigma_{i}\rangle}$
across the $2\text{-BIS}$ and is proportional to $h_{F,i}$~\citep{g_field}.
Here $\mathcal{N}_{k}$ is a normalization factor and $k_{\perp}$
denotes the direction pointing from the region $h_{F,z}<0$ to $h_{F,z}>0$.
As shown in Figs.~\ref{fig:figure1}(b) and \ref{fig:figure1}(d),
the opposite $g_{x,y}$ on the $2\text{-BIS}_{0,\pi}$ manifests the
nontrivial $Z_{2}$ Floquet invariant $\nu_{0}=\nu_{\pi}=-1$, consistent
with the edge modes in each quasienergy gap.

\section{Generic dynamical characterization schemes\label{sec:generic dynamical characterization schemes}}

With the above 1D illustration, we now develop in this section the dynamical characterization scheme
for a generic $d$D $Z_{2}$ Floquet topological system of the tenfold-way symmetry classes and show its
application to a 2D time-reversal invariant Floquet topological model.

\subsection{Generic $Z_{2}$ Floquet topological phases}

We consider the following periodically driven Hamiltonian in the Altland-Zirnbauer
tenfold symmetry classes~\citep{Roy2017,Yao2017}
\begin{equation}
H[\boldsymbol{k};\boldsymbol{\lambda}(t)]=\sum_{i=0}^{d}h_{i}[\boldsymbol{k};\boldsymbol{\lambda}(t)]\gamma_{i}+\sum_{i=d+1}^{d'}h_{i}[\boldsymbol{k};\boldsymbol{\lambda}(t)]\gamma_{i},\label{eq:generic_Z2_Floquet_topological_phases}
\end{equation}
which is parameterized by a collection of parameters $\lambda_{\ell}(t)=\lambda_{\ell}(t+T)$,
such as the driven magnetic field and hopping coefficients. The Clifford
algebra $\gamma$ satisfies $\{\gamma_{i},\gamma_{j}\}=2\delta_{ij}$
and mimics a (pseudo) spin of dimensionality $n_{d'}=2^{\left\lceil d'/2\right\rceil }$,
where $\left\lceil \cdot\right\rceil $ denotes the ceiling function.
{This model covers all the $Z_{2}$-classified symmetry classes in the tenfold way of different dimensionality~\citep{Zhang2022}.}
In the absence of periodic driving, the static Hamiltonian $H(\boldsymbol{k};\boldsymbol{\lambda})$
with $d'=d+1$ (or $d+2$) describes a $d$D first (second) descendant
$Z_{2}$ topological phase~\citep{Z2_descendants,Qi2008,Ryu2010}.
The driven parameters $\boldsymbol{\lambda}(t)$ further bring this
system into the interesting Floquet topological regime.

For the $Z_{2}$ Floquet topological phases, the symmetries play an
important role. Particularly, it has been shown that under the symmetries
of Altland-Zirnbauer tenfold classes, the Hamiltonian coefficient
is either odd or even with respect to the momentum $\boldsymbol{k}$ {(at most after certain basis change)}~\citep{Zhang2022}.
Here we assume $h_{0}$ to be an even function without loss of generality.
With this, we can prove that the periodically driven parameters need
to satisfy $\lambda_{\ell}(t)=\lambda_{\ell}(-t)$ for the systems
with time-reversal symmetry $\Theta H(\boldsymbol{k},t)\Theta^{-1}=H(-\boldsymbol{k},-t)$
and/or chiral symmetry $\Pi H(\boldsymbol{k},t)\Pi^{-1}=-H(\boldsymbol{k},-t)$;
see Appendix \ref{sec:symmetry_constraints_on_the_periodic_driving}.
This constraint applies to most of the Floquet $Z_{2}$ classifications
under physical dimensions (i.e., $d=1,2,3$)~\citep{Roy2017,Yao2017},
except for the 1D topological phases of class D, for which the only
particle-hole symmetry does not impose any constraint on the driven
parameters $\boldsymbol{\lambda}(t)$. On the other hand, for the
Floquet systems with synthetic dimensions $d>3$~\citep{Price2015,Ozawa2016,Yuan2018}
and only with the particle-hole symmetry, we also assume in this work
the above requirement to be satisfied for simplicity, which has covered
a broad range of topological states.

\begin{figure}
\includegraphics{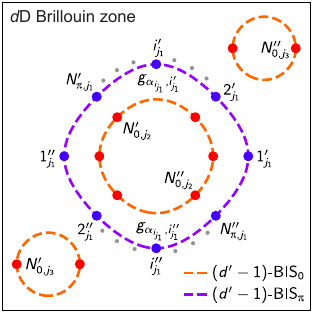}

\caption{Schematic diagram for the BIS Floquet $Z_{2}$ invariant. The higher-order
$1$D $(d'-1)$-BISs are categorized into BISs belonging to the $0$
quasienergy gap (orange dashed lines) or the $\pi$ quasienergy gap
(purple dashed line) according to the Floquet band structure. On the
$j$-th $(d'-1)$-BIS in the $q=0$ or $\pi$ gap, the corresponding
$0$D $d'$-BIS points (red or blue dots) are grouped into $N_{q,j}$
symmetric point pairs $(i'_{j},i''_{j})$, to each of which a nonzero
dynamical field $g_{\alpha_{i_{j}}}\in\{g_{d},\dots,g_{d'}\}$ is
assigned, giving the Floquet $Z_{2}$ invariant $\nu_{q}$ {[}see
Eq.~\eqref{eq:generic_Z2_Floquet_invariant}{]}. Here the middle
rings are single-connected $(d'-1)$-BISs with $N_{\pi,j_{1}}$ and
$N_{0,j_{2}}$ point pairs, respectively, while the other two circles
represent the $j_{3}$-th $(d'-1)$-BIS with two disconnected but
symmetrically related parts.\label{fig:figure2}}
\end{figure}

\subsection{Dynamical characterization}

We now generalize the dynamical characterization theory shown in Sec.~\ref{sec:dynamical_characterization_for_1D_Floquet_topological_phases}
to the generic $d$D $Z_{2}$ Floquet topological phase. Similar to
the 1D periodically driven system, we consider the quench dynamics
starting from an initial state fully polarized in the $\gamma_{0}$
axis. Then the $Z_{2}$ Floquet topology can be characterized by the
emergent dynamical topology on the highest-order BISs. 

For the generic topological phase \eqref{eq:generic_Z2_Floquet_topological_phases},
the Floquet Hamiltonian is given as (see Appendix \ref{sec:Floquet_Hamiltonian})
\begin{equation}
H_{F}(\boldsymbol{k})=\sum_{i=0}^{d}h_{F,i}(\boldsymbol{k})\gamma_{i}+\sum_{i=d+1}^{d'}h_{F,i}(\boldsymbol{k})\gamma_{i}.
\end{equation}
The corresponding higher-order $n\text{-BIS}\equiv\{\boldsymbol{k}\vert h_{F,i}=0,i=0,1,\dots,n+d-d'-1\}$
is a $(d'-n)$D symmetric subspace in the Brillouin zone~\citep{Zhang2022}
and can be detected by the stroboscopic time-averaged spin textures
$\overline{\langle\gamma_{i}(\boldsymbol{k})\rangle}=-h_{F,0}h_{F,i}/|\boldsymbol{h}_{F}|^{2}$.
We first note that the $(d-1)$D lowest-order $(d'-d+1)\text{-BIS}^{(i)}$
with only $h_{F,i}=0$ is determined by the momenta where $\overline{\langle\gamma_{\alpha}\rangle}=0$
for all $\alpha$ if $i=0$ or by the momenta with $\overline{\langle\gamma_{i}\rangle}=0$
while not on the $(d'-d+1)\text{-BIS}^{(0)}$ for $i\neq0$. With
these basic BISs, the higher-order BISs can be dynamically constructed
as
\begin{equation}
n\text{-BIS}=\bigcap_{i=0}^{n+d-d'-1}(d'-d+1)\text{-BIS}^{(i)}.\label{eq:higher-order BIS}
\end{equation}
{Which quasienergy gap the BIS belongs to can be identified as in the $1$D case; see Sec.~\ref{sec:dynamical_characterization_for_1D_Floquet_topological_phases}. Particularly, for the cases with weak spin-orbit coupling and periodic driving, as considered in this work, the $0$ and $\pi$ gap BISs can also be determined from the Floquet band structure.}
We would like to mention that the definition
of higher-order BISs is actually not unique and the $n$-BIS can also
be defined through any other $(n+d-d')$ Floquet Hamiltonian coefficients
and the corresponding $(d'-d+1)\text{-BISs}$, which does not change
the result of dynamical characterization. 
On the other hand, as which quasienergy gap the higher-order BIS belongs to is determined by the 
Floquet band structure, which is symmetric with respect to the momentum, the $(d-1)$D
basic BISs used to construct the higher-order BISs should include
those defined by parity even Hamiltonian coefficients, from which
the $0$ or $\pi$ gap $n$-BISs can be most easily identified. This
is also necessary for the trivial $Z_{2}$ phases with multiple parity
even Hamiltonian coefficients. {We note that in general, there could be multiple band crossings (i.e., BISs) in each quasienergy gap; see Fig.~\ref{fig:figure2} for an illustration.}

To characterize the $Z_{2}$ Floquet topology, we further introduce the dynamical field $g_{i}(\boldsymbol{k})=-(1/\mathcal{N}_{\boldsymbol{k}})\partial_{k_{\perp}}\overline{\langle\gamma_{i}(\boldsymbol{k})\rangle}$
on the $0$D highest-order $d'\text{-BISs}$, where $k_{\perp}$ is
perpendicular to the $(d'-d+1)\text{-BIS}^{(0)}$ and points to the
side with $h_{F,0}>0$. The dynamical field $g_{i}$ is proportional
to $h_{F,i}$. Then the Floquet topological phases can be fully characterized
by the $Z_{2}$ invariants defined in the $q=0$ and $\pi$ quasienergy
gaps and captured by the dynamical topological pattern introduced
in Ref.~\citep{Zhang2022}
\begin{equation}
\nu_{q}=\prod_{j}\prod_{\substack{i_{j}\in d'\text{-BISs on the}\\
(d'-1)\text{-BIS}_{q,j}
}
}^{N_{q,j}}(-1)^{\frac{1}{2}[\mathrm{sgn}(g_{\alpha_{i_{j}},i'_{j}})+\eta_{q,j}\mathrm{sgn}(g_{\alpha_{i_{j}},i''_{j}})]}\label{eq:generic_Z2_Floquet_invariant}
\end{equation}
with nonzero $g_{\alpha_{i_{j}}}\in\{g_{d},\dots,g_{d'}\}$ of the
same parity, which uniquely determines the Floquet boundary modes in the corresponding quasienergy gap. 
Here the first product is performed over all of the $1$D
$(d'-1)$-th order BISs in the quasienergy gap $q$, while the second
product is performed over the $N_{q,j}$ pairs of $d'$-BIS points
$(i'_{j},i''_{j})$ located symmetrically on the $j$-th single-connected
$(d'-1)\text{-BIS}_{q,j}$ or the $(d'-1)\text{-BIS}_{q,j}$ with
two disconnected but symmetrically related parts (see Fig.~\ref{fig:figure2}),
for which we have $\eta_{q,j}=(-1)^{N_{q,j}}$ or $\eta_{q,j}=-1$,
respectively. 

{This invariant can be considered as a generalization of Eq.~\eqref{eq:1D Floquet Z2 invariant} and applies to the cases with multiple BISs and higher dimensions. The corresponding product structure follows the $Z_{2}$ nature of the topological phases, for which the combination of a nontrivial band crossing and a trivial one is still nontrivial, while the combination of two trivial or nontrivial band crossings will lead to a trivial phase.}
On the other hand, as the invariant \eqref{eq:generic_Z2_Floquet_invariant} involves only minimal
information about the Floquet system that can be detected in the quantum
quench dynamics, it provides a highly feasible scheme to characterize $Z_{2}$ Floquet topological phases in real experiments. {Particularly, together with
the integer Floquet topological phases studied in Ref.~\citep{Zhang2020}, the above results further complete the dynamical
characterization for the full tenfold classes of Floquet topological phases.}

{We also note that our dynamical characterization scheme is not affected by the basis change of Hamiltonians, if the resulting Hamiltonian coefficients are still either odd or even with respect to the momentum $\boldsymbol{k}$. In this case, the positions of BISs and/or the dynamical fields may be different, but the underlying dynamical $Z_{2}$ invariants \eqref{eq:generic_Z2_Floquet_invariant} remain unchanged; see Appendix~\ref{subsec:basis change} for an example. This can be demonstrated from the derivation of BIS $Z_{2}$ invariants shown in Ref.~\cite{Zhang2022}, where the $Z_{2}$ topology is derived from the higher-dimensional integer topological phases by dimension reduction and the dynamical characterization for the latter is unaffected by the basis change~\cite{Zhang2018}. On the other hand, there also exist unitary transformations that can mix the odd and even coefficients. For these situations, our dynamical theory may not work, since the property of the system that the momenta $\boldsymbol{k}$ and $-\boldsymbol{k}$ are related by the symmetries will not be reflected on the BISs and in the spin textures, which is important to identify the $Z_{2}$ topology. However, this limitation will not affect the application of our theory in practice, as most of the topological systems in experiments possess either odd or even Hamiltonian coefficients (at most after certain basis change).}

\subsection{Application to the 2D time-reversal invariant Floquet topological
phases\label{subsec:2D_time-reversal_invariant_Floquet_topological_phase}}

\begin{figure}
\includegraphics{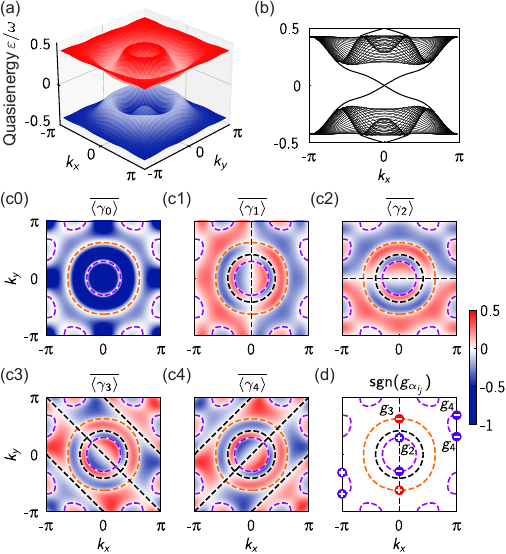}

\caption{Detecting the 2D time-reversal invariant Floquet topological phase.
(a), (b) Quasienergy spectra in the periodic boundary conditions (a)
and for a cylindrical geometry with open boundary condition in the
$y$ direction (b), where the helical gapless boundary modes on one
of the edges in the $0$ and $\pi$ quasienergy gaps are presented.
(c0)-(c4) Stroboscopic time-averaged spin textures. Here six ring-shaped
structures with $\overline{\langle\gamma_{\alpha}\rangle}=0$ for
all $\alpha$ emerge, identifying the 1D basic $3\text{-BISs}^{(0)}$
with vanishing $h_{F,0}$ in the 0 quasienergy gap (orange dashed
line) and $\pi$ quasienergy gap (purple dashed lines), respectively,
according to the Floquet band structure (a). Besides, the spin polarization
$\overline{\langle\gamma_{i}\rangle}$ with $i=1,2,3,4$ also vanishes
on the black dashed lines in (c1)-(c4), giving the corresponding $3\text{-BISs}^{(i)}$
with $h_{F,i}=0$. (d) Dynamical topological pattern. The 0D highest-order
$4\text{-BISs}$ (red or blue dots) are constructed as the intersections
of $3\text{-BISs}^{(0)}$ and $3\text{-BISs}^{(1)}$. We assign a
nonzero dynamical field $g_{\alpha_{i_{j}}}\in\{g_{2},g_{3},g_{4}\}$
with the sign indicated by ``$\pm$'' to each symmetric point pair
of the $4\text{-BISs}$, which gives the Floquet $Z_{2}$ invariant
$\nu_{0}=-1$ and $\nu_{\pi}=-1$. Here we set $m_{0}=t_{0}$, $m_{\mathrm{d}}=3t_{0}$,
$t'_{0}=0.5t_{0}$, $t_{\mathrm{so}}=0.5t_{0}$, and $\omega=7t_{0}$.\label{fig:figure3}}
\end{figure}

We now utilize the above generic dynamical scheme to characterize
the following 2D Floquet topological phase
\begin{equation}
H(\boldsymbol{k},t)=\boldsymbol{h}(\boldsymbol{k},t)\cdot\boldsymbol{\gamma}\label{eq:2D_TRI_Floquet_topological_phases}
\end{equation}
with
\[
\begin{aligned}h_{0}(\boldsymbol{k},t) & =m(t)-2t_{0}\cos k_{x}-2t_{0}\cos k_{y}\\
 & \qquad-2t'_{0}\cos(k_{x}+k_{y})-2t'_{0}\cos(k_{x}-k_{y}),\\
h_{1,2}(\boldsymbol{k}) & =2t_{\mathrm{so}}\sin k_{x,y},\qquad h_{3,4}(\boldsymbol{k})=2t_{\mathrm{so}}\sin(k_{x}\pm k_{y}).
\end{aligned}
\]
Here $t_{0},t'_{0}$ (or $t_{\mathrm{so}}$) denote the spin-conserved
(-flipped) hopping coefficients. Instead of the simple harmonic driving,
here we consider the polychromatic square-wave magnetization, given
by $m(t)=m_{0}+2m_{\mathrm{d}}/\pi$ (or $m_{0}-2m_{\mathrm{d}}/\pi$)
for $-T/4\leq t<T/4$ (or $T/4\leq t<3T/4$), to show the wide validity
of our theory. We take $\gamma_{0}=\sigma_{0}\otimes\tau_{z}$, $\gamma_{1}=\sigma_{z}\otimes\tau_{x}$,
$\gamma_{2}=\sigma_{0}\otimes\tau_{y}$, $\gamma_{3}=\sigma_{x}\otimes\tau_{x}$,
and $\gamma_{4}=\sigma_{y}\otimes\tau_{x}$, with $\sigma_{i}$ and
$\tau_{i}$ being Pauli matrices. The 2D Floquet phase \eqref{eq:2D_TRI_Floquet_topological_phases}
possesses the time-reversal symmetry $\Theta H(\boldsymbol{k},t)\Theta^{-1}=H(-\boldsymbol{k},-t)$
with $\Theta=-\mathrm{i}\sigma_{y}\otimes\tau_{0}K$ and belongs to
class AII with $Z_{2}$ invariant. In Figs.~\ref{fig:figure3}(a)
and \ref{fig:figure3}(b), we show the quasienergy spectra in the
periodic boundary conditions and for a cylindrical geometry with open
boundary condition in the $y$ direction, respectively, with $m_{0}=t_{0}$,
$m_{\mathrm{d}}=3t_{0}$, and $\omega=7t_{0}$. There exist protected
gapless helical edge modes both in the $0$ and $\pi$ quasienergy
gaps.

We study the quench dynamics from a fully polarized initial state
in the $\gamma_{0}$ axis by setting $m_{0}\gg t_{0}$ and $\omega=0$
to the above nontrivial Floquet topological regime. The stroboscopic
time-averaged spin textures are shown in Figs.~\ref{fig:figure3}(c0)-\ref{fig:figure3}(c4),
where six ring-shaped structures with vanishing spin polarizations
emerge in all the plots of $\overline{\langle\gamma_{\alpha}\rangle}$
for $\alpha=0,1,\dots,4$, identifying the $1$D lowest-order $3\text{-BISs}^{(0)}$
with $h_{F,0}=0$. Besides, there are additional lines with $\overline{\langle\gamma_{i}\rangle}=0$
for $i\ne0$ in the corresponding spin texture, capturing the $3\text{-BISs}^{(i)}$
with $h_{F,i}=0$. In Fig.~\ref{fig:figure3}(d), we use $3\text{-BISs}^{(0)}$
and $3\text{-BISs}^{(1)}$ to construct the $0$D highest-order $4\text{-BISs}$.
According to the Floquet band structure {[}Fig.~\ref{fig:figure3}(a){]},
the $4\text{-BIS}$ points on the outer $3\text{-BIS}^{(0)}$ that
surrounds the $\boldsymbol{k}=(0,0)$ point belong to the $0$ quasienergy
gap, while the others represent the $4\text{-BISs}_{\pi}$. We assign
a nonzero dynamical field $g_{\alpha_{i}}$ to each symmetric point
pair of these $4$-BISs, with which we can obtain the $Z_{2}$ invariant
$\nu_{0}=-1$ and $\nu_{\pi}=-1$, characterizing the Floquet topological
phase.

We would like to mention that the dynamical characterization scheme
built on the BISs requires the Floquet bands in the FBZ to be gapped.
For the special 2D time-reversal invariant driven system with only
two bands as considered in Ref.~\citep{Nathan2015}, the bulk Floquet
bands are related by the time-reversal symmetry and the corresponding
gap necessitates to close at high symmetry momenta, leading to a
gapless Floquet Hamiltonian $H_{F}$, to which our approach is not
applicable.

\section{Floquet topological systems with generic reference time\label{sec:floquet topological systems with generic reference time}}

In the above discussions, the periodically driven Hamiltonians are
assumed to be symmetric about the trigger time $t=0$ of the quantum
quench dynamics. In this section, we further discuss the dynamical
characterization for Floquet topological phases $H(\boldsymbol{k},t;t_{*})$
that are symmetric about a generic reference time $t_{*}$, i.e.,
$\Theta H(\boldsymbol{k},t;t_{*})\Theta^{-1}=H(-\boldsymbol{k},2t_{*}-t;t_{*})$
and $\Pi H(\boldsymbol{k},t;t_{*})\Pi^{-1}=-H(\boldsymbol{k},2t_{*}-t;t_{*})$.
As before, we use the notation $H(\boldsymbol{k},t)$ for the Floquet
Hamiltonian with $t_{*}=0$ and have the relation $H(\boldsymbol{k},t;t_{*})=H(\boldsymbol{k},t-t_{*})$
for a generic reference time $t_{*}$.

\subsection{Direct measurements}

We first consider the dynamical characterization scheme proposed above
without any modifications. The time evolution operator over one period
now reads
\begin{equation}
\begin{aligned}U(T;t_{*}) & \equiv\mathcal{T}e^{-\mathrm{i}\int_{0}^{T}\mathrm{d}\tau\,H(\tau;t_{*})}\\
 & =e^{-\mathrm{i}H''(T-2t_{*})}\cdot e^{-\mathrm{i}H'\cdot2t_{*}}=e^{-\mathrm{i}H_{F}^{(t_{*})}T},
\end{aligned}
\end{equation}
where we have $H'=(\mathrm{i}/2t_{*})\log U(t_{*},-t_{*})$ and $H''=[\mathrm{i}/(T-2t_{*})]\log U(T-t_{*},t_{*})$
with $U(t_{\mathrm{f}},t_{\mathrm{i}})\equiv\mathcal{T}\exp[-\mathrm{i}\int_{t_{\mathrm{i}}}^{t_{\mathrm{f}}}\mathrm{d}\tau\,H(\tau)]$.
Similar to the proof in Appendix \ref{sec:Floquet_Hamiltonian}, one
can readily show that $H'$ and $H''$ take the form of Dirac Hamiltonians,
i.e., $H'=\boldsymbol{h}'\cdot\boldsymbol{\gamma}$ and $H''=\boldsymbol{h}''\cdot\boldsymbol{\gamma}$.
Therefore, the time evolution operator in general is given by $U(T;t_{*})=u_{c}-\mathrm{i}\sum_{i}u_{i}\gamma_{i}-\sum_{i<j}u_{ij}\gamma_{i}\gamma_{j}$
for certain coefficients $u_{c}$, $u_{i}$ and $u_{ij}$. On the
other hand, since $H_{F}^{(t_{*})}=U^{\dagger}(0,-t_{*})H_{F}U(0,-t_{*})$,
we have $[H_{F}^{(t_{*})}]^{2}=|\boldsymbol{h}_{F}|^{2}$ and $U(T;t_{*})=\cos(|\boldsymbol{h}_{F}|T)-\mathrm{i}\sin(|\boldsymbol{h}_{F}|T)H_{F}^{(t_{*})}/|\boldsymbol{h}_{F}|$.
Thus the Floquet Hamiltonian with reference time $t_{*}$ takes the
form of
\begin{equation}
H_{F}^{(t_{*})}(\boldsymbol{k})=\sum_{0\leq i\leq d'}h_{F,i}^{(t_{*})}(\boldsymbol{k})\gamma_{i}+\sum_{0\leq i<j\leq d'}h_{F,ij}^{(t_{*})}(\boldsymbol{k})\mathrm{i}\gamma_{i}\gamma_{j}.
\end{equation}
For this Hamiltonian, our dynamical characterization may not be directly
applicable in general, since measuring the stroboscopic spin polarization
$\langle\gamma_{i}(nT)\rangle=\mathrm{Tr}[\rho_{0}U^{\dagger}(nT;t_{*})\gamma_{i}U(nT;t_{*})]$
is equivalent to detecting the observable $U(0,-t_{*})\gamma_{i}U^{\dagger}(0,-t_{*})$
in the quantum quench dynamics governed by Floquet Hamiltonian $H_{F}$
and with initial state $\rho_{0}^{(t_{*})}=U(0,-t_{*})\rho_{0}U^{\dagger}(0,-t_{*})$,
which is quite complicated and from which the BISs are hard to be
extracted.

Nevertheless, there exist a broad range of driving protocols with
$h_{i}[\boldsymbol{k};\boldsymbol{\lambda}(t)]=f(t)$$h_{i}(\boldsymbol{k};\boldsymbol{\lambda})$
for all $i>0$ and certain periodic function $f$, such as the models
with hopping coefficients being driven by the same function, to which
our dynamical characterization still can be applied directly. Here
the quench process is still along the axis $\gamma_{0}$. For these
protocols, we have $H'(\boldsymbol{k})=h'_{0}(\boldsymbol{k})\gamma_{0}+\chi'(\boldsymbol{k})\sum_{i>0}h_{i}(\boldsymbol{k})\gamma_{i}$
and $H''(\boldsymbol{k})=h''_{0}(\boldsymbol{k})\gamma_{0}+\chi''(\boldsymbol{k})\sum_{i>0}h_{i}(\boldsymbol{k})\gamma_{i}$
by using the trick shown in Appendix \ref{sec:Floquet_Hamiltonian},
and the corresponding Floquet Hamiltonian satisfies $h_{F,i}^{(t_{*})}(\boldsymbol{k})=\chi^{(t_{*})}(\boldsymbol{k})h_{i}(\boldsymbol{k})$
and $h_{F,ij}^{(t_{*})}(\boldsymbol{k})=0$ for $0<i<j$, where $\chi$'s
are certain even functions. Although the Floquet Hamiltonian still
takes a complicated form with nonzero $h_{F,0i}^{(t_{*})}$, the corresponding
stroboscopic time-averaged spin polarization now simplifies to $\overline{\langle\gamma_{i}(\boldsymbol{k})\rangle}=-h_{F,0}^{(t_{*})}h_{F,i}^{(t_{*})}/|\boldsymbol{h}_{F}|^{2}$,
which effectively measures the $Z_{2}$ topology of the Floquet Hamiltonian
\begin{equation}
\tilde{H}_{F}^{(t_{*})}(\boldsymbol{k})=h_{F,0}^{(t_{*})}(\boldsymbol{k})\gamma_{0}+\sum_{i>0}^{d'}h_{F,i}^{(t_{*})}(\boldsymbol{k})\gamma_{i}.\label{eq:effective Hamiltonian}
\end{equation}
This Floquet phase may be gapless when both $h_{F,0}^{(t_{*})}$ and
$\chi^{(t_{*})}$ vanish at certain momenta for certain $t_{*}$.
However, we can prove that the effective Hamiltonian $\tilde{H}_{F}^{(t_{*})}$
has the same Floquet $Z_{2}$ invariant as $H(\boldsymbol{k},t)$
whenever it is gapped; see Appendix \ref{sec:effective_Hamiltonian_for_direct_measurements}.
Therefore, the dynamical topological patterns emerging in the spin
texture $\overline{\langle\gamma_{i}(\boldsymbol{k})\rangle}$ still
characterize the desired $Z_{2}$ Floquet topology. Note that the
$1$D Floquet topological phases of class D are very special, for
which the Floquet Hamiltonian always takes the form of Dirac Hamiltonian
and the above requirement for the periodic driving is not necessary.

\begin{figure}
\includegraphics{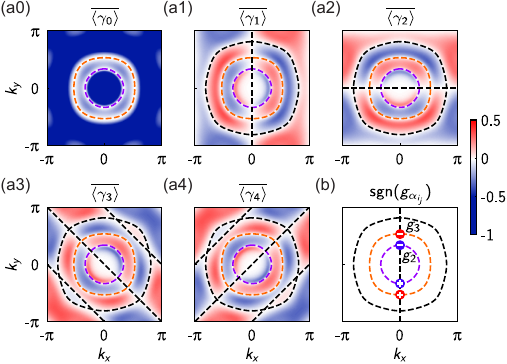}

\caption{Direct measurement of the 2D time-reversal invariant Floquet topological
phase with reference time $t_{*}=0.45T$. (a0)-(a4) Stroboscopic time-averaged
spin textures, from which the 1D $3\text{-BISs}^{(0)}$ with vanishing
$h_{F,0}^{(t_{*})}$ in the $0$ quasienergy gap (orange dashed line)
and $\pi$ quasienergy gap (purple dashed line) as well as the $3\text{-BISs}^{(i)}$
with $h_{F,i}^{(t_{*})}=0$ for $i>0$ (black dashed lines) can be
identified. (b) The 0D $4$-BISs (red and blue dots) constructed from
$3\text{-BISs}^{(0)}$ and $3\text{-BISs}^{(1)}$ and the corresponding
dynamical field $g_{\alpha_{i_{j}}}$. The resulting Floquet $Z_{2}$
invariants are given by $\nu_{0}=-1$ and $\nu_{\pi}=-1$. The parameters
are the same as in Fig.~\ref{fig:figure3}.\label{fig:figure4}}
\end{figure}

As an example, we consider the 2D time-reversal invariant Floquet
topological phase studied in Sec.~\ref{subsec:2D_time-reversal_invariant_Floquet_topological_phase}
but now with reference time $t_{*}=0.45T$. It is clear that the above
requirement for the driving protocol is satisfied. The directly measured
stroboscopic time-averaged spin textures are shown in Figs.~\ref{fig:figure4}(a0)-\ref{fig:figure4}(a4).
Compared with the results for $t_{*}=0$ (cf. Fig.~\ref{fig:figure3}),
the $1$D $3\text{-BISs}^{(0)}$ with $h_{F,0}=0$ in the $\pi$ quasienergy
gap and across the boundaries of Brillouin zone now disappear. As
these BISs does not encircle any topological charge momentum with
$h_{F,i}=0$ for all $i>0$, this change does not close the quasienergy
gaps, and the $Z_{2}$ Floquet topology is unaffected. On the other
hand, for the $3\text{-BISs}^{(0)}$ encircling the zero momentum,
only the positions are slightly deformed. We note that $\tilde{H}_{F}^{(t_{*})}$
is the exact Floquet Hamiltonian when $t_{*}=0$ or $T/2$. When increasing
$t_{*}$ from zero to $T/2$, the $3\text{-BISs}^{(0)}$ evolve from
the one at $t_{*}=0$ into those at $t_{*}=T/2$, during which the
relative positions of BISs in the $0$ and $\pi$ gaps in general
do not change. 

The $3\text{-BISs}^{(i)}$ with vanishing $h_{F,i}^{(t_{*})}$ for
$i>0$ can also be identified from the spin textures. Now the momenta
with $\chi^{(t_{*})}=0$ (black dashed circles in Fig.~\ref{fig:figure4})
have moved from the region between $3\text{-BIS}_{0}^{(0)}$ and $3\text{-BIS}_{\pi}^{(0)}$
(cf. Fig.~\ref{fig:figure3}) to the outside region, manifesting
that there exists reference time $0<t'_{*}<t_{*}$ such that $h_{F,0}^{(t'_{*})}=\chi^{(t'_{*})}=0$
for the considered model and the corresponding effective Hamiltonian
$\tilde{H}_{F}^{(t'_{*})}$ becomes gapless {[}note that $\tilde{H}_{F}^{(t_{*})}$
is gapped{]}. The range for these $t'_{*}$ is quite small from the
numerical calculations, which does not affect most of the dynamical
characterizations. Indeed, the dynamical topological pattern shown
in Fig.~\ref{fig:figure4}(b) characterizes the desired $Z_{2}$
Floquet topology.

\begin{figure}
\includegraphics{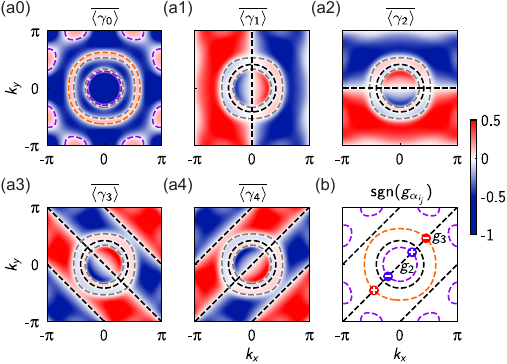}

\caption{Characterizing the 2D time-reversal invariant Floquet topological
phase with reference time $t_{*}=0.45T$ via modified stroboscopic
time averages. (a0-a4) The modified stroboscopic time-averaged spin
textures. Here the gray dashed lines represent the dBISs, on which
the spin polarizations $\overline{\langle\gamma_{\alpha}(\boldsymbol{k};t_{*})\rangle}$
vanish for all $\alpha$. In each spin texture $\overline{\langle\gamma_{i}(\boldsymbol{k};t_{*})\rangle}$
for $i=0,1,\dots,4$, other dashed lines (orange, purple, or black)
represent the $3\text{-BISs}^{(i)}$ with $h_{F,i}=0$. (b) Topological
pattern. Here the highest-order $4\text{-BISs}$ (red or blue dots)
are constructed from the $3\text{-BISs}^{(0)}$ in $0$ (orange) or
$\pi$ (purple) quasienergy gaps and the $3\text{-BISs}^{(4)}$. The
corresponding dynamical field $g_{3}$ and $g_{2}$ give us the Floquet
$Z_{2}$ invariants $\nu_{0}=-1$ and $\nu_{\pi}=-1$. The parameters
are the same as in Fig.~\ref{fig:figure3}.\label{fig:figure5}}
\end{figure}

\subsection{Modified stroboscopic time averages}

For more general Floquet topological systems with reference time $t_{*}$,
although the direct measurements are no longer applicable, our dynamical
characterization scheme still works after slight modifications. First,
instead of measuring from time $t=0$, we consider the modified stroboscopic
time-averaged spin polarizations~\citep{Zhang2020}
\begin{equation}
\overline{\langle\gamma_{i}(\boldsymbol{k};t_{*})\rangle}\equiv\lim_{N\to\infty}\frac{1}{N}\sum_{n=0}^{N-1}\langle\gamma_{i}(\boldsymbol{k},t=t_{*}+nT)\rangle.
\end{equation}
Note that due to $U(t_{*}+nT;t_{*})=U(nT,0)\cdot U(0,-t_{*})$, the
spin polarization $\langle\gamma_{i}(t=t_{*}+nT)\rangle$ is equivalent
to measuring the operator $\gamma_{i}$ at $t=nT$ under the time
evolution of Floquet Hamiltonian $H_{F}$ with initial state $\rho_{0}^{(t_{*})}=U(0,-t_{*})\rho_{0}U^{\dagger}(0,-t_{*})$.
Hence the time-averaged spin textures are given by
\begin{equation}
\overline{\langle\gamma_{i}(\boldsymbol{k};t_{*})\rangle}=h_{F,i}(\boldsymbol{k})\mathrm{Tr}[\rho_{0}^{(t_{*})}(\boldsymbol{k})H_{F}(\boldsymbol{k})]/|\boldsymbol{h}_{F}(\boldsymbol{k})|^{2}.
\end{equation}
This is similar to the shallow quench from an incompletely polarized
initial state studied in Ref.~\citep{Zhang2022}. 

The $Z_{2}$ Floquet topology can be identified by slightly modifying
the definition of BISs and dynamical fields. One can see that each
spin texture $\overline{\langle\gamma_{i}(\boldsymbol{k};t_{*})\rangle}$
vanishes on the momentum subspace given by either $\mathrm{Tr}[\rho_{0}^{(t_{*})}(\boldsymbol{k})H_{F}(\boldsymbol{k})]=0$
or $h_{F,i}(\boldsymbol{k})=0$. Here we define the former as the
dynamical band-inversion surface (dBIS) with the following characteristic
\[
\mathrm{dBIS}=\{\boldsymbol{k}\vert\overline{\langle\gamma_{\alpha}(\boldsymbol{k};t_{*})\rangle}=0,\forall\alpha\},
\]
which is induced by the quantum quench dynamics for Floquet topological
phases with nonzero $t_{*}$ instead of the Floquet Hamiltonian. Obviously,
the dynamical band-inversion surface coincides with the $(d-1)$D
lowest-order $(d'-d+1)\text{-BIS}^{(0)}$ with vanishing $h_{F,0}$
for the special case $t_{*}=0$. However, they are different in general.
The $(d'-d+1)\text{-BIS}^{(i)}$ with $h_{F,i}=0$ ($i=0,1,\dots,d'$)
now is captured by the momenta with vanishing $\overline{\langle\gamma_{i}(\boldsymbol{k};t_{*})\rangle}$
while not on the dBIS. With these basic BISs, the higher-order $n$-BISs
can be constructed in the same way as shown in Eq.~\eqref{eq:higher-order BIS}.
Further, since the spin polarization $\overline{\langle\gamma_{i}(\boldsymbol{k};t_{*})\rangle}$
does not always vanish on the highest-order BISs, the dynamical field
can be modified as
\begin{equation}
g_{i}(\boldsymbol{k})=\begin{cases}
-(1/\mathcal{N}_{\boldsymbol{k}})\partial_{k_{\perp}}\overline{\langle\gamma_{i}(\boldsymbol{k};t_{*})\rangle} & \text{if \ensuremath{\boldsymbol{k}} is also on dBIS},\\
\zeta_{\boldsymbol{k}}(1/\mathcal{N}_{\boldsymbol{k}})\overline{\langle\gamma_{i}(\boldsymbol{k};t_{*})\rangle} & \text{otherwise},
\end{cases}
\end{equation}
for $\boldsymbol{k}\in d'\text{-BIS}$. Here $k_{\perp}$ is perpendicular
to the dBIS and points to the side with negative $\zeta_{\boldsymbol{k}}=\mathrm{sgn}(\mathrm{Tr}[\rho_{0}^{(t_{*})}H_{F}])$.
Given these modifications, the $Z_{2}$ Floquet invariant \eqref{eq:generic_Z2_Floquet_invariant}
remains unchanged.

In Figs.~\ref{fig:figure5}(a0)-\ref{fig:figure5}(a4), we show the
modified stroboscopic time-averaged spin textures for the example
of 2D time-reversal invariant Floquet topological phase with reference
time $t_{*}=0.45T$. Both the dBIS and $3\text{-BISs}^{(i)}$ with
$h_{F,i}=0$ for $i=0,1,\dots,4$ can be identified. Recall that choosing
which Floquet Hamiltonian coefficient to define the highest-order
BISs is actually not unique. Here we choose the $3\text{-BISs}^{(0)}$
and $3\text{-BISs}^{(4)}$ to construct the highest-order $4\text{-BISs}$
{[}Fig.~\ref{fig:figure5}(b){]}, which have different positions
compared with those shown in Fig.~\ref{fig:figure3}. Nevertheless,
the dynamical Floquet $Z_{2}$ invariants $\nu_{0}$ and $\nu_{\pi}$
remain unchanged, where the corresponding dynamical fields $g_{\alpha_{i_{j}}}$
now are chosen from the set $\{g_{1},g_{2},g_{3}\}$.

\section{Conclusion\label{sec:conclusion}}

In conclusion, we have established the first full and unified dynamical characterization theory for $Z_{2}$ Floquet
topological phases of different dimensionality and tenfold-way symmetry classes using the minimal information about the Floquet bands.
We show that the Floquet $Z_{2}$ topology
can be completely characterized by the topological patterns emerging
in the quantum dynamics induced by quenching the system from a trivial
and static initial state to the Floquet topological regime. Particularly,
the dynamical $Z_{2}$ invariants defined on the 0D highest-order
BISs in the $0$ and $\pi$ quasienergy gaps uniquely capture the
corresponding Floquet boundary modes, hence providing a full dynamical characterization for the $Z_{2}$ Floquet topological phases both in the conventional and anomalous sense. 

Our theory can be applied to a broad range of periodically driven
systems, and the measured quantities are easily accessible in current
experiments. Therefore, our work shall advance the experimental studies
of $Z_{2}$ Floquet topological phases. Especially, for the quantum
simulation platforms based on ultracold atoms or solid-state spin
systems, where the boundary physics is hard to simulate and measure,
our dynamical characterization scheme based on the BISs is extremely
useful and provides a highly feasible method to detect the bulk topology.
On the other hand, our work completes the dynamical characterization for the full tenfold classes
of Floquet topological phases, which shall advance this research topic. Especially, it would be interesting to further generalize our theory into more
broad Floquet topological systems, such as those protected by the
crystalline or space-time symmetries~\citep{Morimoto2017,Xu2018,Yu2021b},
which is a meaningful future direction.
\begin{acknowledgments}
We acknowledge support from: ERC AdG NOQIA; MCIN/AEI (PGC2018-0910.13039/501100011033, CEX2019-000910-S/10.13039/501100011033, Plan National FIDEUA PID2019-106901GB-I00, Plan National STAMEENA PID2022-139099NB-I00 project funded by MCIN/AEI/10.13039/501100011033 and by the ``European Union NextGenerationEU/PRTR'' (PRTR-C17.I1), FPI); QUANTERA MAQS PCI2019-111828-2); QUANTERA DYNAMITE PCI2022-132919 (QuantERA II Programme co-funded by European Union's Horizon 2020 program under Grant Agreement No 101017733), Ministry of Economic Affairs and Digital Transformation of the Spanish Government through the QUANTUM ENIA project call -- Quantum Spain project, and by the European Union through the Recovery, Transformation, and Resilience Plan -- NextGenerationEU within the framework of the Digital Spain 2026 Agenda; Fundaci{\' o} Cellex; Fundaci{\' o} Mir-Puig; Generalitat de Catalunya (European Social Fund FEDER and CERCA program, AGAUR Grant No. 2021 SGR 01452, QuantumCAT \textbackslash{} U16-011424, co-funded by ERDF Operational Program of Catalonia 2014-2020); Barcelona Supercomputing Center MareNostrum (FI-2023-1-0013); EU Quantum Flagship (PASQuanS2.1, 101113690); EU Horizon 2020 FET-OPEN OPTOlogic (Grant No 899794); EU Horizon Europe Program (Grant Agreement 101080086 -- NeQST), ICFO Internal ``QuantumGaudi'' project; European Union's Horizon 2020 program under the Marie-Sklodowska-Curie grant agreement No 847648; ``La Caixa'' Junior Leaders fellowships, ``La Caixa'' Foundation (ID 100010434): CF/BQ/PR23/11980043. Views and opinions expressed are, however, those of the author(s) only and do not necessarily reflect those of the European Union, European Commission, European Climate, Infrastructure and Environment Executive Agency (CINEA), or any other granting authority. Neither the European Union nor any granting authority can be held responsible for them.

\end{acknowledgments}

\appendix

{\section{Strong periodic driving or spin-orbit coupling\label{sec:strong periodic driving or spin-orbit coupling}}
In this Appendix, we provide examples to show that our dynamical characterization theory of $Z_{2}$ Floquet topological phases based on the BISs can be applied to the cases with weak (strong) spin-orbit coupling but strong (weak) periodic driving.

\begin{figure}
  \includegraphics{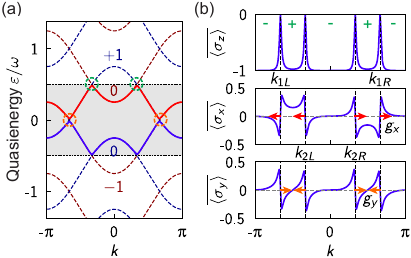}
  
  \caption{Dynamical characterization of the $1$D Floquet topological phase \eqref{eq:1D_Hamiltonian_of_class_D} with weak spin-orbit coupling but strong periodic driving. (a) Quasienergy band structure. The solid lines represent the Floquet bands, while the red and blue dashed lines are the copied and shifted decoupled $h_{z}$-bands labeled by $m=0,\pm 1,\dots$ and with $\sigma_{z}=\pm1$, respectively. The orange (green) dashed circles highlight the BISs in the $0$ ($\pi$) quasienergy gap. (b) Stroboscopic time-averaged spin textures. The vanishing spin polarizations in all directions determine the $\text{2-BIS}_{0}$ points $k_{1L(R)}$ and the $\text{2-BIS}_{\pi}$ points $k_{2L(R)}$. The corresponding nonzero dynamical fields $g_{x,y}$ are shown in red (or orange) arrows. In $\overline{\langle\sigma_{z}\rangle}$, we also label the regions with $h_{F,z}\gtrless 0$ using the symbols ``$\pm$'', respectively. Here we set $\mu_{0}=3t_{0}$, $\mu_{\mathrm{d}}=25t_{0}$, $\Delta=0.2t_{0}$, and $\omega=4t_{0}$. \label{fig:figure6}}
\end{figure}

\subsection{Weak spin-orbit coupling but strong periodic driving\label{subsec:strong periodic driving}}

We first consider the cases with weak spin-orbit coupling but strong periodic driving. As an example, in Fig.~\ref{fig:figure6} we show  numerical results for the $1$D Floquet topological phase \eqref{eq:1D_Hamiltonian_of_class_D} with spin-orbit coupling $\Delta=0.2t_{0}$ and periodic driving strength $\mu_{\mathrm{d}}=25t_{0}$. By comparing the band crossings in the Floquet band structure [Fig.~\ref{fig:figure6}(a)] and the BISs observed in the stroboscopic time-averaged spin polarizations [Fig.~\ref{fig:figure6}(b)], we can conclude that the $0$ and $\pi$ gap BISs still can be determined by the decoupled $h_{z}$-bands for the cases with weak spin-orbit coupling but strong periodic driving. Moreover, the opposite dynamical fields $g_{x,y}$ on the $\text{2-BIS}_{0,\pi}$ manifest the nontrivial $Z_{2}$ Floquet invariant $\nu_{0}=\nu_{\pi}=-1$, consistent with the edge modes in each quasienergy gap as we numerically checked. This demonstrates the validity of our dynamical characterization theory in this case.  

\begin{figure}
  \includegraphics{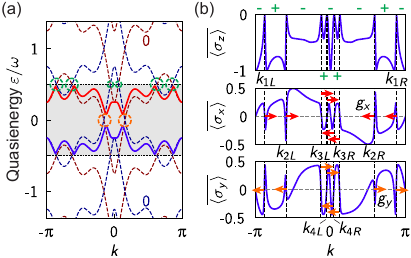}
  
  \caption{Dynamical detection of the $1$D Floquet topological phase \eqref{eq:1D_Hamiltonian_of_class_D} with strong spin-orbit coupling but weak periodic driving. (a) Quasienergy band structure. The solid lines represent the Floquet bands, while the red (blue) dashed lines are the copied and shifted upper (lower) static bands with spin-orbit coupling taken into account. Here the orange and green dashed circles highlight the band crossings in the $0$ and $\pi$ quasienergy gaps, respectively. (b) Stroboscopic time-averaged spin textures. The vanishing spin polarizations in all directions identify the $\text{2-BIS}_{0}$ points $k_{3L(R)}$ and the $\text{2-BIS}_{\pi}$ points $k_{1,2,4L(R)}$. The corresponding dynamical fields $g_{x,y}$ are shown in red (orange) arrows. We also label the regions with $h_{F,z}\gtrless 0$ in $\overline{\langle\sigma_{z}\rangle}$ using the symbols ``$\pm$'', respectively. Here we set $\mu_{0}=3t_{0}$, $\mu_{\mathrm{d}}=3t_{0}$, $\Delta=2t_{0}$, and $\omega=4t_{0}$. \label{fig:figure7}}
\end{figure}

\subsection{Strong spin-orbit coupling but weak periodic driving\label{subsec:strong spin-orbit coupling}}

We now study the cases with strong spin-orbit coupling but weak periodic driving by considering the $1$D Floquet topological phase \eqref{eq:1D_Hamiltonian_of_class_D} with $\Delta=2t_{0}$ and $\mu_{d}=3t_{0}$. The numerical results are shown in Fig.~\ref{fig:figure7}. Clearly, the BISs observed in the quantum quench dynamics [Fig.~\ref{fig:figure7}(b)] cannot be identified from the decoupled $h_{z}$ bands but are determined by the copied and shifted static bands with spin-orbit coupling taken into account [see Fig.~\ref{fig:figure7}(a)], as the spin-orbit coupling strongly deforms the decoupled bands and induces additional $\pi$ gap BIS momentum points $k_{1,2L(R)}$. Since there is no band crossings in the static Hamiltonian for our parameters, using the copied and shifted static bands is enough to determine all the BISs. Otherwise, we have to use the decoupled bands to identify the $0$ gap BISs for the static Hamiltonian, which are opened by the spin-orbit coupling when considering the static bands. Using the general $Z_{2}$ Floquet invariant \eqref{eq:generic_Z2_Floquet_invariant}, we have $\nu_{0}=\nu_{\pi}=-1$ according to the dynamical fields $g_{x,y}$ on the BISs [see Fig.~\ref{fig:figure7}(b)]. We checked that this is consistent with the edge modes in each quasienergy gap. Hence our dynamical characterization theory is also valid for the cases with strong spin-orbit coupling but weak periodic driving. 
}

\section{Symmetry constraints on the periodic driving\label{sec:symmetry_constraints_on_the_periodic_driving}}

In this Appendix, we discuss the symmetry constraints imposed on the
periodically driven parameters $\lambda_{\ell}(t)$. Here we denote
$h_{o(e)}(\boldsymbol{k};\boldsymbol{\lambda})$ to be one of the
static Hamiltonian coefficient of parity odd (even) with respect to
the momentum $\boldsymbol{k}$.

We consider the time-reversal symmetry $\Theta$, the particle-hole
symmetry $\Xi$, and the chiral symmetry $\Pi$, satisfying
\begin{equation}
\begin{aligned}\Theta H(\boldsymbol{k},t)\Theta^{-1} & =H(-\boldsymbol{k},-t),\\
\Xi H(\boldsymbol{k},t)\Xi^{-1} & =-H(-\boldsymbol{k},t),\\
\Pi H(\boldsymbol{k},t)\Pi^{-1} & =-H(\boldsymbol{k},-t).
\end{aligned}
\end{equation}
Here $\Theta$ and $\Xi$ are anti-unitary operators, while $\Pi$
is a unitary operator. Under these symmetries, the matrix $\gamma_{i}$
transforms as
\begin{equation}
\begin{aligned}\Theta: & \quad\gamma_{o}\to-\gamma_{o}, & \gamma_{e} & \to\gamma_{e},\\
\Xi: & \quad\gamma_{o}\to\gamma_{o}, & \gamma_{e} & \to-\gamma_{e},\\
\Pi: & \quad\gamma_{o}\to-\gamma_{o}, & \gamma_{e} & \to-\gamma_{e},
\end{aligned}
\end{equation}
which can be identified from the static Hamiltonian. Therefore, in
the Floquet regime we have
\begin{equation}
\begin{aligned}h_{o}[\boldsymbol{k};\boldsymbol{\lambda}(t)] & =-h_{o}[-\boldsymbol{k};\boldsymbol{\lambda}(-t)]=h_{o}[\boldsymbol{k};\boldsymbol{\lambda}(-t)],\\
h_{e}[\boldsymbol{k};\boldsymbol{\lambda}(t)] & =h_{e}[-\boldsymbol{k};\boldsymbol{\lambda}(-t)]=h_{e}[\boldsymbol{k};\boldsymbol{\lambda}(-t)]
\end{aligned}
\end{equation}
for the time-reversal symmetry, and 
\begin{equation}
h_{o(e)}[\boldsymbol{k};\boldsymbol{\lambda}(t)]=h_{o(e)}[\boldsymbol{k};\boldsymbol{\lambda}(-t)]
\end{equation}
for the chiral symmetry, leading to the result
\begin{equation}
\boldsymbol{\lambda}(t)=\boldsymbol{\lambda}(-t).
\end{equation}
On the other hand, the particle-hole symmetry only requires $h_{o}[\boldsymbol{k};\boldsymbol{\lambda}(t)]=-h_{o}[-\boldsymbol{k};\boldsymbol{\lambda}(t)]$
and $h_{e}[\boldsymbol{k};\boldsymbol{\lambda}(t)=h_{e}[-\boldsymbol{k};\boldsymbol{\lambda}(t)]$,
which does not impose any constraint on the periodic driving $\boldsymbol{\lambda}(t)$.
Nevertheless, we also assume the above requirement for Floquet topological
phases only with the particle-hole symmetry and of dimensionality
$d>3$ in this work, which has covered a broad range of topological
states.

\section{Floquet Hamiltonian $H_{F}$\label{sec:Floquet_Hamiltonian}}

In this Appendix, we show that the Floquet Hamiltonian for the periodically
driven model \eqref{eq:generic_Z2_Floquet_topological_phases} can
be written as $H_{F}=\boldsymbol{h}_{F}\cdot\boldsymbol{\gamma}$.
The proof is straightforward. We note that the Floquet Hamiltonian
is given by $H_{F}=(\mathrm{i}/T)\log U(T)$ with

\begin{equation}
  \begin{aligned}U(t) & =e^{-\mathrm{i}H(T)\delta T}\{\cdots[e^{-\mathrm{i}H(T/2+2\delta t)\delta t}(e^{-\mathrm{i}H(T/2+\delta t)\delta t}\\
   & \quad\times e^{-\mathrm{i}H(T/2)\delta t}e^{-\mathrm{i}H(T/2-\delta t)\delta t})e^{-\mathrm{i}H(T-2\delta t)\delta t}]\cdots\}\\
   & \quad\times e^{-\mathrm{i}H(0)\delta t},
  \end{aligned}
\end{equation}
where $\delta\tau$ is an infinitesimal time interval and $\exp(-\mathrm{i}H\delta\tau)=\cos(|\boldsymbol{h}|\delta\tau)-\mathrm{i}\sin(|\boldsymbol{h}|\delta\tau)H/|\boldsymbol{h}|$.
After some algebra, one can readily show that
\begin{equation}
  \begin{aligned} & \hspace*{-2em}e^{-\mathrm{i}H(T/2+\delta t)\delta t}e^{-\mathrm{i}H(T/2)\delta t}e^{-\mathrm{i}H(T/2-\delta t)\delta t}\\
  = & u_{c}-\mathrm{i}\sum_{i}u_{i}\gamma_{i}\equiv e^{-\mathrm{i}(\tilde{\boldsymbol{h}}\cdot\boldsymbol{\gamma})\cdot3\delta t}
  \end{aligned}
\end{equation}
for the 1D Floquet topological phases of class D or for the symmetry
constrained periodic driving $h_{i}(\boldsymbol{k},t)=h_{i}(\boldsymbol{k},-t)$,
where $u_{c}$ and $u_{i}$ are certain functions of the Hamiltonian
coefficients $h_{i}(T/2)$ and $h_{i}(T/2\pm\delta\tau)$. Repeating
this procedure, i.e., calculating $\exp[-\mathrm{i}H(T/2+2\delta\tau)\delta\tau]\exp[-\mathrm{i}(\tilde{\boldsymbol{h}}\cdot\boldsymbol{\gamma})\cdot3\delta\tau]\exp[-\mathrm{i}H(T/2-2\delta\tau)\delta\tau]$
and so on, we can obtain $H_{F}=\boldsymbol{h}_{F}\cdot\boldsymbol{\gamma}$.

{

\begin{figure}
  \includegraphics{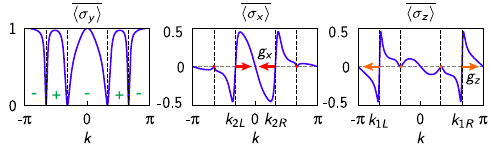}
  
  \caption{Stroboscopic time-averaged spin textures for the rotated Hamiltonian \eqref{eq:rotated_1D_Hamiltonian_of_class_D}. The vanishing spin polarizations in all directions determine the $\text{2-BIS}_{0}$ points $k_{1L(R)}$ and the $\text{2-BIS}_{\pi}$ points $k_{2L(R)}$; cf. Fig.~\ref{fig:figure1}(a). The corresponding dynamical fields $g_{x,z}$ are shown in red (or orange) arrows, while the dots indicate that the dynamical field vanishes on these momentum points. In $\overline{\langle\sigma_{y}\rangle}$, we also label the regions with $h_{F,y}\gtrless 0$ using the symbols ``$\pm$'', respectively. Here the parameters are the same as in Fig.~\ref{fig:figure1}(d). \label{fig:figure8}}
\end{figure}

\section{Basis change of Hamiltonians\label{subsec:basis change}}
Here we show that our dynamical characterization is not affected by the basis change of Hamiltonians, if the resulting Hamiltonian coefficients are still either odd or even with respect to the momentum $\boldsymbol{k}$. As an example, we consider the Hamiltonian \eqref{eq:1D_Hamiltonian_of_class_D} rotated about the $\sigma_{x}$ axis by $90\text{\textdegree}$ followed by a rotation about the $\sigma_{y}$ axis by $45\text{\textdegree}$. The corresponding Hamiltonian is given by
\begin{equation}\label{eq:rotated_1D_Hamiltonian_of_class_D}
  \begin{aligned}
    H(k,t) & =\sqrt{2}\Delta(\sin k+\sin2k)\sigma_{x}-[\mu(t)-2t_{0}\cos k]\sigma_{y}\\
    & \quad -\sqrt{2}\Delta(\sin k-\sin2k)\sigma_{z}.
  \end{aligned}
\end{equation}
Compared with Eq.~\eqref{eq:1D_Hamiltonian_of_class_D}, the Hamiltonian coefficients are rotated among each other but still satisfy the even or odd properties. Similar to Fig.~\ref{fig:figure1}, we consider the quantum dynamics induced by quenching the fully polarized initial state with $\mu_{0}\gg t_{0}, \omega$ (now along the $\sigma_{y}$ axis) to the Floquet topological regime with $\mu_{0}=3t_{0}$, $\mu_{\mathrm{d}}=3t_{0}$, $\Delta=0.2t_{0}$, and $\omega=4t_{0}$. The results are presented in Fig.~\ref{fig:figure8}. Although the spin textures are totally different from those shown in Fig.~\ref{fig:figure1}(d), the BISs still can be identified from the momentum points with vanishing spin polarizations in all directions. Moreover, although $g_{x}$ and $g_{z}$ may vanish on certain BIS points, the nonzero and opposite dynamical fields $g_{z}$ ($g_{x}$) on the $\text{2-BIS}_{0}$ points $k_{1L(R)}$ [$\text{2-BIS}_{\pi}$ points $k_{2L(R)}$] still characterize correctly the nontrivial $Z_{2}$ topology in the $0$ ($\pi$) quasienergy gap, respectively; cf. Fig.~\ref{fig:figure1}(b). Hence our dynamical characterization theory is not affected by the basis change of Hamiltonians.

}

\section{Effective Hamiltonian $\tilde{H}_{F}^{(t_{*})}$ for direct measurements\label{sec:effective_Hamiltonian_for_direct_measurements}}

In this Appendix, we study the quasienergy gap of the Floquet phase
associated with $\tilde{H}_{F}^{(t_{*})}(\boldsymbol{k})=h_{F,0}^{(t_{*})}(\boldsymbol{k})\gamma_{0}+\sum_{i>0}^{d'}h_{F,i}^{(t_{*})}(\boldsymbol{k})\gamma_{i}$
and its topological properties, where we have $h_{F,i}^{(t_{*})}(\boldsymbol{k})=\chi^{(t_{*})}(\boldsymbol{k})h_{i}(\boldsymbol{k})$
for $i>0$ with $\chi^{(t_{*})}(\boldsymbol{k})$ being certain even
function {[}cf. Eq.~\eqref{eq:effective Hamiltonian}{]}. In the
following, we denote $\tilde{\boldsymbol{h}}_{F}^{(t_{*})}\equiv(h_{F,0}^{(t_{*})},h_{F,1}^{(t_{*})},\dots,h_{F,d'}^{(t_{*})})$
for brevity.

We note that the Floquet Hamiltonian for reference time $t_{*}$ is
given by $H_{F}^{(t_{*})}(\boldsymbol{k})=\sum_{0\leq i\leq d'}h_{F,i}^{(t_{*})}(\boldsymbol{k})\gamma_{i}+\sum_{i>0}h_{F,0i}^{(t_{*})}(\boldsymbol{k})\mathrm{i}\gamma_{0}\gamma_{i}$.
Since $|\tilde{\boldsymbol{h}}_{F}^{(t_{*})}|$ is smaller than the
quasienergy of Floquet Hamiltonian, $(|\tilde{\boldsymbol{h}}_{F}^{(t_{*})}|^{2}+\sum_{0<i\leq d'}[h_{F,0i}^{(t_{*})}]^{2})^{1/2}<\pi/T$,
the Floquet phase associated with $\tilde{H}_{F}^{(t_{*})}$ is indeed
gapped in the $\pi$ quasienergy gap. On the other hand, for the $0$
quasienergy gap, we notice that $h_{F,0}^{(t_{*})}$ equals $h_{F,0}$
and is nonzero at the momenta where $h_{i}=0$ for all $i>0$. Thus
$\tilde{\boldsymbol{h}}_{F}^{(t_{*})}$ is always finite whenever
$\chi^{(t_{*})}\ne0$. The effective Hamiltonian $\tilde{H}_{F}^{(t_{*})}$
can be gapless only when both $h_{F,0}^{(t_{*})}$ and $\chi^{(t_{*})}$
vanish at certain momenta for certain $t_{*}$. In general, the range
of these $t_{*}$ is quite small, and $\tilde{H}_{F}^{(t_{*})}$ is
fully gapped for most cases.

To study the topological properties, we consider a family of Hamiltonians
$\tilde{H}_{F}^{(t_{*})}$ parameterized by $t_{*}$. For $t_{*}=0$,
the effective Hamiltonian gives the exact Floquet Hamiltonian $H_{F}$.
Clearly, $\tilde{H}_{F}^{(t_{*})}$ and $H_{F}$ possess the same
symmetries. If the $0$ and $\pi$ quasienergy gaps are always gapped
for all $t_{*}$, then the Floquet phase associated with $\tilde{H}_{F}^{(t_{*})}$
has the same $Z_{2}$ topology as the original phase $H(\boldsymbol{k},t)$.
This is also valid for the cases whenever $\tilde{H}_{F}^{(t_{*})}$
is gapped, although there may be certain reference time between $0$
and $t_{*}$ at which the effective Hamiltonian becomes gapless. The
point is that since both $h_{F,0}^{(t_{*})}$ and $\chi^{(t_{*})}$
are even functions of $\boldsymbol{k}$, the close and reopening of
$0$ quasienergy gap for each Floquet band must occur at pairs of
symmetric momentum points, which shall not affect the $Z_{2}$ topology~\citep{Bernevig2013}.
This completes the proof.


\begin{thebibliography}{105}%
  \makeatletter
  \providecommand \@ifxundefined [1]{%
   \@ifx{#1\undefined}
  }%
  \providecommand \@ifnum [1]{%
   \ifnum #1\expandafter \@firstoftwo
   \else \expandafter \@secondoftwo
   \fi
  }%
  \providecommand \@ifx [1]{%
   \ifx #1\expandafter \@firstoftwo
   \else \expandafter \@secondoftwo
   \fi
  }%
  \providecommand \natexlab [1]{#1}%
  \providecommand \enquote  [1]{``#1''}%
  \providecommand \bibnamefont  [1]{#1}%
  \providecommand \bibfnamefont [1]{#1}%
  \providecommand \citenamefont [1]{#1}%
  \providecommand \href@noop [0]{\@secondoftwo}%
  \providecommand \href [0]{\begingroup \@sanitize@url \@href}%
  \providecommand \@href[1]{\@@startlink{#1}\@@href}%
  \providecommand \@@href[1]{\endgroup#1\@@endlink}%
  \providecommand \@sanitize@url [0]{\catcode `\\12\catcode `\$12\catcode
    `\&12\catcode `\#12\catcode `\^12\catcode `\_12\catcode `\%12\relax}%
  \providecommand \@@startlink[1]{}%
  \providecommand \@@endlink[0]{}%
  \providecommand \url  [0]{\begingroup\@sanitize@url \@url }%
  \providecommand \@url [1]{\endgroup\@href {#1}{\urlprefix }}%
  \providecommand \urlprefix  [0]{URL }%
  \providecommand \Eprint [0]{\href }%
  \providecommand \doibase [0]{https://doi.org/}%
  \providecommand \selectlanguage [0]{\@gobble}%
  \providecommand \bibinfo  [0]{\@secondoftwo}%
  \providecommand \bibfield  [0]{\@secondoftwo}%
  \providecommand \translation [1]{[#1]}%
  \providecommand \BibitemOpen [0]{}%
  \providecommand \bibitemStop [0]{}%
  \providecommand \bibitemNoStop [0]{.\EOS\space}%
  \providecommand \EOS [0]{\spacefactor3000\relax}%
  \providecommand \BibitemShut  [1]{\csname bibitem#1\endcsname}%
  \let\auto@bib@innerbib\@empty
  \bibitem [{\citenamefont {Laughlin}(1981)}]{Laughlin1981}%
    \BibitemOpen
    \bibfield  {author} {\bibinfo {author} {\bibfnamefont {R.~B.}\ \bibnamefont
    {Laughlin}},\ }\bibfield  {title} {\bibinfo {title} {Quantized {H}all
    conductivity in two dimensions},\ }\href
    {https://doi.org/10.1103/physrevb.23.5632} {\bibfield  {journal} {\bibinfo
    {journal} {Phys. Rev. B}\ }\textbf {\bibinfo {volume} {23}},\ \bibinfo
    {pages} {5632} (\bibinfo {year} {1981})}\BibitemShut {NoStop}%
  \bibitem [{\citenamefont {Tsui}\ \emph {et~al.}(1982)\citenamefont {Tsui},
    \citenamefont {Stormer},\ and\ \citenamefont {Gossard}}]{Tsui1982}%
    \BibitemOpen
    \bibfield  {author} {\bibinfo {author} {\bibfnamefont {D.~C.}\ \bibnamefont
    {Tsui}}, \bibinfo {author} {\bibfnamefont {H.~L.}\ \bibnamefont {Stormer}},\
    and\ \bibinfo {author} {\bibfnamefont {A.~C.}\ \bibnamefont {Gossard}},\
    }\bibfield  {title} {\bibinfo {title} {Two-{D}imensional {M}agnetotransport
    in the {E}xtreme {Q}uantum {L}imit},\ }\href
    {https://doi.org/10.1103/physrevlett.48.1559} {\bibfield  {journal} {\bibinfo
     {journal} {Phys. Rev. Lett.}\ }\textbf {\bibinfo {volume} {48}},\ \bibinfo
    {pages} {1559} (\bibinfo {year} {1982})}\BibitemShut {NoStop}%
  \bibitem [{\citenamefont {Laughlin}(1983)}]{Laughlin1983}%
    \BibitemOpen
    \bibfield  {author} {\bibinfo {author} {\bibfnamefont {R.~B.}\ \bibnamefont
    {Laughlin}},\ }\bibfield  {title} {\bibinfo {title} {Anomalous {Q}uantum
    {H}all {E}ffect: An {I}ncompressible {Q}uantum {F}luid with {F}ractionally
    {C}harged {E}xcitations},\ }\href
    {https://doi.org/10.1103/physrevlett.50.1395} {\bibfield  {journal} {\bibinfo
     {journal} {Phys. Rev. Lett.}\ }\textbf {\bibinfo {volume} {50}},\ \bibinfo
    {pages} {1395} (\bibinfo {year} {1983})}\BibitemShut {NoStop}%
  \bibitem [{\citenamefont {v.~Klitzing}\ \emph {et~al.}(1980)\citenamefont
    {v.~Klitzing}, \citenamefont {Dorda},\ and\ \citenamefont
    {Pepper}}]{Klitzing1980}%
    \BibitemOpen
    \bibfield  {author} {\bibinfo {author} {\bibfnamefont {K.}~\bibnamefont
    {v.~Klitzing}}, \bibinfo {author} {\bibfnamefont {G.}~\bibnamefont {Dorda}},\
    and\ \bibinfo {author} {\bibfnamefont {M.}~\bibnamefont {Pepper}},\
    }\bibfield  {title} {\bibinfo {title} {New {M}ethod for {H}igh-{A}ccuracy
    {D}etermination of the {F}ine-{S}tructure {C}onstant {B}ased on {Q}uantized
    {H}all {R}esistance},\ }\href {https://doi.org/10.1103/physrevlett.45.494}
    {\bibfield  {journal} {\bibinfo  {journal} {Phys. Rev. Lett.}\ }\textbf
    {\bibinfo {volume} {45}},\ \bibinfo {pages} {494} (\bibinfo {year}
    {1980})}\BibitemShut {NoStop}%
  \bibitem [{\citenamefont {Thouless}\ \emph {et~al.}(1982)\citenamefont
    {Thouless}, \citenamefont {Kohmoto}, \citenamefont {Nightingale},\ and\
    \citenamefont {den Nijs}}]{Thouless1982}%
    \BibitemOpen
    \bibfield  {author} {\bibinfo {author} {\bibfnamefont {D.~J.}\ \bibnamefont
    {Thouless}}, \bibinfo {author} {\bibfnamefont {M.}~\bibnamefont {Kohmoto}},
    \bibinfo {author} {\bibfnamefont {M.~P.}\ \bibnamefont {Nightingale}},\ and\
    \bibinfo {author} {\bibfnamefont {M.}~\bibnamefont {den Nijs}},\ }\bibfield
    {title} {\bibinfo {title} {Quantized {H}all {C}onductance in a
    {T}wo-{D}imensional {P}eriodic {P}otential},\ }\href
    {https://doi.org/10.1103/physrevlett.49.405} {\bibfield  {journal} {\bibinfo
    {journal} {Phys. Rev. Lett.}\ }\textbf {\bibinfo {volume} {49}},\ \bibinfo
    {pages} {405} (\bibinfo {year} {1982})}\BibitemShut {NoStop}%
  \bibitem [{\citenamefont {Landau}\ and\ \citenamefont
    {Lifshitz}(2013)}]{Landau2013}%
    \BibitemOpen
    \bibfield  {author} {\bibinfo {author} {\bibfnamefont {L.~D.}\ \bibnamefont
    {Landau}}\ and\ \bibinfo {author} {\bibfnamefont {E.~M.}\ \bibnamefont
    {Lifshitz}},\ }\href {https://doi.org/10.1016/C2009-0-24487-4} {\emph
    {\bibinfo {title} {Statistical {P}hysics}}}\ (\bibinfo  {publisher}
    {Elsevier},\ \bibinfo {year} {2013})\BibitemShut {NoStop}%
  \bibitem [{\citenamefont {Kane}\ and\ \citenamefont
    {Mele}(2005{\natexlab{a}})}]{Kane2005}%
    \BibitemOpen
    \bibfield  {author} {\bibinfo {author} {\bibfnamefont {C.~L.}\ \bibnamefont
    {Kane}}\ and\ \bibinfo {author} {\bibfnamefont {E.~J.}\ \bibnamefont
    {Mele}},\ }\bibfield  {title} {\bibinfo {title} {Quantum {S}pin {H}all
    {E}ffect in {G}raphene},\ }\href
    {https://doi.org/10.1103/physrevlett.95.226801} {\bibfield  {journal}
    {\bibinfo  {journal} {Phys. Rev. Lett.}\ }\textbf {\bibinfo {volume} {95}},\
    \bibinfo {pages} {226801} (\bibinfo {year} {2005}{\natexlab{a}})}\BibitemShut
    {NoStop}%
  \bibitem [{\citenamefont {Kane}\ and\ \citenamefont
    {Mele}(2005{\natexlab{b}})}]{Kane2005a}%
    \BibitemOpen
    \bibfield  {author} {\bibinfo {author} {\bibfnamefont {C.~L.}\ \bibnamefont
    {Kane}}\ and\ \bibinfo {author} {\bibfnamefont {E.~J.}\ \bibnamefont
    {Mele}},\ }\bibfield  {title} {\bibinfo {title} {${Z}_{2}$ {T}opological
    {O}rder and the {Q}uantum {S}pin {H}all {E}ffect},\ }\href
    {https://doi.org/10.1103/physrevlett.95.146802} {\bibfield  {journal}
    {\bibinfo  {journal} {Phys. Rev. Lett.}\ }\textbf {\bibinfo {volume} {95}},\
    \bibinfo {pages} {146802} (\bibinfo {year} {2005}{\natexlab{b}})}\BibitemShut
    {NoStop}%
  \bibitem [{\citenamefont {Bernevig}\ and\ \citenamefont
    {Zhang}(2006)}]{Bernevig2006}%
    \BibitemOpen
    \bibfield  {author} {\bibinfo {author} {\bibfnamefont {B.~A.}\ \bibnamefont
    {Bernevig}}\ and\ \bibinfo {author} {\bibfnamefont {S.-C.}\ \bibnamefont
    {Zhang}},\ }\bibfield  {title} {\bibinfo {title} {Quantum {S}pin {H}all
    {E}ffect},\ }\href {https://doi.org/10.1103/physrevlett.96.106802} {\bibfield
     {journal} {\bibinfo  {journal} {Phys. Rev. Lett.}\ }\textbf {\bibinfo
    {volume} {96}},\ \bibinfo {pages} {106802} (\bibinfo {year}
    {2006})}\BibitemShut {NoStop}%
  \bibitem [{\citenamefont {K{\" o}nig}\ \emph {et~al.}(2007)\citenamefont {K{\"
    o}nig}, \citenamefont {Wiedmann}, \citenamefont {Br{\" u}une}, \citenamefont
    {Roth}, \citenamefont {Buhmann}, \citenamefont {Molenkamp}, \citenamefont
    {Qi},\ and\ \citenamefont {Zhang}}]{Konig2007}%
    \BibitemOpen
    \bibfield  {author} {\bibinfo {author} {\bibfnamefont {M.}~\bibnamefont {K{\"
    o}nig}}, \bibinfo {author} {\bibfnamefont {S.}~\bibnamefont {Wiedmann}},
    \bibinfo {author} {\bibfnamefont {C.}~\bibnamefont {Br{\" u}une}}, \bibinfo
    {author} {\bibfnamefont {A.}~\bibnamefont {Roth}}, \bibinfo {author}
    {\bibfnamefont {H.}~\bibnamefont {Buhmann}}, \bibinfo {author} {\bibfnamefont
    {L.~W.}\ \bibnamefont {Molenkamp}}, \bibinfo {author} {\bibfnamefont {X.-L.}\
    \bibnamefont {Qi}},\ and\ \bibinfo {author} {\bibfnamefont {S.-C.}\
    \bibnamefont {Zhang}},\ }\bibfield  {title} {\bibinfo {title} {Quantum {S}pin
    {H}all {I}nsulator {S}tate in {HgTe} {Q}uantum {W}ells},\ }\href
    {https://doi.org/10.1126/science.1148047} {\bibfield  {journal} {\bibinfo
    {journal} {Science}\ }\textbf {\bibinfo {volume} {318}},\ \bibinfo {pages}
    {766} (\bibinfo {year} {2007})}\BibitemShut {NoStop}%
  \bibitem [{\citenamefont {Altland}\ and\ \citenamefont
    {Zirnbauer}(1997)}]{Altland1997}%
    \BibitemOpen
    \bibfield  {author} {\bibinfo {author} {\bibfnamefont {A.}~\bibnamefont
    {Altland}}\ and\ \bibinfo {author} {\bibfnamefont {M.~R.}\ \bibnamefont
    {Zirnbauer}},\ }\bibfield  {title} {\bibinfo {title} {Nonstandard symmetry
    classes in mesoscopic normal-superconducting hybrid structures},\ }\href
    {https://doi.org/10.1103/physrevb.55.1142} {\bibfield  {journal} {\bibinfo
    {journal} {Phys. Rev. B}\ }\textbf {\bibinfo {volume} {55}},\ \bibinfo
    {pages} {1142} (\bibinfo {year} {1997})}\BibitemShut {NoStop}%
  \bibitem [{\citenamefont {Schnyder}\ \emph {et~al.}(2008)\citenamefont
    {Schnyder}, \citenamefont {Ryu}, \citenamefont {Furusaki},\ and\
    \citenamefont {Ludwig}}]{Schnyder2008}%
    \BibitemOpen
    \bibfield  {author} {\bibinfo {author} {\bibfnamefont {A.~P.}\ \bibnamefont
    {Schnyder}}, \bibinfo {author} {\bibfnamefont {S.}~\bibnamefont {Ryu}},
    \bibinfo {author} {\bibfnamefont {A.}~\bibnamefont {Furusaki}},\ and\
    \bibinfo {author} {\bibfnamefont {A.~W.~W.}\ \bibnamefont {Ludwig}},\
    }\bibfield  {title} {\bibinfo {title} {Classification of topological
    insulators and superconductors in three spatial dimensions},\ }\href
    {https://doi.org/10.1103/physrevb.78.195125} {\bibfield  {journal} {\bibinfo
    {journal} {Phys. Rev. B}\ }\textbf {\bibinfo {volume} {78}},\ \bibinfo
    {pages} {195125} (\bibinfo {year} {2008})}\BibitemShut {NoStop}%
  \bibitem [{\citenamefont {Kitaev}(2009)}]{Kitaev2009}%
    \BibitemOpen
    \bibfield  {author} {\bibinfo {author} {\bibfnamefont {A.}~\bibnamefont
    {Kitaev}},\ }\bibfield  {title} {\bibinfo {title} {Periodic table for
    topological insulators and superconductors},\ }\href
    {https://doi.org/10.1063/1.3149495} {\bibfield  {journal} {\bibinfo
    {journal} {AIP Conf. Proc.}\ }\textbf {\bibinfo {volume} {1134}},\ \bibinfo
    {pages} {22} (\bibinfo {year} {2009})}\BibitemShut {NoStop}%
  \bibitem [{\citenamefont {Ryu}\ \emph {et~al.}(2010)\citenamefont {Ryu},
    \citenamefont {Schnyder}, \citenamefont {Furusaki},\ and\ \citenamefont
    {Ludwig}}]{Ryu2010}%
    \BibitemOpen
    \bibfield  {author} {\bibinfo {author} {\bibfnamefont {S.}~\bibnamefont
    {Ryu}}, \bibinfo {author} {\bibfnamefont {A.~P.}\ \bibnamefont {Schnyder}},
    \bibinfo {author} {\bibfnamefont {A.}~\bibnamefont {Furusaki}},\ and\
    \bibinfo {author} {\bibfnamefont {A.~W.~W.}\ \bibnamefont {Ludwig}},\
    }\bibfield  {title} {\bibinfo {title} {Topological insulators and
    superconductors: tenfold way and dimensional hierarchy},\ }\href
    {https://doi.org/10.1088/1367-2630/12/6/065010} {\bibfield  {journal}
    {\bibinfo  {journal} {New J. Phys.}\ }\textbf {\bibinfo {volume} {12}},\
    \bibinfo {pages} {065010} (\bibinfo {year} {2010})}\BibitemShut {NoStop}%
  \bibitem [{\citenamefont {Hasan}\ and\ \citenamefont {Kane}(2010)}]{Hasan2010}%
    \BibitemOpen
    \bibfield  {author} {\bibinfo {author} {\bibfnamefont {M.~Z.}\ \bibnamefont
    {Hasan}}\ and\ \bibinfo {author} {\bibfnamefont {C.~L.}\ \bibnamefont
    {Kane}},\ }\bibfield  {title} {\bibinfo {title} {Colloquium: Topological
    insulators},\ }\href {https://doi.org/10.1103/revmodphys.82.3045} {\bibfield
    {journal} {\bibinfo  {journal} {Rev. Mod. Phys.}\ }\textbf {\bibinfo {volume}
    {82}},\ \bibinfo {pages} {3045} (\bibinfo {year} {2010})}\BibitemShut
    {NoStop}%
  \bibitem [{\citenamefont {Qi}\ and\ \citenamefont {Zhang}(2011)}]{Qi2011}%
    \BibitemOpen
    \bibfield  {author} {\bibinfo {author} {\bibfnamefont {X.-L.}\ \bibnamefont
    {Qi}}\ and\ \bibinfo {author} {\bibfnamefont {S.-C.}\ \bibnamefont {Zhang}},\
    }\bibfield  {title} {\bibinfo {title} {Topological insulators and
    superconductors},\ }\href {https://doi.org/10.1103/revmodphys.83.1057}
    {\bibfield  {journal} {\bibinfo  {journal} {Rev. Mod. Phys.}\ }\textbf
    {\bibinfo {volume} {83}},\ \bibinfo {pages} {1057} (\bibinfo {year}
    {2011})}\BibitemShut {NoStop}%
  \bibitem [{\citenamefont {Chiu}\ \emph {et~al.}(2016)\citenamefont {Chiu},
    \citenamefont {Teo}, \citenamefont {Schnyder},\ and\ \citenamefont
    {Ryu}}]{Chiu2016}%
    \BibitemOpen
    \bibfield  {author} {\bibinfo {author} {\bibfnamefont {C.-K.}\ \bibnamefont
    {Chiu}}, \bibinfo {author} {\bibfnamefont {J.~C.~Y.}\ \bibnamefont {Teo}},
    \bibinfo {author} {\bibfnamefont {A.~P.}\ \bibnamefont {Schnyder}},\ and\
    \bibinfo {author} {\bibfnamefont {S.}~\bibnamefont {Ryu}},\ }\bibfield
    {title} {\bibinfo {title} {Classification of topological quantum matter with
    symmetries},\ }\href {https://doi.org/10.1103/revmodphys.88.035005}
    {\bibfield  {journal} {\bibinfo  {journal} {Rev. Mod. Phys.}\ }\textbf
    {\bibinfo {volume} {88}},\ \bibinfo {pages} {035005} (\bibinfo {year}
    {2016})}\BibitemShut {NoStop}%
  \bibitem [{\citenamefont {Chang}\ \emph {et~al.}(2013)\citenamefont {Chang},
    \citenamefont {Zhang}, \citenamefont {Feng}, \citenamefont {Shen},
    \citenamefont {Zhang}, \citenamefont {Guo}, \citenamefont {Li}, \citenamefont
    {Ou}, \citenamefont {Wei}, \citenamefont {Wang}, \citenamefont {Ji},
    \citenamefont {Feng}, \citenamefont {Ji}, \citenamefont {Chen}, \citenamefont
    {Jia}, \citenamefont {Dai}, \citenamefont {Fang}, \citenamefont {Zhang},
    \citenamefont {He}, \citenamefont {Wang}, \citenamefont {Lu}, \citenamefont
    {Ma},\ and\ \citenamefont {Xue}}]{Chang2013}%
    \BibitemOpen
    \bibfield  {author} {\bibinfo {author} {\bibfnamefont {C.-Z.}\ \bibnamefont
    {Chang}}, \bibinfo {author} {\bibfnamefont {J.}~\bibnamefont {Zhang}},
    \bibinfo {author} {\bibfnamefont {X.}~\bibnamefont {Feng}}, \bibinfo {author}
    {\bibfnamefont {J.}~\bibnamefont {Shen}}, \bibinfo {author} {\bibfnamefont
    {Z.}~\bibnamefont {Zhang}}, \bibinfo {author} {\bibfnamefont
    {M.}~\bibnamefont {Guo}}, \bibinfo {author} {\bibfnamefont {K.}~\bibnamefont
    {Li}}, \bibinfo {author} {\bibfnamefont {Y.}~\bibnamefont {Ou}}, \bibinfo
    {author} {\bibfnamefont {P.}~\bibnamefont {Wei}}, \bibinfo {author}
    {\bibfnamefont {L.-L.}\ \bibnamefont {Wang}}, \bibinfo {author}
    {\bibfnamefont {Z.-Q.}\ \bibnamefont {Ji}}, \bibinfo {author} {\bibfnamefont
    {Y.}~\bibnamefont {Feng}}, \bibinfo {author} {\bibfnamefont {S.}~\bibnamefont
    {Ji}}, \bibinfo {author} {\bibfnamefont {X.}~\bibnamefont {Chen}}, \bibinfo
    {author} {\bibfnamefont {J.}~\bibnamefont {Jia}}, \bibinfo {author}
    {\bibfnamefont {X.}~\bibnamefont {Dai}}, \bibinfo {author} {\bibfnamefont
    {Z.}~\bibnamefont {Fang}}, \bibinfo {author} {\bibfnamefont {S.-C.}\
    \bibnamefont {Zhang}}, \bibinfo {author} {\bibfnamefont {K.}~\bibnamefont
    {He}}, \bibinfo {author} {\bibfnamefont {Y.}~\bibnamefont {Wang}}, \bibinfo
    {author} {\bibfnamefont {L.}~\bibnamefont {Lu}}, \bibinfo {author}
    {\bibfnamefont {X.-C.}\ \bibnamefont {Ma}},\ and\ \bibinfo {author}
    {\bibfnamefont {Q.-K.}\ \bibnamefont {Xue}},\ }\bibfield  {title} {\bibinfo
    {title} {Experimental {O}bservation of the {Q}uantum {A}nomalous {H}all
    {E}ffect in a {M}agnetic {T}opological {I}nsulator},\ }\href
    {https://doi.org/10.1126/science.1234414} {\bibfield  {journal} {\bibinfo
    {journal} {Science}\ }\textbf {\bibinfo {volume} {340}},\ \bibinfo {pages}
    {167} (\bibinfo {year} {2013})}\BibitemShut {NoStop}%
  \bibitem [{\citenamefont {He}\ \emph {et~al.}(2017)\citenamefont {He},
    \citenamefont {Pan}, \citenamefont {Stern}, \citenamefont {Burks},
    \citenamefont {Che}, \citenamefont {Yin}, \citenamefont {Wang}, \citenamefont
    {Lian}, \citenamefont {Zhou}, \citenamefont {Choi}, \citenamefont {Murata},
    \citenamefont {Kou}, \citenamefont {Chen}, \citenamefont {Nie}, \citenamefont
    {Shao}, \citenamefont {Fan}, \citenamefont {Zhang}, \citenamefont {Liu},
    \citenamefont {Xia},\ and\ \citenamefont {Wang}}]{He2017}%
    \BibitemOpen
    \bibfield  {author} {\bibinfo {author} {\bibfnamefont {Q.~L.}\ \bibnamefont
    {He}}, \bibinfo {author} {\bibfnamefont {L.}~\bibnamefont {Pan}}, \bibinfo
    {author} {\bibfnamefont {A.~L.}\ \bibnamefont {Stern}}, \bibinfo {author}
    {\bibfnamefont {E.~C.}\ \bibnamefont {Burks}}, \bibinfo {author}
    {\bibfnamefont {X.}~\bibnamefont {Che}}, \bibinfo {author} {\bibfnamefont
    {G.}~\bibnamefont {Yin}}, \bibinfo {author} {\bibfnamefont {J.}~\bibnamefont
    {Wang}}, \bibinfo {author} {\bibfnamefont {B.}~\bibnamefont {Lian}}, \bibinfo
    {author} {\bibfnamefont {Q.}~\bibnamefont {Zhou}}, \bibinfo {author}
    {\bibfnamefont {E.~S.}\ \bibnamefont {Choi}}, \bibinfo {author}
    {\bibfnamefont {K.}~\bibnamefont {Murata}}, \bibinfo {author} {\bibfnamefont
    {X.}~\bibnamefont {Kou}}, \bibinfo {author} {\bibfnamefont {Z.}~\bibnamefont
    {Chen}}, \bibinfo {author} {\bibfnamefont {T.}~\bibnamefont {Nie}}, \bibinfo
    {author} {\bibfnamefont {Q.}~\bibnamefont {Shao}}, \bibinfo {author}
    {\bibfnamefont {Y.}~\bibnamefont {Fan}}, \bibinfo {author} {\bibfnamefont
    {S.-C.}\ \bibnamefont {Zhang}}, \bibinfo {author} {\bibfnamefont
    {K.}~\bibnamefont {Liu}}, \bibinfo {author} {\bibfnamefont {J.}~\bibnamefont
    {Xia}},\ and\ \bibinfo {author} {\bibfnamefont {K.~L.}\ \bibnamefont
    {Wang}},\ }\bibfield  {title} {\bibinfo {title} {Chiral {M}ajorana fermion
    modes in a quantum anomalous {H}all insulator-superconductor structure},\
    }\href {https://doi.org/10.1126/science.aag2792} {\bibfield  {journal}
    {\bibinfo  {journal} {Science}\ }\textbf {\bibinfo {volume} {357}},\ \bibinfo
    {pages} {294} (\bibinfo {year} {2017})}\BibitemShut {NoStop}%
  \bibitem [{\citenamefont {Hsieh}\ \emph {et~al.}(2008)\citenamefont {Hsieh},
    \citenamefont {Qian}, \citenamefont {Wray}, \citenamefont {Xia},
    \citenamefont {Hor}, \citenamefont {Cava},\ and\ \citenamefont
    {Hasan}}]{Hsieh2008}%
    \BibitemOpen
    \bibfield  {author} {\bibinfo {author} {\bibfnamefont {D.}~\bibnamefont
    {Hsieh}}, \bibinfo {author} {\bibfnamefont {D.}~\bibnamefont {Qian}},
    \bibinfo {author} {\bibfnamefont {L.}~\bibnamefont {Wray}}, \bibinfo {author}
    {\bibfnamefont {Y.}~\bibnamefont {Xia}}, \bibinfo {author} {\bibfnamefont
    {Y.~S.}\ \bibnamefont {Hor}}, \bibinfo {author} {\bibfnamefont {R.~J.}\
    \bibnamefont {Cava}},\ and\ \bibinfo {author} {\bibfnamefont {M.~Z.}\
    \bibnamefont {Hasan}},\ }\bibfield  {title} {\bibinfo {title} {A topological
    {D}irac insulator in a quantum spin {H}all phase},\ }\href
    {https://doi.org/10.1038/nature06843} {\bibfield  {journal} {\bibinfo
    {journal} {Nature}\ }\textbf {\bibinfo {volume} {452}},\ \bibinfo {pages}
    {970} (\bibinfo {year} {2008})}\BibitemShut {NoStop}%
  \bibitem [{\citenamefont {Chen}\ \emph {et~al.}(2009)\citenamefont {Chen},
    \citenamefont {Analytis}, \citenamefont {Chu}, \citenamefont {Liu},
    \citenamefont {Mo}, \citenamefont {Qi}, \citenamefont {Zhang}, \citenamefont
    {Lu}, \citenamefont {Dai}, \citenamefont {Fang}, \citenamefont {Zhang},
    \citenamefont {Fisher}, \citenamefont {Hussain},\ and\ \citenamefont
    {Shen}}]{Chen2009}%
    \BibitemOpen
    \bibfield  {author} {\bibinfo {author} {\bibfnamefont {Y.~L.}\ \bibnamefont
    {Chen}}, \bibinfo {author} {\bibfnamefont {J.~G.}\ \bibnamefont {Analytis}},
    \bibinfo {author} {\bibfnamefont {J.-H.}\ \bibnamefont {Chu}}, \bibinfo
    {author} {\bibfnamefont {Z.~K.}\ \bibnamefont {Liu}}, \bibinfo {author}
    {\bibfnamefont {S.-K.}\ \bibnamefont {Mo}}, \bibinfo {author} {\bibfnamefont
    {X.~L.}\ \bibnamefont {Qi}}, \bibinfo {author} {\bibfnamefont {H.~J.}\
    \bibnamefont {Zhang}}, \bibinfo {author} {\bibfnamefont {D.~H.}\ \bibnamefont
    {Lu}}, \bibinfo {author} {\bibfnamefont {X.}~\bibnamefont {Dai}}, \bibinfo
    {author} {\bibfnamefont {Z.}~\bibnamefont {Fang}}, \bibinfo {author}
    {\bibfnamefont {S.~C.}\ \bibnamefont {Zhang}}, \bibinfo {author}
    {\bibfnamefont {I.~R.}\ \bibnamefont {Fisher}}, \bibinfo {author}
    {\bibfnamefont {Z.}~\bibnamefont {Hussain}},\ and\ \bibinfo {author}
    {\bibfnamefont {Z.-X.}\ \bibnamefont {Shen}},\ }\bibfield  {title} {\bibinfo
    {title} {Experimental {R}ealization of a {T}hree-{D}imensional {T}opological
    {I}nsulator, {Bi$_{2}$Te$_{3}$}},\ }\href
    {https://doi.org/10.1126/science.1173034} {\bibfield  {journal} {\bibinfo
    {journal} {Science}\ }\textbf {\bibinfo {volume} {325}},\ \bibinfo {pages}
    {178} (\bibinfo {year} {2009})}\BibitemShut {NoStop}%
  \bibitem [{\citenamefont {Xia}\ \emph {et~al.}(2009)\citenamefont {Xia},
    \citenamefont {Qian}, \citenamefont {Hsieh}, \citenamefont {Wray},
    \citenamefont {Pal}, \citenamefont {Lin}, \citenamefont {Bansil},
    \citenamefont {Grauer}, \citenamefont {Hor}, \citenamefont {Cava},\ and\
    \citenamefont {Hasan}}]{Xia2009}%
    \BibitemOpen
    \bibfield  {author} {\bibinfo {author} {\bibfnamefont {Y.}~\bibnamefont
    {Xia}}, \bibinfo {author} {\bibfnamefont {D.}~\bibnamefont {Qian}}, \bibinfo
    {author} {\bibfnamefont {D.}~\bibnamefont {Hsieh}}, \bibinfo {author}
    {\bibfnamefont {L.}~\bibnamefont {Wray}}, \bibinfo {author} {\bibfnamefont
    {A.}~\bibnamefont {Pal}}, \bibinfo {author} {\bibfnamefont {H.}~\bibnamefont
    {Lin}}, \bibinfo {author} {\bibfnamefont {A.}~\bibnamefont {Bansil}},
    \bibinfo {author} {\bibfnamefont {D.}~\bibnamefont {Grauer}}, \bibinfo
    {author} {\bibfnamefont {Y.~S.}\ \bibnamefont {Hor}}, \bibinfo {author}
    {\bibfnamefont {R.~J.}\ \bibnamefont {Cava}},\ and\ \bibinfo {author}
    {\bibfnamefont {M.~Z.}\ \bibnamefont {Hasan}},\ }\bibfield  {title} {\bibinfo
    {title} {Observation of a large-gap topological-insulator class with a single
    {D}irac cone on the surface},\ }\href {https://doi.org/10.1038/nphys1274}
    {\bibfield  {journal} {\bibinfo  {journal} {Nat. Phys.}\ }\textbf {\bibinfo
    {volume} {5}},\ \bibinfo {pages} {398} (\bibinfo {year} {2009})}\BibitemShut
    {NoStop}%
  \bibitem [{\citenamefont {Oka}\ and\ \citenamefont {Aoki}(2009)}]{Oka2009}%
    \BibitemOpen
    \bibfield  {author} {\bibinfo {author} {\bibfnamefont {T.}~\bibnamefont
    {Oka}}\ and\ \bibinfo {author} {\bibfnamefont {H.}~\bibnamefont {Aoki}},\
    }\bibfield  {title} {\bibinfo {title} {Photovoltaic {H}all effect in
    graphene},\ }\href {https://doi.org/10.1103/physrevb.79.081406} {\bibfield
    {journal} {\bibinfo  {journal} {Phys. Rev. B}\ }\textbf {\bibinfo {volume}
    {79}},\ \bibinfo {pages} {081406} (\bibinfo {year} {2009})}\BibitemShut
    {NoStop}%
  \bibitem [{\citenamefont {Kitagawa}\ \emph
    {et~al.}(2010{\natexlab{a}})\citenamefont {Kitagawa}, \citenamefont {Rudner},
    \citenamefont {Berg},\ and\ \citenamefont {Demler}}]{Kitagawa2010}%
    \BibitemOpen
    \bibfield  {author} {\bibinfo {author} {\bibfnamefont {T.}~\bibnamefont
    {Kitagawa}}, \bibinfo {author} {\bibfnamefont {M.~S.}\ \bibnamefont
    {Rudner}}, \bibinfo {author} {\bibfnamefont {E.}~\bibnamefont {Berg}},\ and\
    \bibinfo {author} {\bibfnamefont {E.}~\bibnamefont {Demler}},\ }\bibfield
    {title} {\bibinfo {title} {Exploring topological phases with quantum walks},\
    }\href {https://doi.org/10.1103/physreva.82.033429} {\bibfield  {journal}
    {\bibinfo  {journal} {Phys. Rev. A}\ }\textbf {\bibinfo {volume} {82}},\
    \bibinfo {pages} {033429} (\bibinfo {year} {2010}{\natexlab{a}})}\BibitemShut
    {NoStop}%
  \bibitem [{\citenamefont {Kitagawa}\ \emph
    {et~al.}(2010{\natexlab{b}})\citenamefont {Kitagawa}, \citenamefont {Berg},
    \citenamefont {Rudner},\ and\ \citenamefont {Demler}}]{Kitagawa2010a}%
    \BibitemOpen
    \bibfield  {author} {\bibinfo {author} {\bibfnamefont {T.}~\bibnamefont
    {Kitagawa}}, \bibinfo {author} {\bibfnamefont {E.}~\bibnamefont {Berg}},
    \bibinfo {author} {\bibfnamefont {M.}~\bibnamefont {Rudner}},\ and\ \bibinfo
    {author} {\bibfnamefont {E.}~\bibnamefont {Demler}},\ }\bibfield  {title}
    {\bibinfo {title} {Topological characterization of periodically driven
    quantum systems},\ }\href {https://doi.org/10.1103/physrevb.82.235114}
    {\bibfield  {journal} {\bibinfo  {journal} {Phys. Rev. B}\ }\textbf {\bibinfo
    {volume} {82}},\ \bibinfo {pages} {235114} (\bibinfo {year}
    {2010}{\natexlab{b}})}\BibitemShut {NoStop}%
  \bibitem [{\citenamefont {ichi Inoue}\ and\ \citenamefont
    {Tanaka}(2010)}]{Inoue2010}%
    \BibitemOpen
    \bibfield  {author} {\bibinfo {author} {\bibfnamefont {J.}~\bibnamefont {ichi
    Inoue}}\ and\ \bibinfo {author} {\bibfnamefont {A.}~\bibnamefont {Tanaka}},\
    }\bibfield  {title} {\bibinfo {title} {Photoinduced {T}ransition between
    {C}onventional and {T}opological {I}nsulators in {T}wo-{D}imensional
    {E}lectronic {S}ystems},\ }\href
    {https://doi.org/10.1103/physrevlett.105.017401} {\bibfield  {journal}
    {\bibinfo  {journal} {Phys. Rev. Lett.}\ }\textbf {\bibinfo {volume} {105}},\
    \bibinfo {pages} {017401} (\bibinfo {year} {2010})}\BibitemShut {NoStop}%
  \bibitem [{\citenamefont {Jiang}\ \emph {et~al.}(2011)\citenamefont {Jiang},
    \citenamefont {Kitagawa}, \citenamefont {Alicea}, \citenamefont {Akhmerov},
    \citenamefont {Pekker}, \citenamefont {Refael}, \citenamefont {Cirac},
    \citenamefont {Demler}, \citenamefont {Lukin},\ and\ \citenamefont
    {Zoller}}]{Jiang2011}%
    \BibitemOpen
    \bibfield  {author} {\bibinfo {author} {\bibfnamefont {L.}~\bibnamefont
    {Jiang}}, \bibinfo {author} {\bibfnamefont {T.}~\bibnamefont {Kitagawa}},
    \bibinfo {author} {\bibfnamefont {J.}~\bibnamefont {Alicea}}, \bibinfo
    {author} {\bibfnamefont {A.~R.}\ \bibnamefont {Akhmerov}}, \bibinfo {author}
    {\bibfnamefont {D.}~\bibnamefont {Pekker}}, \bibinfo {author} {\bibfnamefont
    {G.}~\bibnamefont {Refael}}, \bibinfo {author} {\bibfnamefont {J.~I.}\
    \bibnamefont {Cirac}}, \bibinfo {author} {\bibfnamefont {E.}~\bibnamefont
    {Demler}}, \bibinfo {author} {\bibfnamefont {M.~D.}\ \bibnamefont {Lukin}},\
    and\ \bibinfo {author} {\bibfnamefont {P.}~\bibnamefont {Zoller}},\
    }\bibfield  {title} {\bibinfo {title} {Majorana {F}ermions in {E}quilibrium
    and in {D}riven {C}old-{A}tom {Q}uantum {W}ires},\ }\href
    {https://doi.org/10.1103/physrevlett.106.220402} {\bibfield  {journal}
    {\bibinfo  {journal} {Phys. Rev. Lett.}\ }\textbf {\bibinfo {volume} {106}},\
    \bibinfo {pages} {220402} (\bibinfo {year} {2011})}\BibitemShut {NoStop}%
  \bibitem [{\citenamefont {Lindner}\ \emph {et~al.}(2011)\citenamefont
    {Lindner}, \citenamefont {Refael},\ and\ \citenamefont
    {Galitski}}]{Lindner2011}%
    \BibitemOpen
    \bibfield  {author} {\bibinfo {author} {\bibfnamefont {N.~H.}\ \bibnamefont
    {Lindner}}, \bibinfo {author} {\bibfnamefont {G.}~\bibnamefont {Refael}},\
    and\ \bibinfo {author} {\bibfnamefont {V.}~\bibnamefont {Galitski}},\
    }\bibfield  {title} {\bibinfo {title} {Floquet topological insulator in
    semiconductor quantum wells},\ }\href {https://doi.org/10.1038/nphys1926}
    {\bibfield  {journal} {\bibinfo  {journal} {Nat. Phys.}\ }\textbf {\bibinfo
    {volume} {7}},\ \bibinfo {pages} {490} (\bibinfo {year} {2011})}\BibitemShut
    {NoStop}%
  \bibitem [{\citenamefont {Kitagawa}\ \emph {et~al.}(2011)\citenamefont
    {Kitagawa}, \citenamefont {Oka}, \citenamefont {Brataas}, \citenamefont
    {Fu},\ and\ \citenamefont {Demler}}]{Kitagawa2011}%
    \BibitemOpen
    \bibfield  {author} {\bibinfo {author} {\bibfnamefont {T.}~\bibnamefont
    {Kitagawa}}, \bibinfo {author} {\bibfnamefont {T.}~\bibnamefont {Oka}},
    \bibinfo {author} {\bibfnamefont {A.}~\bibnamefont {Brataas}}, \bibinfo
    {author} {\bibfnamefont {L.}~\bibnamefont {Fu}},\ and\ \bibinfo {author}
    {\bibfnamefont {E.}~\bibnamefont {Demler}},\ }\bibfield  {title} {\bibinfo
    {title} {Transport properties of nonequilibrium systems under the application
    of light: Photoinduced quantum {H}all insulators without {L}andau levels},\
    }\href {https://doi.org/10.1103/physrevb.84.235108} {\bibfield  {journal}
    {\bibinfo  {journal} {Phys. Rev. B}\ }\textbf {\bibinfo {volume} {84}},\
    \bibinfo {pages} {235108} (\bibinfo {year} {2011})}\BibitemShut {NoStop}%
  \bibitem [{\citenamefont {Wang}\ \emph {et~al.}(2013)\citenamefont {Wang},
    \citenamefont {Steinberg}, \citenamefont {Jarillo-Herrero},\ and\
    \citenamefont {Gedik}}]{Wang2013}%
    \BibitemOpen
    \bibfield  {author} {\bibinfo {author} {\bibfnamefont {Y.~H.}\ \bibnamefont
    {Wang}}, \bibinfo {author} {\bibfnamefont {H.}~\bibnamefont {Steinberg}},
    \bibinfo {author} {\bibfnamefont {P.}~\bibnamefont {Jarillo-Herrero}},\ and\
    \bibinfo {author} {\bibfnamefont {N.}~\bibnamefont {Gedik}},\ }\bibfield
    {title} {\bibinfo {title} {Observation of {F}loquet-{B}loch {S}tates on the
    {S}urface of a {T}opological {I}nsulator},\ }\href
    {https://doi.org/10.1126/science.1239834} {\bibfield  {journal} {\bibinfo
    {journal} {Science}\ }\textbf {\bibinfo {volume} {342}},\ \bibinfo {pages}
    {453} (\bibinfo {year} {2013})}\BibitemShut {NoStop}%
  \bibitem [{\citenamefont {Mahmood}\ \emph {et~al.}(2016)\citenamefont
    {Mahmood}, \citenamefont {Chan}, \citenamefont {Alpichshev}, \citenamefont
    {Gardner}, \citenamefont {Lee}, \citenamefont {Lee},\ and\ \citenamefont
    {Gedik}}]{Mahmood2016}%
    \BibitemOpen
    \bibfield  {author} {\bibinfo {author} {\bibfnamefont {F.}~\bibnamefont
    {Mahmood}}, \bibinfo {author} {\bibfnamefont {C.-K.}\ \bibnamefont {Chan}},
    \bibinfo {author} {\bibfnamefont {Z.}~\bibnamefont {Alpichshev}}, \bibinfo
    {author} {\bibfnamefont {D.}~\bibnamefont {Gardner}}, \bibinfo {author}
    {\bibfnamefont {Y.}~\bibnamefont {Lee}}, \bibinfo {author} {\bibfnamefont
    {P.~A.}\ \bibnamefont {Lee}},\ and\ \bibinfo {author} {\bibfnamefont
    {N.}~\bibnamefont {Gedik}},\ }\bibfield  {title} {\bibinfo {title} {Selective
    scattering between {F}loquet-{B}loch and {V}olkov states in a topological
    insulator},\ }\href {https://doi.org/10.1038/nphys3609} {\bibfield  {journal}
    {\bibinfo  {journal} {Nat. Phys.}\ }\textbf {\bibinfo {volume} {12}},\
    \bibinfo {pages} {306} (\bibinfo {year} {2016})}\BibitemShut {NoStop}%
  \bibitem [{\citenamefont {Jotzu}\ \emph {et~al.}(2014)\citenamefont {Jotzu},
    \citenamefont {Messer}, \citenamefont {Desbuquois}, \citenamefont {Lebrat},
    \citenamefont {Uehlinger}, \citenamefont {Greif},\ and\ \citenamefont
    {Esslinger}}]{Jotzu2014}%
    \BibitemOpen
    \bibfield  {author} {\bibinfo {author} {\bibfnamefont {G.}~\bibnamefont
    {Jotzu}}, \bibinfo {author} {\bibfnamefont {M.}~\bibnamefont {Messer}},
    \bibinfo {author} {\bibfnamefont {R.}~\bibnamefont {Desbuquois}}, \bibinfo
    {author} {\bibfnamefont {M.}~\bibnamefont {Lebrat}}, \bibinfo {author}
    {\bibfnamefont {T.}~\bibnamefont {Uehlinger}}, \bibinfo {author}
    {\bibfnamefont {D.}~\bibnamefont {Greif}},\ and\ \bibinfo {author}
    {\bibfnamefont {T.}~\bibnamefont {Esslinger}},\ }\bibfield  {title} {\bibinfo
    {title} {Experimental realization of the topological {H}aldane model with
    ultracold fermions},\ }\href {https://doi.org/10.1038/nature13915} {\bibfield
     {journal} {\bibinfo  {journal} {Nature}\ }\textbf {\bibinfo {volume}
    {515}},\ \bibinfo {pages} {237} (\bibinfo {year} {2014})}\BibitemShut
    {NoStop}%
  \bibitem [{\citenamefont {Fl{\" a}schner}\ \emph {et~al.}(2016)\citenamefont
    {Fl{\" a}schner}, \citenamefont {Rem}, \citenamefont {Tarnowski},
    \citenamefont {Vogel}, \citenamefont {L{\" u}hmann}, \citenamefont
    {Sengstock},\ and\ \citenamefont {Weitenberg}}]{Flaeschner2016}%
    \BibitemOpen
    \bibfield  {author} {\bibinfo {author} {\bibfnamefont {N.}~\bibnamefont
    {Fl{\" a}schner}}, \bibinfo {author} {\bibfnamefont {B.~S.}\ \bibnamefont
    {Rem}}, \bibinfo {author} {\bibfnamefont {M.}~\bibnamefont {Tarnowski}},
    \bibinfo {author} {\bibfnamefont {D.}~\bibnamefont {Vogel}}, \bibinfo
    {author} {\bibfnamefont {D.-S.}\ \bibnamefont {L{\" u}hmann}}, \bibinfo
    {author} {\bibfnamefont {K.}~\bibnamefont {Sengstock}},\ and\ \bibinfo
    {author} {\bibfnamefont {C.}~\bibnamefont {Weitenberg}},\ }\bibfield  {title}
    {\bibinfo {title} {Experimental reconstruction of the {B}erry curvature in a
    {F}loquet {B}loch band},\ }\href {https://doi.org/10.1126/science.aad4568}
    {\bibfield  {journal} {\bibinfo  {journal} {Science}\ }\textbf {\bibinfo
    {volume} {352}},\ \bibinfo {pages} {1091} (\bibinfo {year}
    {2016})}\BibitemShut {NoStop}%
  \bibitem [{\citenamefont {Wintersperger}\ \emph {et~al.}(2020)\citenamefont
    {Wintersperger}, \citenamefont {Braun}, \citenamefont {{\" U}nal},
    \citenamefont {Eckardt}, \citenamefont {Liberto}, \citenamefont {Goldman},
    \citenamefont {Bloch},\ and\ \citenamefont
    {Aidelsburger}}]{Wintersperger2020}%
    \BibitemOpen
    \bibfield  {author} {\bibinfo {author} {\bibfnamefont {K.}~\bibnamefont
    {Wintersperger}}, \bibinfo {author} {\bibfnamefont {C.}~\bibnamefont
    {Braun}}, \bibinfo {author} {\bibfnamefont {F.~N.}\ \bibnamefont {{\"
    U}nal}}, \bibinfo {author} {\bibfnamefont {A.}~\bibnamefont {Eckardt}},
    \bibinfo {author} {\bibfnamefont {M.~D.}\ \bibnamefont {Liberto}}, \bibinfo
    {author} {\bibfnamefont {N.}~\bibnamefont {Goldman}}, \bibinfo {author}
    {\bibfnamefont {I.}~\bibnamefont {Bloch}},\ and\ \bibinfo {author}
    {\bibfnamefont {M.}~\bibnamefont {Aidelsburger}},\ }\bibfield  {title}
    {\bibinfo {title} {Realization of an anomalous {F}loquet topological system
    with ultracold atoms},\ }\href {https://doi.org/10.1038/s41567-020-0949-y}
    {\bibfield  {journal} {\bibinfo  {journal} {Nat. Phys.}\ }\textbf {\bibinfo
    {volume} {16}},\ \bibinfo {pages} {1058} (\bibinfo {year}
    {2020})}\BibitemShut {NoStop}%
  \bibitem [{\citenamefont {Slager}\ \emph {et~al.}()\citenamefont {Slager},
    \citenamefont {Bouhon},\ and\ \citenamefont {{\" U}nal}}]{Slager2022}%
    \BibitemOpen
    \bibfield  {author} {\bibinfo {author} {\bibfnamefont {R.-J.}\ \bibnamefont
    {Slager}}, \bibinfo {author} {\bibfnamefont {A.}~\bibnamefont {Bouhon}},\
    and\ \bibinfo {author} {\bibfnamefont {F.~N.}\ \bibnamefont {{\" U}nal}},\
    }\href@noop {} {\bibinfo {title} {{Floquet multi-gap topology: Non-Abelian
    braiding and anomalous Dirac string phase}}},\ \Eprint
    {https://arxiv.org/abs/2208.12824} {arXiv:2208.12824} \BibitemShut {NoStop}%
  \bibitem [{\citenamefont {Cayssol}\ \emph {et~al.}(2013)\citenamefont
    {Cayssol}, \citenamefont {D{\'{o}}ra}, \citenamefont {Simon},\ and\
    \citenamefont {Moessner}}]{Cayssol2013}%
    \BibitemOpen
    \bibfield  {author} {\bibinfo {author} {\bibfnamefont {J.}~\bibnamefont
    {Cayssol}}, \bibinfo {author} {\bibfnamefont {B.}~\bibnamefont {D{\'{o}}ra}},
    \bibinfo {author} {\bibfnamefont {F.}~\bibnamefont {Simon}},\ and\ \bibinfo
    {author} {\bibfnamefont {R.}~\bibnamefont {Moessner}},\ }\bibfield  {title}
    {\bibinfo {title} {Floquet topological insulators},\ }\href
    {https://doi.org/10.1002/pssr.201206451} {\bibfield  {journal} {\bibinfo
    {journal} {Phys. Status Solidi RRL}\ }\textbf {\bibinfo {volume} {7}},\
    \bibinfo {pages} {101} (\bibinfo {year} {2013})}\BibitemShut {NoStop}%
  \bibitem [{\citenamefont {Nag}\ \emph {et~al.}(2019)\citenamefont {Nag},
    \citenamefont {Slager}, \citenamefont {Higuchi},\ and\ \citenamefont
    {Oka}}]{Nag2019}%
    \BibitemOpen
    \bibfield  {author} {\bibinfo {author} {\bibfnamefont {T.}~\bibnamefont
    {Nag}}, \bibinfo {author} {\bibfnamefont {R.-J.}\ \bibnamefont {Slager}},
    \bibinfo {author} {\bibfnamefont {T.}~\bibnamefont {Higuchi}},\ and\ \bibinfo
    {author} {\bibfnamefont {T.}~\bibnamefont {Oka}},\ }\bibfield  {title}
    {\bibinfo {title} {Dynamical synchronization transition in interacting
    electron systems},\ }\href {https://doi.org/10.1103/PhysRevB.100.134301}
    {\bibfield  {journal} {\bibinfo  {journal} {Phys. Rev. B}\ }\textbf {\bibinfo
    {volume} {100}},\ \bibinfo {pages} {134301} (\bibinfo {year}
    {2019})}\BibitemShut {NoStop}%
  \bibitem [{\citenamefont {Harper}\ \emph {et~al.}(2020)\citenamefont {Harper},
    \citenamefont {Roy}, \citenamefont {Rudner},\ and\ \citenamefont
    {Sondhi}}]{Harper2020}%
    \BibitemOpen
    \bibfield  {author} {\bibinfo {author} {\bibfnamefont {F.}~\bibnamefont
    {Harper}}, \bibinfo {author} {\bibfnamefont {R.}~\bibnamefont {Roy}},
    \bibinfo {author} {\bibfnamefont {M.~S.}\ \bibnamefont {Rudner}},\ and\
    \bibinfo {author} {\bibfnamefont {S.~L.}\ \bibnamefont {Sondhi}},\ }\bibfield
     {title} {\bibinfo {title} {Topology and {B}roken {S}ymmetry in {F}loquet
    {S}ystems},\ }\href
    {https://doi.org/10.1146/annurev-conmatphys-031218-013721} {\bibfield
    {journal} {\bibinfo  {journal} {Annu. Rev. Condens. Matter Phys.}\ }\textbf
    {\bibinfo {volume} {11}},\ \bibinfo {pages} {345} (\bibinfo {year}
    {2020})}\BibitemShut {NoStop}%
  \bibitem [{\citenamefont {Rudner}\ and\ \citenamefont
    {Lindner}(2020)}]{Rudner2020}%
    \BibitemOpen
    \bibfield  {author} {\bibinfo {author} {\bibfnamefont {M.~S.}\ \bibnamefont
    {Rudner}}\ and\ \bibinfo {author} {\bibfnamefont {N.~H.}\ \bibnamefont
    {Lindner}},\ }\bibfield  {title} {\bibinfo {title} {Band structure
    engineering and non-equilibrium dynamics in {F}loquet topological
    insulators},\ }\href {https://doi.org/10.1038/s42254-020-0170-z} {\bibfield
    {journal} {\bibinfo  {journal} {Nat. Rev. Phys.}\ }\textbf {\bibinfo {volume}
    {2}},\ \bibinfo {pages} {229} (\bibinfo {year} {2020})}\BibitemShut {NoStop}%
  \bibitem [{\citenamefont {Jangjan}\ and\ \citenamefont
    {Hosseini}(2020)}]{Jangjan2020}%
    \BibitemOpen
    \bibfield  {author} {\bibinfo {author} {\bibfnamefont {M.}~\bibnamefont
    {Jangjan}}\ and\ \bibinfo {author} {\bibfnamefont {M.~V.}\ \bibnamefont
    {Hosseini}},\ }\bibfield  {title} {\bibinfo {title} {Floquet engineering of
    topological metal states and hybridization of edge states with bulk states in
    dimerized two-leg ladders},\ }\href
    {https://doi.org/10.1038/s41598-020-71196-3} {\bibfield  {journal} {\bibinfo
    {journal} {Sci. Rep.}\ }\textbf {\bibinfo {volume} {10}},\ \bibinfo {pages}
    {14256} (\bibinfo {year} {2020})}\BibitemShut {NoStop}%
  {\bibitem [{\citenamefont {Molignini}\ \emph {et~al.}(2018)\citenamefont
    {Molignini}, \citenamefont {Chen},\ and\ \citenamefont
    {Chitra}}]{Molignini2018}%
    \BibitemOpen
    \bibfield  {author} {\bibinfo {author} {\bibfnamefont {P.}~\bibnamefont
    {Molignini}}, \bibinfo {author} {\bibfnamefont {W.}~\bibnamefont {Chen}},\
    and\ \bibinfo {author} {\bibfnamefont {R.}~\bibnamefont {Chitra}},\
    }\bibfield  {title} {\bibinfo {title} {Universal quantum criticality in
    static and floquet-majorana chains},\ }\href
    {https://doi.org/10.1103/PhysRevB.98.125129} {\bibfield  {journal} {\bibinfo
    {journal} {Phys. Rev. B}\ }\textbf {\bibinfo {volume} {98}},\ \bibinfo
    {pages} {125129} (\bibinfo {year} {2018})}\BibitemShut {NoStop}}%
  {\bibitem [{\citenamefont {Molignini}\ \emph {et~al.}(2020)\citenamefont
    {Molignini}, \citenamefont {Chen},\ and\ \citenamefont
    {Chitra}}]{Molignini2020}%
    \BibitemOpen
    \bibfield  {author} {\bibinfo {author} {\bibfnamefont {P.}~\bibnamefont
    {Molignini}}, \bibinfo {author} {\bibfnamefont {W.}~\bibnamefont {Chen}},\
    and\ \bibinfo {author} {\bibfnamefont {R.}~\bibnamefont {Chitra}},\
    }\bibfield  {title} {\bibinfo {title} {Generating quantum multicriticality in
    topological insulators by periodic driving},\ }\href
    {https://doi.org/10.1103/PhysRevB.101.165106} {\bibfield  {journal} {\bibinfo
     {journal} {Phys. Rev. B}\ }\textbf {\bibinfo {volume} {101}},\ \bibinfo
    {pages} {165106} (\bibinfo {year} {2020})}\BibitemShut {NoStop}} %
  {\bibitem [{\citenamefont {Molignini}(2020)}]{Molignini2020a}%
    \BibitemOpen
    \bibfield  {author} {\bibinfo {author} {\bibfnamefont {P.}~\bibnamefont
    {Molignini}},\ }\bibfield  {title} {\bibinfo {title} {Edge mode manipulation
    through commensurate multifrequency driving},\ }\href
    {https://doi.org/10.1103/PhysRevB.102.235143} {\bibfield  {journal} {\bibinfo
     {journal} {Phys. Rev. B}\ }\textbf {\bibinfo {volume} {102}},\ \bibinfo
    {pages} {235143} (\bibinfo {year} {2020})}\BibitemShut {NoStop}}%
  {\bibitem [{\citenamefont {Molignini}\ \emph {et~al.}(2021)\citenamefont
    {Molignini}, \citenamefont {Celades}, \citenamefont {Chitra},\ and\
    \citenamefont {Chen}}]{Molignini2021}%
    \BibitemOpen
    \bibfield  {author} {\bibinfo {author} {\bibfnamefont {P.}~\bibnamefont
    {Molignini}}, \bibinfo {author} {\bibfnamefont {A.~G.}\ \bibnamefont
    {Celades}}, \bibinfo {author} {\bibfnamefont {R.}~\bibnamefont {Chitra}},\
    and\ \bibinfo {author} {\bibfnamefont {W.}~\bibnamefont {Chen}},\ }\bibfield
    {title} {\bibinfo {title} {Crossdimensional universality classes in static
    and periodically driven kitaev models},\ }\href
    {https://doi.org/10.1103/PhysRevB.103.184507} {\bibfield  {journal} {\bibinfo
     {journal} {Phys. Rev. B}\ }\textbf {\bibinfo {volume} {103}},\ \bibinfo
    {pages} {184507} (\bibinfo {year} {2021})}\BibitemShut {NoStop}}%
  \bibitem [{\citenamefont {Nag}\ and\ \citenamefont {Roy}(2021)}]{Nag2021}%
    \BibitemOpen
    \bibfield  {author} {\bibinfo {author} {\bibfnamefont {T.}~\bibnamefont
    {Nag}}\ and\ \bibinfo {author} {\bibfnamefont {B.}~\bibnamefont {Roy}},\
    }\bibfield  {title} {\bibinfo {title} {{Anomalous and normal dislocation
    modes in Floquet topological insulators}},\ }\href
    {https://doi.org/10.1038/s42005-021-00659-4} {\bibfield  {journal} {\bibinfo
    {journal} {Commun. Phys.}\ }\textbf {\bibinfo {volume} {4}},\ \bibinfo
    {pages} {157} (\bibinfo {year} {2021})}\BibitemShut {NoStop}%
  \bibitem [{\citenamefont {Ghosh}\ \emph {et~al.}(2022)\citenamefont {Ghosh},
    \citenamefont {Nag},\ and\ \citenamefont {Saha}}]{Ghosh2022}%
    \BibitemOpen
    \bibfield  {author} {\bibinfo {author} {\bibfnamefont {A.~K.}\ \bibnamefont
    {Ghosh}}, \bibinfo {author} {\bibfnamefont {T.}~\bibnamefont {Nag}},\ and\
    \bibinfo {author} {\bibfnamefont {A.}~\bibnamefont {Saha}},\ }\bibfield
    {title} {\bibinfo {title} {{Systematic generation of the cascade of anomalous
    dynamical first- and higher-order modes in Floquet topological insulators}},\
    }\href {https://doi.org/10.1103/PhysRevB.105.115418} {\bibfield  {journal}
    {\bibinfo  {journal} {Phys. Rev. B}\ }\textbf {\bibinfo {volume} {105}},\
    \bibinfo {pages} {115418} (\bibinfo {year} {2022})}\BibitemShut {NoStop}%
  \bibitem [{\citenamefont {Jangjan}\ \emph {et~al.}(2022)\citenamefont
    {Jangjan}, \citenamefont {Foa~Torres},\ and\ \citenamefont
    {Hosseini}}]{Jangjan2022}%
    \BibitemOpen
    \bibfield  {author} {\bibinfo {author} {\bibfnamefont {M.}~\bibnamefont
    {Jangjan}}, \bibinfo {author} {\bibfnamefont {L.~E.~F.}\ \bibnamefont
    {Foa~Torres}},\ and\ \bibinfo {author} {\bibfnamefont {M.~V.}\ \bibnamefont
    {Hosseini}},\ }\bibfield  {title} {\bibinfo {title} {Floquet topological
    phase transitions in a periodically quenched dimer},\ }\href
    {https://doi.org/10.1103/PhysRevB.106.224306} {\bibfield  {journal} {\bibinfo
     {journal} {Phys. Rev. B}\ }\textbf {\bibinfo {volume} {106}},\ \bibinfo
    {pages} {224306} (\bibinfo {year} {2022})}\BibitemShut {NoStop}%
  \bibitem [{\citenamefont {Bukov}\ \emph {et~al.}(2015)\citenamefont {Bukov},
    \citenamefont {D'Alessio},\ and\ \citenamefont {Polkovnikov}}]{Bukov2015}%
    \BibitemOpen
    \bibfield  {author} {\bibinfo {author} {\bibfnamefont {M.}~\bibnamefont
    {Bukov}}, \bibinfo {author} {\bibfnamefont {L.}~\bibnamefont {D'Alessio}},\
    and\ \bibinfo {author} {\bibfnamefont {A.}~\bibnamefont {Polkovnikov}},\
    }\bibfield  {title} {\bibinfo {title} {Universal high-frequency behavior of
    periodically driven systems: from dynamical stabilization to {F}loquet
    engineering},\ }\href {https://doi.org/10.1080/00018732.2015.1055918}
    {\bibfield  {journal} {\bibinfo  {journal} {Adv. Phys.}\ }\textbf {\bibinfo
    {volume} {64}},\ \bibinfo {pages} {139} (\bibinfo {year} {2015})}\BibitemShut
    {NoStop}%
  \bibitem [{\citenamefont {Eckardt}(2017)}]{Eckardt2017}%
    \BibitemOpen
    \bibfield  {author} {\bibinfo {author} {\bibfnamefont {A.}~\bibnamefont
    {Eckardt}},\ }\bibfield  {title} {\bibinfo {title} {Colloquium: Atomic
    quantum gases in periodically driven optical lattices},\ }\href
    {https://doi.org/10.1103/revmodphys.89.011004} {\bibfield  {journal}
    {\bibinfo  {journal} {Rev. Mod. Phys.}\ }\textbf {\bibinfo {volume} {89}},\
    \bibinfo {pages} {011004} (\bibinfo {year} {2017})}\BibitemShut {NoStop}%
  \bibitem [{\citenamefont {Kitagawa}\ \emph {et~al.}(2012)\citenamefont
    {Kitagawa}, \citenamefont {Broome}, \citenamefont {Fedrizzi}, \citenamefont
    {Rudner}, \citenamefont {Berg}, \citenamefont {Kassal}, \citenamefont
    {Aspuru-Guzik}, \citenamefont {Demler},\ and\ \citenamefont
    {White}}]{Kitagawa2012}%
    \BibitemOpen
    \bibfield  {author} {\bibinfo {author} {\bibfnamefont {T.}~\bibnamefont
    {Kitagawa}}, \bibinfo {author} {\bibfnamefont {M.~A.}\ \bibnamefont
    {Broome}}, \bibinfo {author} {\bibfnamefont {A.}~\bibnamefont {Fedrizzi}},
    \bibinfo {author} {\bibfnamefont {M.~S.}\ \bibnamefont {Rudner}}, \bibinfo
    {author} {\bibfnamefont {E.}~\bibnamefont {Berg}}, \bibinfo {author}
    {\bibfnamefont {I.}~\bibnamefont {Kassal}}, \bibinfo {author} {\bibfnamefont
    {A.}~\bibnamefont {Aspuru-Guzik}}, \bibinfo {author} {\bibfnamefont
    {E.}~\bibnamefont {Demler}},\ and\ \bibinfo {author} {\bibfnamefont {A.~G.}\
    \bibnamefont {White}},\ }\bibfield  {title} {\bibinfo {title} {Observation of
    topologically protected bound states in photonic quantum walks},\ }\href
    {https://doi.org/10.1038/ncomms1872} {\bibfield  {journal} {\bibinfo
    {journal} {Nat. Commun.}\ }\textbf {\bibinfo {volume} {3}},\ \bibinfo {pages}
    {882} (\bibinfo {year} {2012})}\BibitemShut {NoStop}%
  \bibitem [{\citenamefont {Rudner}\ \emph {et~al.}(2013)\citenamefont {Rudner},
    \citenamefont {Lindner}, \citenamefont {Berg},\ and\ \citenamefont
    {Levin}}]{Rudner2013}%
    \BibitemOpen
    \bibfield  {author} {\bibinfo {author} {\bibfnamefont {M.~S.}\ \bibnamefont
    {Rudner}}, \bibinfo {author} {\bibfnamefont {N.~H.}\ \bibnamefont {Lindner}},
    \bibinfo {author} {\bibfnamefont {E.}~\bibnamefont {Berg}},\ and\ \bibinfo
    {author} {\bibfnamefont {M.}~\bibnamefont {Levin}},\ }\bibfield  {title}
    {\bibinfo {title} {Anomalous {E}dge {S}tates and the {B}ulk-{E}dge
    {C}orrespondence for {P}eriodically {D}riven {T}wo-{D}imensional {S}ystems},\
    }\href {https://doi.org/10.1103/physrevx.3.031005} {\bibfield  {journal}
    {\bibinfo  {journal} {Phys. Rev. X}\ }\textbf {\bibinfo {volume} {3}},\
    \bibinfo {pages} {031005} (\bibinfo {year} {2013})}\BibitemShut {NoStop}%
  \bibitem [{\citenamefont {Yao}\ \emph {et~al.}(2017)\citenamefont {Yao},
    \citenamefont {Yan},\ and\ \citenamefont {Wang}}]{Yao2017}%
    \BibitemOpen
    \bibfield  {author} {\bibinfo {author} {\bibfnamefont {S.}~\bibnamefont
    {Yao}}, \bibinfo {author} {\bibfnamefont {Z.}~\bibnamefont {Yan}},\ and\
    \bibinfo {author} {\bibfnamefont {Z.}~\bibnamefont {Wang}},\ }\bibfield
    {title} {\bibinfo {title} {Topological invariants of {F}loquet systems:
    {G}eneral formulation, special properties, and {F}loquet topological
    defects},\ }\href {https://doi.org/10.1103/physrevb.96.195303} {\bibfield
    {journal} {\bibinfo  {journal} {Phys. Rev. B}\ }\textbf {\bibinfo {volume}
    {96}},\ \bibinfo {pages} {195303} (\bibinfo {year} {2017})}\BibitemShut
    {NoStop}%
  \bibitem [{\citenamefont {Xu}\ \emph {et~al.}(2022)\citenamefont {Xu},
    \citenamefont {Zheng},\ and\ \citenamefont {Zhai}}]{Xu2022}%
    \BibitemOpen
    \bibfield  {author} {\bibinfo {author} {\bibfnamefont {P.}~\bibnamefont
    {Xu}}, \bibinfo {author} {\bibfnamefont {W.}~\bibnamefont {Zheng}},\ and\
    \bibinfo {author} {\bibfnamefont {H.}~\bibnamefont {Zhai}},\ }\bibfield
    {title} {\bibinfo {title} {{Topological micromotion of Floquet quantum
    systems}},\ }\href {https://doi.org/10.1103/PhysRevB.105.045139} {\bibfield
    {journal} {\bibinfo  {journal} {Phys. Rev. B}\ }\textbf {\bibinfo {volume}
    {105}},\ \bibinfo {pages} {045139} (\bibinfo {year} {2022})}\BibitemShut
    {NoStop}%
  \bibitem [{\citenamefont {Nathan}\ and\ \citenamefont
    {Rudner}(2015)}]{Nathan2015}%
    \BibitemOpen
    \bibfield  {author} {\bibinfo {author} {\bibfnamefont {F.}~\bibnamefont
    {Nathan}}\ and\ \bibinfo {author} {\bibfnamefont {M.~S.}\ \bibnamefont
    {Rudner}},\ }\bibfield  {title} {\bibinfo {title} {Topological singularities
    and the general classification of {F}loquet-{B}loch systems},\ }\href
    {https://doi.org/10.1088/1367-2630/17/12/125014} {\bibfield  {journal}
    {\bibinfo  {journal} {New J. Phys.}\ }\textbf {\bibinfo {volume} {17}},\
    \bibinfo {pages} {125014} (\bibinfo {year} {2015})}\BibitemShut {NoStop}%
  \bibitem [{\citenamefont {Roy}\ and\ \citenamefont {Harper}(2017)}]{Roy2017}%
    \BibitemOpen
    \bibfield  {author} {\bibinfo {author} {\bibfnamefont {R.}~\bibnamefont
    {Roy}}\ and\ \bibinfo {author} {\bibfnamefont {F.}~\bibnamefont {Harper}},\
    }\bibfield  {title} {\bibinfo {title} {Periodic table for {F}loquet
    topological insulators},\ }\href {https://doi.org/10.1103/physrevb.96.155118}
    {\bibfield  {journal} {\bibinfo  {journal} {Phys. Rev. B}\ }\textbf {\bibinfo
    {volume} {96}},\ \bibinfo {pages} {155118} (\bibinfo {year}
    {2017})}\BibitemShut {NoStop}%
  \bibitem [{\citenamefont {Wang}\ \emph {et~al.}(2017)\citenamefont {Wang},
    \citenamefont {Zhang}, \citenamefont {Chen}, \citenamefont {Yu},\ and\
    \citenamefont {Zhai}}]{Wang2017}%
    \BibitemOpen
    \bibfield  {author} {\bibinfo {author} {\bibfnamefont {C.}~\bibnamefont
    {Wang}}, \bibinfo {author} {\bibfnamefont {P.}~\bibnamefont {Zhang}},
    \bibinfo {author} {\bibfnamefont {X.}~\bibnamefont {Chen}}, \bibinfo {author}
    {\bibfnamefont {J.}~\bibnamefont {Yu}},\ and\ \bibinfo {author}
    {\bibfnamefont {H.}~\bibnamefont {Zhai}},\ }\bibfield  {title} {\bibinfo
    {title} {Scheme to {M}easure the {T}opological {N}umber of a {C}hern
    {I}nsulator from {Q}uench {D}ynamics},\ }\href
    {https://doi.org/10.1103/physrevlett.118.185701} {\bibfield  {journal}
    {\bibinfo  {journal} {Phys. Rev. Lett.}\ }\textbf {\bibinfo {volume} {118}},\
    \bibinfo {pages} {185701} (\bibinfo {year} {2017})}\BibitemShut {NoStop}%
  \bibitem [{\citenamefont {Tarnowski}\ \emph {et~al.}(2019)\citenamefont
    {Tarnowski}, \citenamefont {{\" U}nal}, \citenamefont {Fl{\" a}schner},
    \citenamefont {Rem}, \citenamefont {Eckardt}, \citenamefont {Sengstock},\
    and\ \citenamefont {Weitenberg}}]{Tarnowski2019}%
    \BibitemOpen
    \bibfield  {author} {\bibinfo {author} {\bibfnamefont {M.}~\bibnamefont
    {Tarnowski}}, \bibinfo {author} {\bibfnamefont {F.~N.}\ \bibnamefont {{\"
    U}nal}}, \bibinfo {author} {\bibfnamefont {N.}~\bibnamefont {Fl{\"
    a}schner}}, \bibinfo {author} {\bibfnamefont {B.~S.}\ \bibnamefont {Rem}},
    \bibinfo {author} {\bibfnamefont {A.}~\bibnamefont {Eckardt}}, \bibinfo
    {author} {\bibfnamefont {K.}~\bibnamefont {Sengstock}},\ and\ \bibinfo
    {author} {\bibfnamefont {C.}~\bibnamefont {Weitenberg}},\ }\bibfield  {title}
    {\bibinfo {title} {Measuring topology from dynamics by obtaining the {C}hern
    number from a linking number},\ }\href
    {https://doi.org/10.1038/s41467-019-09668-y} {\bibfield  {journal} {\bibinfo
    {journal} {Nat. Commun.}\ }\textbf {\bibinfo {volume} {10}},\ \bibinfo
    {pages} {1728} (\bibinfo {year} {2019})}\BibitemShut {NoStop}%
  \bibitem [{\citenamefont {{\" U}nal}\ \emph {et~al.}(2020)\citenamefont {{\"
    U}nal}, \citenamefont {Bouhon},\ and\ \citenamefont {Slager}}]{Uenal2020}%
    \BibitemOpen
    \bibfield  {author} {\bibinfo {author} {\bibfnamefont {F.~N.}\ \bibnamefont
    {{\" U}nal}}, \bibinfo {author} {\bibfnamefont {A.}~\bibnamefont {Bouhon}},\
    and\ \bibinfo {author} {\bibfnamefont {R.-J.}\ \bibnamefont {Slager}},\
    }\bibfield  {title} {\bibinfo {title} {{Topological Euler Class as a
    Dynamical Observable in Optical Lattices}},\ }\href
    {https://doi.org/10.1103/PhysRevLett.125.053601} {\bibfield  {journal}
    {\bibinfo  {journal} {Phys. Rev. Lett.}\ }\textbf {\bibinfo {volume} {125}},\
    \bibinfo {pages} {053601} (\bibinfo {year} {2020})}\BibitemShut {NoStop}%
  \bibitem [{\citenamefont {Mizoguchi}\ \emph {et~al.}(2021)\citenamefont
    {Mizoguchi}, \citenamefont {Kuno},\ and\ \citenamefont
    {Hatsugai}}]{Mizoguchi2021}%
    \BibitemOpen
    \bibfield  {author} {\bibinfo {author} {\bibfnamefont {T.}~\bibnamefont
    {Mizoguchi}}, \bibinfo {author} {\bibfnamefont {Y.}~\bibnamefont {Kuno}},\
    and\ \bibinfo {author} {\bibfnamefont {Y.}~\bibnamefont {Hatsugai}},\
    }\bibfield  {title} {\bibinfo {title} {Detecting {B}ulk {T}opology of
    {Q}uadrupolar {P}hase from {Q}uench {D}ynamics},\ }\href
    {https://doi.org/10.1103/physrevlett.126.016802} {\bibfield  {journal}
    {\bibinfo  {journal} {Phys. Rev. Lett.}\ }\textbf {\bibinfo {volume} {126}},\
    \bibinfo {pages} {016802} (\bibinfo {year} {2021})}\BibitemShut {NoStop}%
  \bibitem [{\citenamefont {Budich}\ and\ \citenamefont
    {Heyl}(2016)}]{Budich2016}%
    \BibitemOpen
    \bibfield  {author} {\bibinfo {author} {\bibfnamefont {J.~C.}\ \bibnamefont
    {Budich}}\ and\ \bibinfo {author} {\bibfnamefont {M.}~\bibnamefont {Heyl}},\
    }\bibfield  {title} {\bibinfo {title} {Dynamical topological order parameters
    far from equilibrium},\ }\href {https://doi.org/10.1103/physrevb.93.085416}
    {\bibfield  {journal} {\bibinfo  {journal} {Phys. Rev. B}\ }\textbf {\bibinfo
    {volume} {93}},\ \bibinfo {pages} {085416} (\bibinfo {year}
    {2016})}\BibitemShut {NoStop}%
  \bibitem [{\citenamefont {Yang}\ \emph {et~al.}(2018)\citenamefont {Yang},
    \citenamefont {Li},\ and\ \citenamefont {Chen}}]{Yang2018}%
    \BibitemOpen
    \bibfield  {author} {\bibinfo {author} {\bibfnamefont {C.}~\bibnamefont
    {Yang}}, \bibinfo {author} {\bibfnamefont {L.}~\bibnamefont {Li}},\ and\
    \bibinfo {author} {\bibfnamefont {S.}~\bibnamefont {Chen}},\ }\bibfield
    {title} {\bibinfo {title} {Dynamical topological invariant after a quantum
    quench},\ }\href {https://doi.org/10.1103/physrevb.97.060304} {\bibfield
    {journal} {\bibinfo  {journal} {Phys. Rev. B}\ }\textbf {\bibinfo {volume}
    {97}},\ \bibinfo {pages} {060304} (\bibinfo {year} {2018})}\BibitemShut
    {NoStop}%
  \bibitem [{\citenamefont {Gong}\ and\ \citenamefont {Ueda}(2018)}]{Gong2018}%
    \BibitemOpen
    \bibfield  {author} {\bibinfo {author} {\bibfnamefont {Z.}~\bibnamefont
    {Gong}}\ and\ \bibinfo {author} {\bibfnamefont {M.}~\bibnamefont {Ueda}},\
    }\bibfield  {title} {\bibinfo {title} {Topological {E}ntanglement-{S}pectrum
    {C}rossing in {Q}uench {D}ynamics},\ }\href
    {https://doi.org/10.1103/physrevlett.121.250601} {\bibfield  {journal}
    {\bibinfo  {journal} {Phys. Rev. Lett.}\ }\textbf {\bibinfo {volume} {121}},\
    \bibinfo {pages} {250601} (\bibinfo {year} {2018})}\BibitemShut {NoStop}%
  \bibitem [{\citenamefont {McGinley}\ and\ \citenamefont
    {Cooper}(2018)}]{McGinley2018}%
    \BibitemOpen
    \bibfield  {author} {\bibinfo {author} {\bibfnamefont {M.}~\bibnamefont
    {McGinley}}\ and\ \bibinfo {author} {\bibfnamefont {N.~R.}\ \bibnamefont
    {Cooper}},\ }\bibfield  {title} {\bibinfo {title} {Topology of
    {O}ne-{D}imensional {Q}uantum {S}ystems {O}ut of {E}quilibrium},\ }\href
    {https://doi.org/10.1103/physrevlett.121.090401} {\bibfield  {journal}
    {\bibinfo  {journal} {Phys. Rev. Lett.}\ }\textbf {\bibinfo {volume} {121}},\
    \bibinfo {pages} {090401} (\bibinfo {year} {2018})}\BibitemShut {NoStop}%
  \bibitem [{\citenamefont {Qiu}\ \emph {et~al.}(2019)\citenamefont {Qiu},
    \citenamefont {Deng}, \citenamefont {Hu}, \citenamefont {Xue},\ and\
    \citenamefont {Yi}}]{Qiu2019}%
    \BibitemOpen
    \bibfield  {author} {\bibinfo {author} {\bibfnamefont {X.}~\bibnamefont
    {Qiu}}, \bibinfo {author} {\bibfnamefont {T.-S.}\ \bibnamefont {Deng}},
    \bibinfo {author} {\bibfnamefont {Y.}~\bibnamefont {Hu}}, \bibinfo {author}
    {\bibfnamefont {P.}~\bibnamefont {Xue}},\ and\ \bibinfo {author}
    {\bibfnamefont {W.}~\bibnamefont {Yi}},\ }\bibfield  {title} {\bibinfo
    {title} {Fixed {P}oints and {D}ynamic {T}opological {P}henomena in a
    {P}arity-{T}ime-{S}ymmetric {Q}uantum {Q}uench},\ }\href
    {https://doi.org/10.1016/j.isci.2019.09.037} {\bibfield  {journal} {\bibinfo
    {journal} {{iScience}}\ }\textbf {\bibinfo {volume} {20}},\ \bibinfo {pages}
    {392} (\bibinfo {year} {2019})}\BibitemShut {NoStop}%
  \bibitem [{\citenamefont {Wang}\ \emph
    {et~al.}(2019{\natexlab{a}})\citenamefont {Wang}, \citenamefont {Qiu},
    \citenamefont {Xiao}, \citenamefont {Zhan}, \citenamefont {Bian},
    \citenamefont {Yi},\ and\ \citenamefont {Xue}}]{Wang2019}%
    \BibitemOpen
    \bibfield  {author} {\bibinfo {author} {\bibfnamefont {K.}~\bibnamefont
    {Wang}}, \bibinfo {author} {\bibfnamefont {X.}~\bibnamefont {Qiu}}, \bibinfo
    {author} {\bibfnamefont {L.}~\bibnamefont {Xiao}}, \bibinfo {author}
    {\bibfnamefont {X.}~\bibnamefont {Zhan}}, \bibinfo {author} {\bibfnamefont
    {Z.}~\bibnamefont {Bian}}, \bibinfo {author} {\bibfnamefont {W.}~\bibnamefont
    {Yi}},\ and\ \bibinfo {author} {\bibfnamefont {P.}~\bibnamefont {Xue}},\
    }\bibfield  {title} {\bibinfo {title} {Simulating {D}ynamic {Q}uantum {P}hase
    {T}ransitions in {P}hotonic {Q}uantum {W}alks},\ }\href
    {https://doi.org/10.1103/physrevlett.122.020501} {\bibfield  {journal}
    {\bibinfo  {journal} {Phys. Rev. Lett.}\ }\textbf {\bibinfo {volume} {122}},\
    \bibinfo {pages} {020501} (\bibinfo {year} {2019}{\natexlab{a}})}\BibitemShut
    {NoStop}%
  \bibitem [{\citenamefont {McGinley}\ and\ \citenamefont
    {Cooper}(2019)}]{McGinley2019}%
    \BibitemOpen
    \bibfield  {author} {\bibinfo {author} {\bibfnamefont {M.}~\bibnamefont
    {McGinley}}\ and\ \bibinfo {author} {\bibfnamefont {N.~R.}\ \bibnamefont
    {Cooper}},\ }\bibfield  {title} {\bibinfo {title} {Classification of
    topological insulators and superconductors out of equilibrium},\ }\href
    {https://doi.org/10.1103/physrevb.99.075148} {\bibfield  {journal} {\bibinfo
    {journal} {Phys. Rev. B}\ }\textbf {\bibinfo {volume} {99}},\ \bibinfo
    {pages} {075148} (\bibinfo {year} {2019})}\BibitemShut {NoStop}%
  \bibitem [{\citenamefont {Hu}\ and\ \citenamefont {Zhao}(2020)}]{Hu2020}%
    \BibitemOpen
    \bibfield  {author} {\bibinfo {author} {\bibfnamefont {H.}~\bibnamefont
    {Hu}}\ and\ \bibinfo {author} {\bibfnamefont {E.}~\bibnamefont {Zhao}},\
    }\bibfield  {title} {\bibinfo {title} {Topological {I}nvariants for {Q}uantum
    {Q}uench {D}ynamics from {U}nitary {E}volution},\ }\href
    {https://doi.org/10.1103/physrevlett.124.160402} {\bibfield  {journal}
    {\bibinfo  {journal} {Phys. Rev. Lett.}\ }\textbf {\bibinfo {volume} {124}},\
    \bibinfo {pages} {160402} (\bibinfo {year} {2020})}\BibitemShut {NoStop}%
  {\bibitem [{\citenamefont {Sim}\ \emph {et~al.}(2022)\citenamefont {Sim},
    \citenamefont {Chitra},\ and\ \citenamefont {Molignini}}]{Sim2022}%
    \BibitemOpen
    \bibfield  {author} {\bibinfo {author} {\bibfnamefont {K.}~\bibnamefont
    {Sim}}, \bibinfo {author} {\bibfnamefont {R.}~\bibnamefont {Chitra}},\ and\
    \bibinfo {author} {\bibfnamefont {P.}~\bibnamefont {Molignini}},\ }\bibfield
    {title} {\bibinfo {title} {Quench dynamics and scaling laws in topological
    nodal loop semimetals},\ }\href {https://doi.org/10.1103/physrevb.106.224302}
    {\bibfield  {journal} {\bibinfo  {journal} {Phys. Rev. B}\ }\textbf {\bibinfo
    {volume} {106}},\ \bibinfo {pages} {224302} (\bibinfo {year}
    {2022})}\BibitemShut {NoStop}}%
  \bibitem [{\citenamefont {Zhang}\ \emph {et~al.}(2020)\citenamefont {Zhang},
    \citenamefont {Zhang},\ and\ \citenamefont {Liu}}]{Zhang2020}%
    \BibitemOpen
    \bibfield  {author} {\bibinfo {author} {\bibfnamefont {L.}~\bibnamefont
    {Zhang}}, \bibinfo {author} {\bibfnamefont {L.}~\bibnamefont {Zhang}},\ and\
    \bibinfo {author} {\bibfnamefont {X.-J.}\ \bibnamefont {Liu}},\ }\bibfield
    {title} {\bibinfo {title} {Unified {T}heory to {C}haracterize {F}loquet
    {T}opological {P}hases by {Q}uench {D}ynamics},\ }\href
    {https://doi.org/10.1103/physrevlett.125.183001} {\bibfield  {journal}
    {\bibinfo  {journal} {Phys. Rev. Lett.}\ }\textbf {\bibinfo {volume} {125}},\
    \bibinfo {pages} {183001} (\bibinfo {year} {2020})}\BibitemShut {NoStop}%
  \bibitem [{\citenamefont {Zhang}\ and\ \citenamefont {Liu}(2022)}]{Zhang2021}%
    \BibitemOpen
    \bibfield  {author} {\bibinfo {author} {\bibfnamefont {L.}~\bibnamefont
    {Zhang}}\ and\ \bibinfo {author} {\bibfnamefont {X.-J.}\ \bibnamefont
    {Liu}},\ }\bibfield  {title} {\bibinfo {title} {{Unconventional Floquet
    Topological Phases from Quantum Engineering of Band-Inversion Surfaces}},\
    }\href {https://doi.org/10.1103/PRXQuantum.3.040312} {\bibfield  {journal}
    {\bibinfo  {journal} {PRX Quantum}\ }\textbf {\bibinfo {volume} {3}},\
    \bibinfo {pages} {040312} (\bibinfo {year} {2022})}\BibitemShut {NoStop}%
  \bibitem [{\citenamefont {Wang}\ and\ \citenamefont {Zhang}()}]{Wang2023}%
    \BibitemOpen
    \bibfield  {author} {\bibinfo {author} {\bibfnamefont {B.-B.}\ \bibnamefont
    {Wang}}\ and\ \bibinfo {author} {\bibfnamefont {L.}~\bibnamefont {Zhang}},\
    }\href@noop {} {\bibinfo {title} {{Characterizing Floquet topological phases
    by quench dynamics: A multiple-subsystem approach}}},\ \Eprint
    {https://arxiv.org/abs/2310.08409} {arXiv:2310.08409} \BibitemShut {NoStop}%
  \bibitem [{\citenamefont {Zhang}\ \emph {et~al.}(2018)\citenamefont {Zhang},
    \citenamefont {Zhang}, \citenamefont {Niu},\ and\ \citenamefont
    {Liu}}]{Zhang2018}%
    \BibitemOpen
    \bibfield  {author} {\bibinfo {author} {\bibfnamefont {L.}~\bibnamefont
    {Zhang}}, \bibinfo {author} {\bibfnamefont {L.}~\bibnamefont {Zhang}},
    \bibinfo {author} {\bibfnamefont {S.}~\bibnamefont {Niu}},\ and\ \bibinfo
    {author} {\bibfnamefont {X.-J.}\ \bibnamefont {Liu}},\ }\bibfield  {title}
    {\bibinfo {title} {Dynamical classification of topological quantum phases},\
    }\href {https://doi.org/10.1016/j.scib.2018.09.018} {\bibfield  {journal}
    {\bibinfo  {journal} {Sci. Bull.}\ }\textbf {\bibinfo {volume} {63}},\
    \bibinfo {pages} {1385} (\bibinfo {year} {2018})}\BibitemShut {NoStop}%
  \bibitem [{\citenamefont {Zhang}\ \emph
    {et~al.}(2019{\natexlab{a}})\citenamefont {Zhang}, \citenamefont {Zhang},\
    and\ \citenamefont {Liu}}]{Zhang2019}%
    \BibitemOpen
    \bibfield  {author} {\bibinfo {author} {\bibfnamefont {L.}~\bibnamefont
    {Zhang}}, \bibinfo {author} {\bibfnamefont {L.}~\bibnamefont {Zhang}},\ and\
    \bibinfo {author} {\bibfnamefont {X.-J.}\ \bibnamefont {Liu}},\ }\bibfield
    {title} {\bibinfo {title} {Dynamical detection of topological charges},\
    }\href {https://doi.org/10.1103/physreva.99.053606} {\bibfield  {journal}
    {\bibinfo  {journal} {Phys. Rev. A}\ }\textbf {\bibinfo {volume} {99}},\
    \bibinfo {pages} {053606} (\bibinfo {year} {2019}{\natexlab{a}})}\BibitemShut
    {NoStop}%
  \bibitem [{\citenamefont {Zhang}\ \emph
    {et~al.}(2019{\natexlab{b}})\citenamefont {Zhang}, \citenamefont {Zhang},\
    and\ \citenamefont {Liu}}]{Zhang2019a}%
    \BibitemOpen
    \bibfield  {author} {\bibinfo {author} {\bibfnamefont {L.}~\bibnamefont
    {Zhang}}, \bibinfo {author} {\bibfnamefont {L.}~\bibnamefont {Zhang}},\ and\
    \bibinfo {author} {\bibfnamefont {X.-J.}\ \bibnamefont {Liu}},\ }\bibfield
    {title} {\bibinfo {title} {Characterizing topological phases by quantum
    quenches: A general theory},\ }\href
    {https://doi.org/10.1103/physreva.100.063624} {\bibfield  {journal} {\bibinfo
     {journal} {Phys. Rev. A}\ }\textbf {\bibinfo {volume} {100}},\ \bibinfo
    {pages} {063624} (\bibinfo {year} {2019}{\natexlab{b}})}\BibitemShut
    {NoStop}%
  \bibitem [{\citenamefont {Yu}\ \emph {et~al.}(2021{\natexlab{a}})\citenamefont
    {Yu}, \citenamefont {Ji}, \citenamefont {Zhang}, \citenamefont {Wang},
    \citenamefont {Wu},\ and\ \citenamefont {Liu}}]{Yu2021}%
    \BibitemOpen
    \bibfield  {author} {\bibinfo {author} {\bibfnamefont {X.-L.}\ \bibnamefont
    {Yu}}, \bibinfo {author} {\bibfnamefont {W.}~\bibnamefont {Ji}}, \bibinfo
    {author} {\bibfnamefont {L.}~\bibnamefont {Zhang}}, \bibinfo {author}
    {\bibfnamefont {Y.}~\bibnamefont {Wang}}, \bibinfo {author} {\bibfnamefont
    {J.}~\bibnamefont {Wu}},\ and\ \bibinfo {author} {\bibfnamefont {X.-J.}\
    \bibnamefont {Liu}},\ }\bibfield  {title} {\bibinfo {title} {Quantum
    {D}ynamical {C}haracterization and {S}imulation of {T}opological {P}hases
    {W}ith {H}igh-{O}rder {B}and {I}nversion {S}urfaces},\ }\href
    {https://doi.org/10.1103/prxquantum.2.020320} {\bibfield  {journal} {\bibinfo
     {journal} {PRX Quantum}\ }\textbf {\bibinfo {volume} {2}},\ \bibinfo {pages}
    {020320} (\bibinfo {year} {2021}{\natexlab{a}})}\BibitemShut {NoStop}%
  \bibitem [{\citenamefont {Zhang}\ \emph {et~al.}(2022)\citenamefont {Zhang},
    \citenamefont {Jia},\ and\ \citenamefont {Liu}}]{Zhang2022}%
    \BibitemOpen
    \bibfield  {author} {\bibinfo {author} {\bibfnamefont {L.}~\bibnamefont
    {Zhang}}, \bibinfo {author} {\bibfnamefont {W.}~\bibnamefont {Jia}},\ and\
    \bibinfo {author} {\bibfnamefont {X.-J.}\ \bibnamefont {Liu}},\ }\bibfield
    {title} {\bibinfo {title} {Universal topological quench dynamics for
    $\mathbb{Z}_2$ topological phases},\ }\href
    {https://doi.org/10.1016/j.scib.2022.04.019} {\bibfield  {journal} {\bibinfo
    {journal} {Sci. Bull.}\ }\textbf {\bibinfo {volume} {67}},\ \bibinfo {pages}
    {1236} (\bibinfo {year} {2022})}\BibitemShut {NoStop}%
  \bibitem [{\citenamefont {Zhou}\ and\ \citenamefont {Gong}(2018)}]{Zhou2018}%
    \BibitemOpen
    \bibfield  {author} {\bibinfo {author} {\bibfnamefont {L.}~\bibnamefont
    {Zhou}}\ and\ \bibinfo {author} {\bibfnamefont {J.}~\bibnamefont {Gong}},\
    }\bibfield  {title} {\bibinfo {title} {Non-{H}ermitian {F}loquet topological
    phases with arbitrarily many real-quasienergy edge states},\ }\href
    {https://doi.org/10.1103/physrevb.98.205417} {\bibfield  {journal} {\bibinfo
    {journal} {Phys. Rev. B}\ }\textbf {\bibinfo {volume} {98}},\ \bibinfo
    {pages} {205417} (\bibinfo {year} {2018})}\BibitemShut {NoStop}%
  \bibitem [{\citenamefont {Zhu}\ \emph {et~al.}(2020)\citenamefont {Zhu},
    \citenamefont {Ke}, \citenamefont {Zhong},\ and\ \citenamefont
    {Lee}}]{Zhu2020}%
    \BibitemOpen
    \bibfield  {author} {\bibinfo {author} {\bibfnamefont {B.}~\bibnamefont
    {Zhu}}, \bibinfo {author} {\bibfnamefont {Y.}~\bibnamefont {Ke}}, \bibinfo
    {author} {\bibfnamefont {H.}~\bibnamefont {Zhong}},\ and\ \bibinfo {author}
    {\bibfnamefont {C.}~\bibnamefont {Lee}},\ }\bibfield  {title} {\bibinfo
    {title} {Dynamic winding number for exploring band topology},\ }\href
    {https://doi.org/10.1103/physrevresearch.2.023043} {\bibfield  {journal}
    {\bibinfo  {journal} {Phys. Rev. Research}\ }\textbf {\bibinfo {volume}
    {2}},\ \bibinfo {pages} {023043} (\bibinfo {year} {2020})}\BibitemShut
    {NoStop}%
  \bibitem [{\citenamefont {Li}\ \emph {et~al.}(2021)\citenamefont {Li},
    \citenamefont {Zhu},\ and\ \citenamefont {Gong}}]{Li2021}%
    \BibitemOpen
    \bibfield  {author} {\bibinfo {author} {\bibfnamefont {L.}~\bibnamefont
    {Li}}, \bibinfo {author} {\bibfnamefont {W.}~\bibnamefont {Zhu}},\ and\
    \bibinfo {author} {\bibfnamefont {J.}~\bibnamefont {Gong}},\ }\bibfield
    {title} {\bibinfo {title} {Direct dynamical characterization of higher-order
    topological phases with nested band inversion surfaces},\ }\href
    {https://doi.org/10.1016/j.scib.2021.04.006} {\bibfield  {journal} {\bibinfo
    {journal} {Sci. Bull.}\ }\textbf {\bibinfo {volume} {66}},\ \bibinfo {pages}
    {1502} (\bibinfo {year} {2021})}\BibitemShut {NoStop}%
  \bibitem [{\citenamefont {Niu}\ \emph {et~al.}(2021)\citenamefont {Niu},
    \citenamefont {Yan}, \citenamefont {Zhou}, \citenamefont {Tao}, \citenamefont
    {Li}, \citenamefont {Liu}, \citenamefont {Zhang}, \citenamefont {Jia},
    \citenamefont {Liu}, \citenamefont {Yan}, \citenamefont {Chen},\ and\
    \citenamefont {Yu}}]{Niu2021}%
    \BibitemOpen
    \bibfield  {author} {\bibinfo {author} {\bibfnamefont {J.}~\bibnamefont
    {Niu}}, \bibinfo {author} {\bibfnamefont {T.}~\bibnamefont {Yan}}, \bibinfo
    {author} {\bibfnamefont {Y.}~\bibnamefont {Zhou}}, \bibinfo {author}
    {\bibfnamefont {Z.}~\bibnamefont {Tao}}, \bibinfo {author} {\bibfnamefont
    {X.}~\bibnamefont {Li}}, \bibinfo {author} {\bibfnamefont {W.}~\bibnamefont
    {Liu}}, \bibinfo {author} {\bibfnamefont {L.}~\bibnamefont {Zhang}}, \bibinfo
    {author} {\bibfnamefont {H.}~\bibnamefont {Jia}}, \bibinfo {author}
    {\bibfnamefont {S.}~\bibnamefont {Liu}}, \bibinfo {author} {\bibfnamefont
    {Z.}~\bibnamefont {Yan}}, \bibinfo {author} {\bibfnamefont {Y.}~\bibnamefont
    {Chen}},\ and\ \bibinfo {author} {\bibfnamefont {D.}~\bibnamefont {Yu}},\
    }\bibfield  {title} {\bibinfo {title} {Simulation of higher-order topological
    phases and related topological phase transitions in a superconducting
    qubit},\ }\href {https://doi.org/10.1016/j.scib.2021.02.035} {\bibfield
    {journal} {\bibinfo  {journal} {Sci. Bull.}\ }\textbf {\bibinfo {volume}
    {66}},\ \bibinfo {pages} {1168} (\bibinfo {year} {2021})}\BibitemShut
    {NoStop}%
  \bibitem [{\citenamefont {Lei}\ \emph {et~al.}(2022)\citenamefont {Lei},
    \citenamefont {Deng},\ and\ \citenamefont {Li}}]{Lei2022}%
    \BibitemOpen
    \bibfield  {author} {\bibinfo {author} {\bibfnamefont {Z.}~\bibnamefont
    {Lei}}, \bibinfo {author} {\bibfnamefont {Y.}~\bibnamefont {Deng}},\ and\
    \bibinfo {author} {\bibfnamefont {L.}~\bibnamefont {Li}},\ }\bibfield
    {title} {\bibinfo {title} {{Topological classification of higher-order
    topological phases with nested band inversion surfaces}},\ }\href
    {https://doi.org/10.1103/PhysRevB.106.245105} {\bibfield  {journal} {\bibinfo
     {journal} {Phys. Rev. B}\ }\textbf {\bibinfo {volume} {106}},\ \bibinfo
    {pages} {245105} (\bibinfo {year} {2022})}\BibitemShut {NoStop}%
  \bibitem [{\citenamefont {Jia}\ \emph {et~al.}(2023)\citenamefont {Jia},
    \citenamefont {Zhou}, \citenamefont {Zhang}, \citenamefont {Zhang},\ and\
    \citenamefont {Liu}}]{Jia2022}%
    \BibitemOpen
    \bibfield  {author} {\bibinfo {author} {\bibfnamefont {W.}~\bibnamefont
    {Jia}}, \bibinfo {author} {\bibfnamefont {X.-C.}\ \bibnamefont {Zhou}},
    \bibinfo {author} {\bibfnamefont {L.}~\bibnamefont {Zhang}}, \bibinfo
    {author} {\bibfnamefont {L.}~\bibnamefont {Zhang}},\ and\ \bibinfo {author}
    {\bibfnamefont {X.-J.}\ \bibnamefont {Liu}},\ }\bibfield  {title} {\bibinfo
    {title} {Unified characterization for higher-order topological phase
    transitions},\ }\href {https://doi.org/10.1103/PhysRevResearch.5.L022032}
    {\bibfield  {journal} {\bibinfo  {journal} {Phys. Rev. Res.}\ }\textbf
    {\bibinfo {volume} {5}},\ \bibinfo {pages} {L022032} (\bibinfo {year}
    {2023})}\BibitemShut {NoStop}%
  \bibitem [{\citenamefont {Sun}\ \emph {et~al.}(2018)\citenamefont {Sun},
    \citenamefont {Yi}, \citenamefont {Wang}, \citenamefont {Zhang},
    \citenamefont {Sanders}, \citenamefont {Xu}, \citenamefont {Wang},
    \citenamefont {Schmiedmayer}, \citenamefont {Deng}, \citenamefont {Liu},
    \citenamefont {Chen},\ and\ \citenamefont {Pan}}]{Sun2018}%
    \BibitemOpen
    \bibfield  {author} {\bibinfo {author} {\bibfnamefont {W.}~\bibnamefont
    {Sun}}, \bibinfo {author} {\bibfnamefont {C.-R.}\ \bibnamefont {Yi}},
    \bibinfo {author} {\bibfnamefont {B.-Z.}\ \bibnamefont {Wang}}, \bibinfo
    {author} {\bibfnamefont {W.-W.}\ \bibnamefont {Zhang}}, \bibinfo {author}
    {\bibfnamefont {B.~C.}\ \bibnamefont {Sanders}}, \bibinfo {author}
    {\bibfnamefont {X.-T.}\ \bibnamefont {Xu}}, \bibinfo {author} {\bibfnamefont
    {Z.-Y.}\ \bibnamefont {Wang}}, \bibinfo {author} {\bibfnamefont
    {J.}~\bibnamefont {Schmiedmayer}}, \bibinfo {author} {\bibfnamefont
    {Y.}~\bibnamefont {Deng}}, \bibinfo {author} {\bibfnamefont {X.-J.}\
    \bibnamefont {Liu}}, \bibinfo {author} {\bibfnamefont {S.}~\bibnamefont
    {Chen}},\ and\ \bibinfo {author} {\bibfnamefont {J.-W.}\ \bibnamefont
    {Pan}},\ }\bibfield  {title} {\bibinfo {title} {Uncover {T}opology by
    {Q}uantum {Q}uench {D}ynamics},\ }\href
    {https://doi.org/10.1103/physrevlett.121.250403} {\bibfield  {journal}
    {\bibinfo  {journal} {Phys. Rev. Lett.}\ }\textbf {\bibinfo {volume} {121}},\
    \bibinfo {pages} {250403} (\bibinfo {year} {2018})}\BibitemShut {NoStop}%
  \bibitem [{\citenamefont {Yi}\ \emph {et~al.}(2019)\citenamefont {Yi},
    \citenamefont {Zhang}, \citenamefont {Zhang}, \citenamefont {Jiao},
    \citenamefont {Cheng}, \citenamefont {Wang}, \citenamefont {Xu},
    \citenamefont {Sun}, \citenamefont {Liu}, \citenamefont {Chen},\ and\
    \citenamefont {Pan}}]{Yi2019}%
    \BibitemOpen
    \bibfield  {author} {\bibinfo {author} {\bibfnamefont {C.-R.}\ \bibnamefont
    {Yi}}, \bibinfo {author} {\bibfnamefont {L.}~\bibnamefont {Zhang}}, \bibinfo
    {author} {\bibfnamefont {L.}~\bibnamefont {Zhang}}, \bibinfo {author}
    {\bibfnamefont {R.-H.}\ \bibnamefont {Jiao}}, \bibinfo {author}
    {\bibfnamefont {X.-C.}\ \bibnamefont {Cheng}}, \bibinfo {author}
    {\bibfnamefont {Z.-Y.}\ \bibnamefont {Wang}}, \bibinfo {author}
    {\bibfnamefont {X.-T.}\ \bibnamefont {Xu}}, \bibinfo {author} {\bibfnamefont
    {W.}~\bibnamefont {Sun}}, \bibinfo {author} {\bibfnamefont {X.-J.}\
    \bibnamefont {Liu}}, \bibinfo {author} {\bibfnamefont {S.}~\bibnamefont
    {Chen}},\ and\ \bibinfo {author} {\bibfnamefont {J.-W.}\ \bibnamefont
    {Pan}},\ }\bibfield  {title} {\bibinfo {title} {Observing {T}opological
    {C}harges and {D}ynamical {B}ulk-{S}urface {C}orrespondence with {U}ltracold
    {A}toms},\ }\href {https://doi.org/10.1103/physrevlett.123.190603} {\bibfield
     {journal} {\bibinfo  {journal} {Phys. Rev. Lett.}\ }\textbf {\bibinfo
    {volume} {123}},\ \bibinfo {pages} {190603} (\bibinfo {year}
    {2019})}\BibitemShut {NoStop}%
  \bibitem [{\citenamefont {Song}\ \emph {et~al.}(2019)\citenamefont {Song},
    \citenamefont {He}, \citenamefont {Niu}, \citenamefont {Zhang}, \citenamefont
    {Ren}, \citenamefont {Liu},\ and\ \citenamefont {Jo}}]{Song2019}%
    \BibitemOpen
    \bibfield  {author} {\bibinfo {author} {\bibfnamefont {B.}~\bibnamefont
    {Song}}, \bibinfo {author} {\bibfnamefont {C.}~\bibnamefont {He}}, \bibinfo
    {author} {\bibfnamefont {S.}~\bibnamefont {Niu}}, \bibinfo {author}
    {\bibfnamefont {L.}~\bibnamefont {Zhang}}, \bibinfo {author} {\bibfnamefont
    {Z.}~\bibnamefont {Ren}}, \bibinfo {author} {\bibfnamefont {X.-J.}\
    \bibnamefont {Liu}},\ and\ \bibinfo {author} {\bibfnamefont {G.-B.}\
    \bibnamefont {Jo}},\ }\bibfield  {title} {\bibinfo {title} {Observation of
    nodal-line semimetal with ultracold fermions in an optical lattice},\ }\href
    {https://doi.org/10.1038/s41567-019-0564-y} {\bibfield  {journal} {\bibinfo
    {journal} {Nat. Phys.}\ }\textbf {\bibinfo {volume} {15}},\ \bibinfo {pages}
    {911} (\bibinfo {year} {2019})}\BibitemShut {NoStop}%
  \bibitem [{\citenamefont {Wang}\ \emph
    {et~al.}(2019{\natexlab{b}})\citenamefont {Wang}, \citenamefont {Ji},
    \citenamefont {Chai}, \citenamefont {Guo}, \citenamefont {Wang},
    \citenamefont {Ye}, \citenamefont {Yu}, \citenamefont {Zhang}, \citenamefont
    {Qin}, \citenamefont {Wang}, \citenamefont {Shi}, \citenamefont {Rong},
    \citenamefont {Lu}, \citenamefont {Liu},\ and\ \citenamefont
    {Du}}]{Wang2019a}%
    \BibitemOpen
    \bibfield  {author} {\bibinfo {author} {\bibfnamefont {Y.}~\bibnamefont
    {Wang}}, \bibinfo {author} {\bibfnamefont {W.}~\bibnamefont {Ji}}, \bibinfo
    {author} {\bibfnamefont {Z.}~\bibnamefont {Chai}}, \bibinfo {author}
    {\bibfnamefont {Y.}~\bibnamefont {Guo}}, \bibinfo {author} {\bibfnamefont
    {M.}~\bibnamefont {Wang}}, \bibinfo {author} {\bibfnamefont {X.}~\bibnamefont
    {Ye}}, \bibinfo {author} {\bibfnamefont {P.}~\bibnamefont {Yu}}, \bibinfo
    {author} {\bibfnamefont {L.}~\bibnamefont {Zhang}}, \bibinfo {author}
    {\bibfnamefont {X.}~\bibnamefont {Qin}}, \bibinfo {author} {\bibfnamefont
    {P.}~\bibnamefont {Wang}}, \bibinfo {author} {\bibfnamefont {F.}~\bibnamefont
    {Shi}}, \bibinfo {author} {\bibfnamefont {X.}~\bibnamefont {Rong}}, \bibinfo
    {author} {\bibfnamefont {D.}~\bibnamefont {Lu}}, \bibinfo {author}
    {\bibfnamefont {X.-J.}\ \bibnamefont {Liu}},\ and\ \bibinfo {author}
    {\bibfnamefont {J.}~\bibnamefont {Du}},\ }\bibfield  {title} {\bibinfo
    {title} {Experimental observation of dynamical bulk-surface correspondence in
    momentum space for topological phases},\ }\href
    {https://doi.org/10.1103/physreva.100.052328} {\bibfield  {journal} {\bibinfo
     {journal} {Phys. Rev. A}\ }\textbf {\bibinfo {volume} {100}},\ \bibinfo
    {pages} {052328} (\bibinfo {year} {2019}{\natexlab{b}})}\BibitemShut
    {NoStop}%
  \bibitem [{\citenamefont {Ji}\ \emph {et~al.}(2020)\citenamefont {Ji},
    \citenamefont {Zhang}, \citenamefont {Wang}, \citenamefont {Zhang},
    \citenamefont {Guo}, \citenamefont {Chai}, \citenamefont {Rong},
    \citenamefont {Shi}, \citenamefont {Liu}, \citenamefont {Wang},\ and\
    \citenamefont {Du}}]{Ji2020}%
    \BibitemOpen
    \bibfield  {author} {\bibinfo {author} {\bibfnamefont {W.}~\bibnamefont
    {Ji}}, \bibinfo {author} {\bibfnamefont {L.}~\bibnamefont {Zhang}}, \bibinfo
    {author} {\bibfnamefont {M.}~\bibnamefont {Wang}}, \bibinfo {author}
    {\bibfnamefont {L.}~\bibnamefont {Zhang}}, \bibinfo {author} {\bibfnamefont
    {Y.}~\bibnamefont {Guo}}, \bibinfo {author} {\bibfnamefont {Z.}~\bibnamefont
    {Chai}}, \bibinfo {author} {\bibfnamefont {X.}~\bibnamefont {Rong}}, \bibinfo
    {author} {\bibfnamefont {F.}~\bibnamefont {Shi}}, \bibinfo {author}
    {\bibfnamefont {X.-J.}\ \bibnamefont {Liu}}, \bibinfo {author} {\bibfnamefont
    {Y.}~\bibnamefont {Wang}},\ and\ \bibinfo {author} {\bibfnamefont
    {J.}~\bibnamefont {Du}},\ }\bibfield  {title} {\bibinfo {title} {Quantum
    {S}imulation for {T}hree-{D}imensional {C}hiral {T}opological {I}nsulator},\
    }\href {https://doi.org/10.1103/physrevlett.125.020504} {\bibfield  {journal}
    {\bibinfo  {journal} {Phys. Rev. Lett.}\ }\textbf {\bibinfo {volume} {125}},\
    \bibinfo {pages} {020504} (\bibinfo {year} {2020})}\BibitemShut {NoStop}%
  \bibitem [{\citenamefont {Xin}\ \emph {et~al.}(2020)\citenamefont {Xin},
    \citenamefont {Li}, \citenamefont {ang Fan}, \citenamefont {Zhu},
    \citenamefont {Zhang}, \citenamefont {Nie}, \citenamefont {Li}, \citenamefont
    {Liu},\ and\ \citenamefont {Lu}}]{Xin2020}%
    \BibitemOpen
    \bibfield  {author} {\bibinfo {author} {\bibfnamefont {T.}~\bibnamefont
    {Xin}}, \bibinfo {author} {\bibfnamefont {Y.}~\bibnamefont {Li}}, \bibinfo
    {author} {\bibfnamefont {Y.}~\bibnamefont {ang Fan}}, \bibinfo {author}
    {\bibfnamefont {X.}~\bibnamefont {Zhu}}, \bibinfo {author} {\bibfnamefont
    {Y.}~\bibnamefont {Zhang}}, \bibinfo {author} {\bibfnamefont
    {X.}~\bibnamefont {Nie}}, \bibinfo {author} {\bibfnamefont {J.}~\bibnamefont
    {Li}}, \bibinfo {author} {\bibfnamefont {Q.}~\bibnamefont {Liu}},\ and\
    \bibinfo {author} {\bibfnamefont {D.}~\bibnamefont {Lu}},\ }\bibfield
    {title} {\bibinfo {title} {Quantum {P}hases of {T}hree-{D}imensional {C}hiral
    {T}opological {I}nsulators on a {S}pin {Q}uantum {S}imulator},\ }\href
    {https://doi.org/10.1103/physrevlett.125.090502} {\bibfield  {journal}
    {\bibinfo  {journal} {Phys. Rev. Lett.}\ }\textbf {\bibinfo {volume} {125}},\
    \bibinfo {pages} {090502} (\bibinfo {year} {2020})}\BibitemShut {NoStop}%
  \bibitem [{\citenamefont {Wang}\ \emph {et~al.}(2021)\citenamefont {Wang},
    \citenamefont {Cheng}, \citenamefont {Wang}, \citenamefont {Zhang},
    \citenamefont {Lu}, \citenamefont {Yi}, \citenamefont {Niu}, \citenamefont
    {Deng}, \citenamefont {Liu}, \citenamefont {Chen},\ and\ \citenamefont
    {Pan}}]{Wang2021}%
    \BibitemOpen
    \bibfield  {author} {\bibinfo {author} {\bibfnamefont {Z.-Y.}\ \bibnamefont
    {Wang}}, \bibinfo {author} {\bibfnamefont {X.-C.}\ \bibnamefont {Cheng}},
    \bibinfo {author} {\bibfnamefont {B.-Z.}\ \bibnamefont {Wang}}, \bibinfo
    {author} {\bibfnamefont {J.-Y.}\ \bibnamefont {Zhang}}, \bibinfo {author}
    {\bibfnamefont {Y.-H.}\ \bibnamefont {Lu}}, \bibinfo {author} {\bibfnamefont
    {C.-R.}\ \bibnamefont {Yi}}, \bibinfo {author} {\bibfnamefont
    {S.}~\bibnamefont {Niu}}, \bibinfo {author} {\bibfnamefont {Y.}~\bibnamefont
    {Deng}}, \bibinfo {author} {\bibfnamefont {X.-J.}\ \bibnamefont {Liu}},
    \bibinfo {author} {\bibfnamefont {S.}~\bibnamefont {Chen}},\ and\ \bibinfo
    {author} {\bibfnamefont {J.-W.}\ \bibnamefont {Pan}},\ }\bibfield  {title}
    {\bibinfo {title} {Realization of an ideal {W}eyl semimetal band in a quantum
    gas with {3D} spin-orbit coupling},\ }\href
    {https://doi.org/10.1126/science.abc0105} {\bibfield  {journal} {\bibinfo
    {journal} {Science}\ }\textbf {\bibinfo {volume} {372}},\ \bibinfo {pages}
    {271} (\bibinfo {year} {2021})}\BibitemShut {NoStop}%
  \bibitem [{\citenamefont {Liang}\ \emph {et~al.}(2023)\citenamefont {Liang},
    \citenamefont {Wei}, \citenamefont {Zhang}, \citenamefont {Wang},
    \citenamefont {Zhang}, \citenamefont {Wang}, \citenamefont {Qi},
    \citenamefont {Liu},\ and\ \citenamefont {Zhang}}]{Liang2021}%
    \BibitemOpen
    \bibfield  {author} {\bibinfo {author} {\bibfnamefont {M.-C.}\ \bibnamefont
    {Liang}}, \bibinfo {author} {\bibfnamefont {Y.-D.}\ \bibnamefont {Wei}},
    \bibinfo {author} {\bibfnamefont {L.}~\bibnamefont {Zhang}}, \bibinfo
    {author} {\bibfnamefont {X.-J.}\ \bibnamefont {Wang}}, \bibinfo {author}
    {\bibfnamefont {H.}~\bibnamefont {Zhang}}, \bibinfo {author} {\bibfnamefont
    {W.-W.}\ \bibnamefont {Wang}}, \bibinfo {author} {\bibfnamefont
    {W.}~\bibnamefont {Qi}}, \bibinfo {author} {\bibfnamefont {X.-J.}\
    \bibnamefont {Liu}},\ and\ \bibinfo {author} {\bibfnamefont {X.}~\bibnamefont
    {Zhang}},\ }\bibfield  {title} {\bibinfo {title} {{Realization of Qi-Wu-Zhang
    model in spin-orbit-coupled ultracold fermions}},\ }\href
    {https://doi.org/10.1103/PhysRevResearch.5.L012006} {\bibfield  {journal}
    {\bibinfo  {journal} {Phys. Rev. Res.}\ }\textbf {\bibinfo {volume} {5}},\
    \bibinfo {pages} {L012006} (\bibinfo {year} {2023})}\BibitemShut {NoStop}%
  \bibitem [{\citenamefont {Yu}\ \emph {et~al.}(2021{\natexlab{b}})\citenamefont
    {Yu}, \citenamefont {Peng}, \citenamefont {Chen}, \citenamefont {Liu},\ and\
    \citenamefont {Yuan}}]{Yu2021a}%
    \BibitemOpen
    \bibfield  {author} {\bibinfo {author} {\bibfnamefont {D.}~\bibnamefont
    {Yu}}, \bibinfo {author} {\bibfnamefont {B.}~\bibnamefont {Peng}}, \bibinfo
    {author} {\bibfnamefont {X.}~\bibnamefont {Chen}}, \bibinfo {author}
    {\bibfnamefont {X.-J.}\ \bibnamefont {Liu}},\ and\ \bibinfo {author}
    {\bibfnamefont {L.}~\bibnamefont {Yuan}},\ }\bibfield  {title} {\bibinfo
    {title} {Topological holographic quench dynamics in a synthetic frequency
    dimension},\ }\href {https://doi.org/10.1038/s41377-021-00646-y} {\bibfield
    {journal} {\bibinfo  {journal} {Light Sci. Appl.}\ }\textbf {\bibinfo
    {volume} {10}},\ \bibinfo {pages} {209} (\bibinfo {year}
    {2021}{\natexlab{b}})}\BibitemShut {NoStop}%
  \bibitem [{\citenamefont {Chen}\ \emph {et~al.}(2020)\citenamefont {Chen},
    \citenamefont {Li}, \citenamefont {Hou}, \citenamefont {Ge}, \citenamefont
    {Zhou}, \citenamefont {Qian}, \citenamefont {Mei}, \citenamefont {Jia},
    \citenamefont {Xu},\ and\ \citenamefont {Shen}}]{Chen2020}%
    \BibitemOpen
    \bibfield  {author} {\bibinfo {author} {\bibfnamefont {B.}~\bibnamefont
    {Chen}}, \bibinfo {author} {\bibfnamefont {S.}~\bibnamefont {Li}}, \bibinfo
    {author} {\bibfnamefont {X.}~\bibnamefont {Hou}}, \bibinfo {author}
    {\bibfnamefont {F.}~\bibnamefont {Ge}}, \bibinfo {author} {\bibfnamefont
    {F.}~\bibnamefont {Zhou}}, \bibinfo {author} {\bibfnamefont {P.}~\bibnamefont
    {Qian}}, \bibinfo {author} {\bibfnamefont {F.}~\bibnamefont {Mei}}, \bibinfo
    {author} {\bibfnamefont {S.}~\bibnamefont {Jia}}, \bibinfo {author}
    {\bibfnamefont {N.}~\bibnamefont {Xu}},\ and\ \bibinfo {author}
    {\bibfnamefont {H.}~\bibnamefont {Shen}},\ }\bibfield  {title} {\bibinfo
    {title} {Digital quantum simulation of {F}loquet topological phases with a
    solid-state quantum simulator},\ }\href {https://doi.org/10.1364/prj.404163}
    {\bibfield  {journal} {\bibinfo  {journal} {Photonics Res.}\ }\textbf
    {\bibinfo {volume} {9}},\ \bibinfo {pages} {81} (\bibinfo {year}
    {2020})}\BibitemShut {NoStop}%
  \bibitem [{\citenamefont {Zhang}\ \emph {et~al.}(2023)\citenamefont {Zhang},
    \citenamefont {Yi}, \citenamefont {Zhang}, \citenamefont {Jiao},
    \citenamefont {Shi}, \citenamefont {Yuan}, \citenamefont {Zhang},
    \citenamefont {Liu}, \citenamefont {Chen},\ and\ \citenamefont
    {Pan}}]{Zhang2023}%
    \BibitemOpen
    \bibfield  {author} {\bibinfo {author} {\bibfnamefont {J.-Y.}\ \bibnamefont
    {Zhang}}, \bibinfo {author} {\bibfnamefont {C.-R.}\ \bibnamefont {Yi}},
    \bibinfo {author} {\bibfnamefont {L.}~\bibnamefont {Zhang}}, \bibinfo
    {author} {\bibfnamefont {R.-H.}\ \bibnamefont {Jiao}}, \bibinfo {author}
    {\bibfnamefont {K.-Y.}\ \bibnamefont {Shi}}, \bibinfo {author} {\bibfnamefont
    {H.}~\bibnamefont {Yuan}}, \bibinfo {author} {\bibfnamefont {W.}~\bibnamefont
    {Zhang}}, \bibinfo {author} {\bibfnamefont {X.-J.}\ \bibnamefont {Liu}},
    \bibinfo {author} {\bibfnamefont {S.}~\bibnamefont {Chen}},\ and\ \bibinfo
    {author} {\bibfnamefont {J.-W.}\ \bibnamefont {Pan}},\ }\bibfield  {title}
    {\bibinfo {title} {{Tuning Anomalous Floquet Topological Bands with Ultracold
    Atoms}},\ }\href {https://doi.org/10.1103/PhysRevLett.130.043201} {\bibfield
    {journal} {\bibinfo  {journal} {Phys. Rev. Lett.}\ }\textbf {\bibinfo
    {volume} {130}},\ \bibinfo {pages} {043201} (\bibinfo {year}
    {2023})}\BibitemShut {NoStop}%
  \bibitem [{2-B()}]{2-BIS}%
    \BibitemOpen
    \href@noop {} {}\bibinfo {note} {For the 2D parent integer phase of a $1$D
    class D $Z_{2}$ static Hamiltonian $H_{{\rm s}}$, the $n$-th order BISs with
    $n$ vanishing Hamiltonian components are of dimensionality $(2-n)$. Hence the
    $0$D highest-order BISs for $H_{{\rm s}}$ are denoted as
    $2\text{-BISs}$.}\BibitemShut {Stop}%
  \bibitem [{\citenamefont {Eckardt}\ and\ \citenamefont
    {Anisimovas}(2015)}]{Eckardt2015}%
    \BibitemOpen
    \bibfield  {author} {\bibinfo {author} {\bibfnamefont {A.}~\bibnamefont
    {Eckardt}}\ and\ \bibinfo {author} {\bibfnamefont {E.}~\bibnamefont
    {Anisimovas}},\ }\bibfield  {title} {\bibinfo {title} {High-frequency
    approximation for periodically driven quantum systems from a {F}loquet-space
    perspective},\ }\href {https://doi.org/10.1088/1367-2630/17/9/093039}
    {\bibfield  {journal} {\bibinfo  {journal} {New J. Phys.}\ }\textbf {\bibinfo
    {volume} {17}},\ \bibinfo {pages} {093039} (\bibinfo {year}
    {2015})}\BibitemShut {NoStop}%
  \bibitem [{g_f()}]{g_field}%
    \BibitemOpen
    \href@noop {} {}\bibinfo {note} {Near the BISs, we can expand the Floquet
    Hamiltonian coefficients in terms of $k_{\perp}$ as $h_{F,z}(k)\approx
    k_{\perp}$ and $h_{F,i}(k)\approx
    h_{F,i}\vert_{2\text{-BIS}}+\mathcal{O}(k_{\perp})$ for $i=x,y$. Then the
    dynamical field is given by
    $g_{i}\propto\lim_{k_{\perp}\to0}(1/2k_{\perp})\{(2k_{\perp})[h_{F,i}+\mathcal{O}(k_{\perp})]/[|\boldsymbol{h}_{F}|^{2}+\mathcal{O}(k_{\perp})]\}=h_{F,i}/|\boldsymbol{h}_{F}|^{2}$.}\BibitemShut
    {Stop}%
  \bibitem [{Z2_()}]{Z2_descendants}%
    \BibitemOpen
    \href@noop {} {}\bibinfo {note} {The $d$D first (second) descendant $Z_{2}$
    topological phases can be derived as lower-dimensional descendants of the
    $d'$D parent $Z$ topological phases with dimensionality being reduced by $1$
    (or $2$); see Refs.~\citep{Qi2008,Ryu2010}.}\BibitemShut {Stop}%
  \bibitem [{\citenamefont {Qi}\ \emph {et~al.}(2008)\citenamefont {Qi},
    \citenamefont {Hughes},\ and\ \citenamefont {Zhang}}]{Qi2008}%
    \BibitemOpen
    \bibfield  {author} {\bibinfo {author} {\bibfnamefont {X.-L.}\ \bibnamefont
    {Qi}}, \bibinfo {author} {\bibfnamefont {T.~L.}\ \bibnamefont {Hughes}},\
    and\ \bibinfo {author} {\bibfnamefont {S.-C.}\ \bibnamefont {Zhang}},\
    }\bibfield  {title} {\bibinfo {title} {Topological field theory of
    time-reversal invariant insulators},\ }\href
    {https://doi.org/10.1103/physrevb.78.195424} {\bibfield  {journal} {\bibinfo
    {journal} {Phys. Rev. B}\ }\textbf {\bibinfo {volume} {78}},\ \bibinfo
    {pages} {195424} (\bibinfo {year} {2008})}\BibitemShut {NoStop}%
  \bibitem [{\citenamefont {Price}\ \emph {et~al.}(2015)\citenamefont {Price},
    \citenamefont {Zilberberg}, \citenamefont {Ozawa}, \citenamefont
    {Carusotto},\ and\ \citenamefont {Goldman}}]{Price2015}%
    \BibitemOpen
    \bibfield  {author} {\bibinfo {author} {\bibfnamefont {H.~M.}\ \bibnamefont
    {Price}}, \bibinfo {author} {\bibfnamefont {O.}~\bibnamefont {Zilberberg}},
    \bibinfo {author} {\bibfnamefont {T.}~\bibnamefont {Ozawa}}, \bibinfo
    {author} {\bibfnamefont {I.}~\bibnamefont {Carusotto}},\ and\ \bibinfo
    {author} {\bibfnamefont {N.}~\bibnamefont {Goldman}},\ }\bibfield  {title}
    {\bibinfo {title} {Four-{D}imensional {Q}uantum {H}all {E}ffect with
    {U}ltracold {A}toms},\ }\href
    {https://doi.org/10.1103/physrevlett.115.195303} {\bibfield  {journal}
    {\bibinfo  {journal} {Phys. Rev. Lett.}\ }\textbf {\bibinfo {volume} {115}},\
    \bibinfo {pages} {195303} (\bibinfo {year} {2015})}\BibitemShut {NoStop}%
  \bibitem [{\citenamefont {Ozawa}\ \emph {et~al.}(2016)\citenamefont {Ozawa},
    \citenamefont {Price}, \citenamefont {Goldman}, \citenamefont {Zilberberg},\
    and\ \citenamefont {Carusotto}}]{Ozawa2016}%
    \BibitemOpen
    \bibfield  {author} {\bibinfo {author} {\bibfnamefont {T.}~\bibnamefont
    {Ozawa}}, \bibinfo {author} {\bibfnamefont {H.~M.}\ \bibnamefont {Price}},
    \bibinfo {author} {\bibfnamefont {N.}~\bibnamefont {Goldman}}, \bibinfo
    {author} {\bibfnamefont {O.}~\bibnamefont {Zilberberg}},\ and\ \bibinfo
    {author} {\bibfnamefont {I.}~\bibnamefont {Carusotto}},\ }\bibfield  {title}
    {\bibinfo {title} {Synthetic dimensions in integrated photonics: {F}rom
    optical isolation to four-dimensional quantum {H}all physics},\ }\href
    {https://doi.org/10.1103/physreva.93.043827} {\bibfield  {journal} {\bibinfo
    {journal} {Phys. Rev. A}\ }\textbf {\bibinfo {volume} {93}},\ \bibinfo
    {pages} {043827} (\bibinfo {year} {2016})}\BibitemShut {NoStop}%
  \bibitem [{\citenamefont {Yuan}\ \emph {et~al.}(2018)\citenamefont {Yuan},
    \citenamefont {Lin}, \citenamefont {Xiao},\ and\ \citenamefont
    {Fan}}]{Yuan2018}%
    \BibitemOpen
    \bibfield  {author} {\bibinfo {author} {\bibfnamefont {L.}~\bibnamefont
    {Yuan}}, \bibinfo {author} {\bibfnamefont {Q.}~\bibnamefont {Lin}}, \bibinfo
    {author} {\bibfnamefont {M.}~\bibnamefont {Xiao}},\ and\ \bibinfo {author}
    {\bibfnamefont {S.}~\bibnamefont {Fan}},\ }\bibfield  {title} {\bibinfo
    {title} {Synthetic dimension in photonics},\ }\href
    {https://doi.org/10.1364/optica.5.001396} {\bibfield  {journal} {\bibinfo
    {journal} {Optica}\ }\textbf {\bibinfo {volume} {5}},\ \bibinfo {pages}
    {1396} (\bibinfo {year} {2018})}\BibitemShut {NoStop}%
  \bibitem [{\citenamefont {Morimoto}\ \emph {et~al.}(2017)\citenamefont
    {Morimoto}, \citenamefont {Po},\ and\ \citenamefont
    {Vishwanath}}]{Morimoto2017}%
    \BibitemOpen
    \bibfield  {author} {\bibinfo {author} {\bibfnamefont {T.}~\bibnamefont
    {Morimoto}}, \bibinfo {author} {\bibfnamefont {H.~C.}\ \bibnamefont {Po}},\
    and\ \bibinfo {author} {\bibfnamefont {A.}~\bibnamefont {Vishwanath}},\
    }\bibfield  {title} {\bibinfo {title} {{Floquet topological phases protected
    by time glide symmetry}},\ }\href
    {https://doi.org/10.1103/PhysRevB.95.195155} {\bibfield  {journal} {\bibinfo
    {journal} {Phys. Rev. B}\ }\textbf {\bibinfo {volume} {95}},\ \bibinfo
    {pages} {195155} (\bibinfo {year} {2017})}\BibitemShut {NoStop}%
  \bibitem [{\citenamefont {Xu}\ and\ \citenamefont {Wu}(2018)}]{Xu2018}%
    \BibitemOpen
    \bibfield  {author} {\bibinfo {author} {\bibfnamefont {S.}~\bibnamefont
    {Xu}}\ and\ \bibinfo {author} {\bibfnamefont {C.}~\bibnamefont {Wu}},\
    }\bibfield  {title} {\bibinfo {title} {{Space-Time Crystal and Space-Time
    Group}},\ }\href {https://doi.org/10.1103/PhysRevLett.120.096401} {\bibfield
    {journal} {\bibinfo  {journal} {Phys. Rev. Lett.}\ }\textbf {\bibinfo
    {volume} {120}},\ \bibinfo {pages} {096401} (\bibinfo {year}
    {2018})}\BibitemShut {NoStop}%
  \bibitem [{\citenamefont {Yu}\ \emph {et~al.}(2021{\natexlab{c}})\citenamefont
    {Yu}, \citenamefont {Zhang},\ and\ \citenamefont {Song}}]{Yu2021b}%
    \BibitemOpen
    \bibfield  {author} {\bibinfo {author} {\bibfnamefont {J.}~\bibnamefont
    {Yu}}, \bibinfo {author} {\bibfnamefont {R.-X.}\ \bibnamefont {Zhang}},\ and\
    \bibinfo {author} {\bibfnamefont {Z.-D.}\ \bibnamefont {Song}},\ }\bibfield
    {title} {\bibinfo {title} {{Dynamical symmetry indicators for Floquet
    crystals}},\ }\href {https://doi.org/10.1038/s41467-021-26092-3} {\bibfield
    {journal} {\bibinfo  {journal} {Nat. Commun.}\ }\textbf {\bibinfo {volume}
    {12}},\ \bibinfo {pages} {5985} (\bibinfo {year}
    {2021}{\natexlab{c}})}\BibitemShut {NoStop}%
  \bibitem [{\citenamefont {Bernevig}\ and\ \citenamefont
    {Hughes}(2013)}]{Bernevig2013}%
    \BibitemOpen
    \bibfield  {author} {\bibinfo {author} {\bibfnamefont {B.~A.}\ \bibnamefont
    {Bernevig}}\ and\ \bibinfo {author} {\bibfnamefont {T.~L.}\ \bibnamefont
    {Hughes}},\ }\href {https://doi.org/10.1515/9781400846733} {\emph {\bibinfo
    {title} {Topological {I}nsulators and {T}opological {S}uperconductors}}}\
    (\bibinfo  {publisher} {Princeton University Press},\ \bibinfo {year}
    {2013})\BibitemShut {NoStop}%
  \end{thebibliography}

%
  
\end{document}